UNIVERSITÀ DEGLI STUDI DI ROMA
TOR VERGATA

Physics Department

PHYSICS PhD PROGRAM
(XXXV cycle)
A.Y. 2022/2023

*Thesis title:*

# Characterization and ground calibration of the Electric Field Detector aboard the CSES-02/LIMADOU mission

**SUPERVISOR**

Prof.ssa Roberta Sparvoli

**CANDIDATE**

Gianmaria Rebustini

**CO-SUPERVISOR**

Dr. Davide Badoni

**COORDINATOR**

Prof. Massimo Bianchi

*“Nihil difficile volenti”*

# INDEX OF THE THESIS

# INTRODUCTION

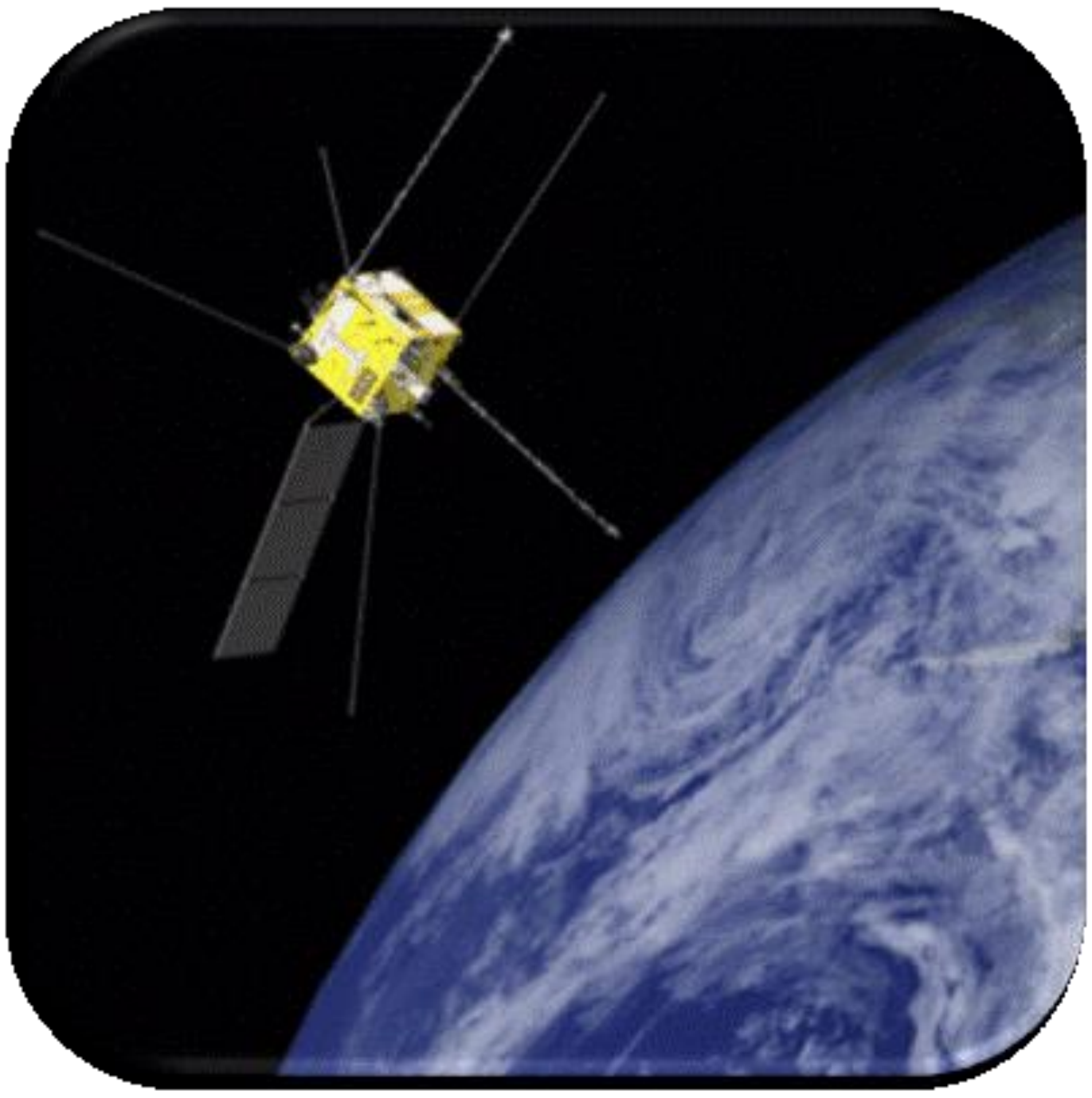

## THE MISSION

Many observations in space have been reported, over the past fifty years, about signals traced back to earthquakes and claimed as possible pre-seismic measurements.

The analyses have been carried out on datasets gathered by satellites non-specifically devoted to these observations. The first mission dedicated to this purpose was the DEMETER mission, which laid the foundations for space-based investigations of seismo-associated phenomena, drawing a baseline for next missions as to instruments, observational strategy, and measurements uncertainty [1].

The China Seismo-Electromagnetic Satellite (CSES) mission, successfully operative since February 2018 with the launch of CSES-01, is a China-Italy collaboration with the participation of Austrian Institutes; the mission aims to study the space environment around the Earth in order to find evidence of possible correlations, both in spatial and temporal terms, between the occurrence of seismic events and the observation of iono-magnetospheric perturbations as well as precipitation of particles from the inner Van Allen belts.

CSES monitors the dynamics of the top-side ionospheric electromagnetic field, plasma and particle distribution, the coupling mechanisms between upper atmosphere, ionosphere and magnetosphere and the temporal variations of the geomagnetic field, in quiet and disturbed conditions.

Data collected by the mission also allows to study solar-terrestrial interactions and phenomena of solar physics, namely Coronal Mass Ejections (CMEs), solar flares and cosmic ray solar modulation. CSES represents the most advanced mission for investigating the near-Earth electromagnetic environment, the extension of seismic monitoring to long time intervals. [1]

The CSES-02 mission foresees the launch of a second satellite, scheduled by the end of 2023, with an expected lifetime of 6 years, after injection in the same sun-synchronous low Earth orbit (LEO) with a phase shift of 180° with respect to the first satellite.

This second launch is meant for optimization of the temporal overlap with the first satellite, and improvement of data quality.

## PURPOSE AND THESIS STATEMENTS

In this doctoral thesis the focus will be on the Electric Field Detector (EFD) which is scheduled for integration in the next CSES-02 satellite, with special emphasis on instrumental characteristics and on-ground calibration procedures. EFD-02 will measure the electric field components in the ionosphere over a wide frequency band (from $DC\ to\ 3.5\ MHz$) and with high sensitivity of about 1 $\mu V/m$ in the Ultra-Low Frequency (ULF) band.

The biggest part of the thesis work has consisted in the evaluation of a matrix of conversion factors (CFs) between analog signals applied to the Electric Field Probes (EFPs) and the digitized output from the acquisition system, considering CF variability upon variable frequency and amplitude of the input signal. This procedure has provided an accurate calibration of the system. These measurements are extremely important since they enable data interpretation at a scientific level, when passing from raw to higher-level data (electric field). To make them more consistent, verification tests carried out in a realistic ionospheric plasma environment for a satellite in LEO orbit have been accurately defined.

In the following, a thorough description of the entire set of measurements mentioned above is reported for all instrumental stages from prototype to final flight model (FM).

## THESIS OUTLINE

The manuscript is organized as follows: the first chapter, accommodates a general description of some of the results from earlier missions dedicated to the monitoring of earthquake precursors from space. The most popular theories to associate possible seismic precursors measured in space with the occurrence of earthquakes are highlighted. Also, at the end of the chapter, a brief recall of the main theoretical concepts regarding sensors immersed in plasma are exposed.

In chapter 2, a detailed description of the CSES mission, both CSES-01 and especially CSES-02, is given, as well as of the essential components of EFD-02 to underline current novelties with respect to previous experiments that had the same target.

The first part of Chapter 3 describes the functionality tests, designed to guarantee the correct electronic operation of the instrument under different conditions of temperature, humidity and gamma irradiation. The various performance tests are shown and discussed in the second part of Chapter 3, to highlight the characteristics of the instrument from the point of view of: Linearity, Dynamic Range, Bandwidth, Signal / Noise ratio.

Results from calibration using a signal generator in a dedicated lab at INFN-Tor Vergata, as well as the characterization of the instrument with a plasma source – in Plasma Chamber at INAF-IAPS, are reported in Chapter 4. In this last chapter, the variability of the calibration factor will also be analyzed, considering various control parameters. A final section explains how and why the tests and measurements described here have a fundamental value for the CSES-02 mission, at both an operational and scientific level.

## AUTHOR'S CONTRIBUTION

In my 3-year PhD activity, within the CSES-Limadou project, the characterization and calibration of the EFD-02 electric field detector has played a central role.

In the first year of my PhD, I joined the final design phase of the Analog Processing Unit (APU) board, using simulation software to verify the functionality of some analog sections of the board and I directly took part in the various tests devised to characterize the board's electronics and to study in detail the various arrangements of the analog chains already designed. Over that period, I performed tests in a Climatic Chamber to verify the electronic stability of one of the analog chains, the most susceptible one to changing temperature.

Once this phase was completed, I helped design the tests intended to verify the functionality and performance of the APU board sent to production, and I personally performed most of the acceptance tests of the boards produced.

In Dec 2020, I coauthored a publication about the first satellite, entitled "The Electric Field Detector on Board the China Seismo Electromagnetic Satellite. In-Orbit Results and Validation" [69].

In the second year of my doctoral program, I participated in the 107th national congress of the Italian Physical Society (SIF) with a presentation of my work, which was selected for publication in the journal "Il Nuovo Cimento" [84].

Still in this year, I carried out radiation resistance tests of the APU EM prototype by gamma irradiation at the Calliope Gamma Irradiation Facility of ENEA Casaccia to obtain a heritage useful for space missions and I conducted a test campaign for board calibration of the EM model first and the QM model later.

In the last year of my PhD, I have characterized and calibrated the complete QM and FM system including all the board and parts that make up the EFD instrument, such as the EFPs. I have also helped develop some parts of the software needed for the digital processing of signal from the DPU board. In this third year, I have taken part in the measurements carried out in the plasma chamber at INAF-IAPS, which have proven crucial in assessing instrumental performance at the scientific level.

# CHAPTER 1

## GENERAL OVERVIEW OF THE RESULTS FROM PREVIOUS MISSIONS

*In this chapter, it will be presented a general description of some major result from earlier missions dedicated to the study of seismic precursors from space.*

*Since the end of the 70's, growing evidence has been collected in space of signals related to earthquakes and considered as possible pre-seismic measurements. The observables under investigation have been including electromagnetic field components (in a large band of frequencies), plasmas parameters, particle fluxes, thermal anomalies, etc. The earliest datasets were gathered by satellites not properly devoted to seismic investigations. The DEMETER mission, instead, was the first one dedicated to this purpose and it laid the foundations for space-based investigations of seismo-associated phenomena, thus paving the way for next missions as to instruments and observational strategy.*

*The best way to study processes that lead to the preparation of an earthquake is strictly related to the possibility of identifying seismic precursors. In addition to a classification as a function of time delay with respect to the seismic event, precursors can be further distinguished on the spatial scale as a function of the detection distance and their localization or diffusion. Indeed, some alleged precursors can be detected around the seismic focal area (local precursors) even though eventually at significant distance. Due to the topology of the geomagnetic field, other possible precursors can be detected not only over the epicenter but also near to its magnetically conjugated region or along the field line with footprint over the epicenter (diffused precursors).*

*Finally, a further class of precursors could be constituted by fluctuations detectable not only along a geomagnetic flux tube associated to the epicenter but spread in a suitable iono-magnetospheric "shell" (distributed precursors). Whereas the co-seismic effects in the atmosphere are well-established [2] the possible pre-earthquake phenomena on the surface as well as the coupling between lithosphere, atmosphere, and ionosphere (called LAIC for short, in the following sections) are still disputed (e.g., [3, 4, 5] and references therein).*

*The physical processes involved to explain the various observations are still unable to describe the whole mechanism. The statistical validity of these results is still much debated, and a full consensus is still lacking.*

## 1.1 SOME CONSIDERATIONS FROM MISSIONS NON-DEDICATED TO SEISMIC INVESTIGATIONS

Seismic events are the last stage of a long preparation process generated by a continuous and variable tectonic stress. Many attempts have been done to monitor on ground the earthquake preparation phase and the underlying physical processes on specific fault systems, but the involved processes are deep, slow, and complex.

The possibility of remote sensing earthquakes, through their effects in near-Earth space, has been explored for decades.

To explain the effects of the LAIC, there is a model [7] - based on the rising of gas and fluid toward the surface in the seismic preparation phase that considers phenomena belonging to the last stages of the long-term seismic phase.

Other conjectures ([1] and reference therein) propose a mechanism - successfully tested in a laboratory environment [8] - which is based on the theory of positive holes (lack of electrons) that could locally ionize the lower atmosphere and create instability in the ionosphere. Finally, the authors of [9, 10] propose a coupling mechanism that, via the effect of the geomagnetic field, would induce perturbation in the ionosphere.

Many authors have reported measurements of seismic electromagnetic precursors detected on ground or in space, mainly focused on analyzing electric and magnetic field variations. Other studies have discussed fluctuations of plasma parameters, precipitation of high-energy charged particles from the inner Van Allen belt, etc. (mentioned later).

## 1.2 EARLY OBSERVATIONS

The first results from satellite surveys of low-frequency electromagnetic emissions before earthquakes were obtained via the Intercosmos 19 and Aureol 3 missions, while the detection of VLF noise in the region magnetically conjugate to an earthquake zone was reported by the geostationary GEOS 1 and GEOS 2 satellites. The data from Intercosmos 19 satellite, were detected at 800 and 4,650 Hz, from about 8 h before up to about 3 h after any quake, within 2 degrees of latitude and 60 degrees of longitude around the epicenters. The Intercosmos-Bulgaria 1300 satellite measured an anomalous variation of 3–7 mV/m in the quasi-DC component of the vertical electric field at the altitudes of about 800 km over the magnetically conjugate zone of the epicenter, about 15 min before an event of magnitude 4.8 in the Pacific Ocean ([1] and reference therein).

A few years later, COSMOS-1809 detected anomalous electromagnetic emissions at frequencies below 450 Hz, up to a few hours before the seismic event, in more than 92% of the satellite traces, within 6 degrees of longitude from the epicenter and about 4–10° of latitude south of Armenia in 1988.

The AUREOL-3 satellite data confirmed such observation. By studying the seismic sequence of the Armenia earthquake of 1988, some anomalous fluctuations in the ULF magnetic and electric field measurements recorded on ground have been reported about 200 km far from the epicenter and some hours before the main event and some aftershocks.

More recently, by re-analyzing data from Intercosmos-Bulgaria 1300 for hundreds of earthquakes (Gousheva et al.,2008; Gousheva et al., 2009) [66, 67], it has been estimated that the amplitude of pre-seismic quasi-DC electric field disturbances in space was of the order of 10 mV/m over seismic regions both in land and in sea.

## 1.3 THE RADON COUPLING MECHANISM

From space, several seismo-associated parameters can be measured, such as lithospheric deformations, temperature fluctuation, gas and aerosol exhalation and variations in the local electromagnetic field. The pre-seismic deformations are lower and more complex to investigate from space with respect to their co-seismic counterparts.

Remote sensing observations allow to measure concentrations of gases and aerosol in the atmosphere potentially involved in pre-seismic phenomena.

One of the most debated issues in the physics of earthquake precursors includes the role of the seismo-induced radon exhalation in the generation of electromagnetic disturbances [12, 13, 14]. It has been proposed that the enhancement of the total rock surface due to failure would increase the emissions of radon and other gases from grains and migration [15]. This would agree with:

1) the enhancement of radon concentration observed in aftershocks.
2) some laboratory experiments [16] aimed at verifying the growth of radon emissions of granites under compressional stress.

It has been claimed that, due to radon exhalation from the soil, local fair-weather conductivity could increase up to 50%, whereas the electric field could decrease by 30% [17].

According to this LAIC model, the relative movement of tectonic blocks leads to the generation of tectonic stresses with the release of gases (including radon) along seismically active faults. Radon can generate local ionization in the lower layers of the atmosphere that can facilitate water vapor condensation with:

1) release of latent heat (which could explain thermal fluctuations).
2) local variations in the conductivity, which would impact the global electric circuit over the earthquake preparation zone, thus generating the observed seismo-associated ionospheric anomalies.

## 1.4 IONOSPHERIC DISTURBANCES

The large variety of atmospheric, ionospheric, and magnetospheric anomalies - claimed as possibly related to earthquakes - shows the importance of both ground-based measurements of ionospheric parameters and satellite-based remote sensing for the investigation of earthquake precursors.

Anyway, it must be highlighted that many studies of seismo-ionospheric precursors are case studies (seldom reproduced in further investigations carried out in “similar” conditions) and that the statistical significance is low in most cases.

## 1.5 Thermal Anomalies

Several authors have suggested that the earthquake preparation process can generate temperature variations (induced by flow/exhalation of geochemical fluids in the deep lithosphere and/or by secondary effects of the friction and displacement along seismic faults) that can affect the energy budget in the LAIC. Pre-seismic processes could result in the release of radon and optically active gases (including carbon dioxide and methane) whose concentration could influence the thermal radiation emitted from the ground [18].

In recent years, the possibility to identify, on a global scale, seismo-associated thermal anomalies has been enormously facilitated by continuous satellite monitoring. Phenomena generically called thermal anomalies usually refer to anomalous fluctuations in several different parameters such as the atmospheric temperature (at various altitudes), Brightness Temperature (BT), Surface Latent Heat Flux (SLHF), Outgoing Longwave Radiation (OLR), etc.

## 1.6 Acoustic Gravity Waves

It has been proposed that acoustic (AW) and acoustic gravity waves (AGW) could be responsible for the coupling between lithospheric processes and tropo-ionospheric disturbances [19].

Therefore, it is particularly relevant that several studies support the common idea that at the bottom of the ionosphere, above the epicenter of an impending earthquake, there is a statistically relevant excess of ionization. Changes in ground motion and/or temperature and pressure would induce oscillations in the atmosphere over the earthquake preparation zone, which can propagate up to the ionosphere.

The hypothesis of a correlation between pre-seismic processes, tropo-ionospheric oscillations and thermal fluctuation has been supported also by the studies reported in [20].

The authors have found three correlated phenomena:

1) an anomalous fluctuation of the OLR satellite data (which can be related to thermal anomalies).
2) the occurrence of an AGW with a period of about 1 h.
3) ground based detection of VLF disturbances.

## 1.7 Electromagnetic Field Measurements

In [21] two series of electric field measurements before the Wenchuan earthquake of May 12, 2008, have been analyzed: a long time series (since March 2008) detected on ground, and a shorter series (1–2 days) measured by ICE on board the DEMETER satellite [82]. The time occurrence and spatial distribution of ground- and space-based measurements are mutually consistent for long and brief time series. The authors have found that the amplitude of measured electric field anomalies (detected a few days before the earthquake was: from about $3\,mV/km$ up to $100\,mV/km$ on ground, and about $3\text{–}5\,mV/m$ at frequencies $< 0.5\,Hz$ (relative variation $> 4\%$) from space [21]. The amplitudes differ largely between ground and satellite observations, so that it is difficult to reconcile ground and space-based measurements via a direct propagation of electromagnetic waves from the lithosphere through the atmosphere up to the ionosphere.

## 1.8 Particle Precipitations from the Van Allen Belts

In the Earth's magnetosphere, the magnetic field can trap charged particles (electron, positrons, protons, and ions) up to energies of tens or hundreds of MeV, generating the so-called inner and outer radiation belts separated by the slot region. This latter zone is originated by particle flux depletion due to interactions with whistler waves.

Particles are defined trapped if they are bound in the dipolar terrestrial field because of mirroring in the "magnetic bottle" (see Figure 1). The motion of trapped particles consists of 3 periodic and simultaneous movements: gyration around a magnetic field line, north-south bouncing between conjugate points along a field line, and slow longitudinal drift around Earth, towards west and east for respectively positive and negative charged particles. The resulting trajectory lies on a toroidal surface, called drift shell, centered on the Earth's dipole center. Particles confined in a drift shell can also remain there for long periods, even years for protons at altitude of some thousand kilometers.

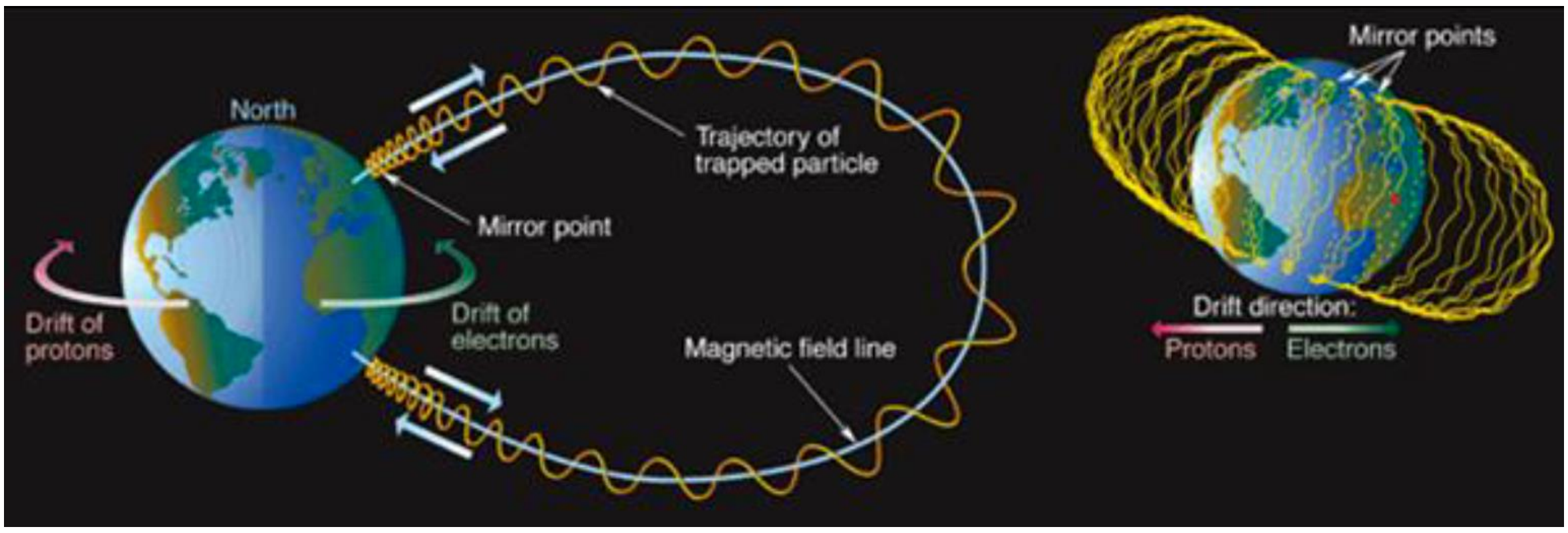

Figure 1 - Particle drift in Van Allen belts [21].

The inner Van Allen belt is mainly constituted by protons through the decay of albedo neutrons [21] originated by cosmic rays impinging the upper atmosphere (CRAND process).

Electrons are the main constituent of the outer belts occasionally energized through wave-particle interactions. Whistlers are the main source of particles precipitation via the so-called Whistler Electron Precipitation (WEP) due to the resonant interaction between circularly polarized VLF (3–30 kHz) waves traveling along the geomagnetic field lines and trapped electrons resulting in particle deflection in the loss cone and the consequent precipitation.

The Sun is the main driver of the magnetospheric particle dynamics, but precipitation can be induced also by nuclear explosions (that can originate also long-term artificial belts) and the already cited VLF emissions generated by lightning, artificial radio signals and possible electromagnetic emission due to seismic activity. Particle precipitations can be observed by satellite detectors as sudden increases in local particle flux on the time scale from few up to tens of seconds.

It has been suggested that the stable motion of high-energy trapped and quasi-trapped Van Allen particles can be also perturbed by seismo-associated electromagnetic emissions [22, 23]. These authors suggest that electromagnetic emissions possibly generated in the preparation phase of an earthquake could modify the pitch angle of any particle, inducing the lowering of its mirror points and final precipitation, which is detected as a sudden particle flux increase by LEO detectors. During precipitation, such bursts of particles could still partially follow their longitudinal drift, which would increase the satellite capability of detection in space not only over the hypo-central zones, but also far from the area of the earthquake preparation.

## 1.9 Remote Sensing and Multi-Parametric Approach

Remote sensing makes it possible to monitor precursor signatures via the simultaneous variation of several physical variables (above the epicenter, around it and in its conjugate zone), but also to survey large areas that could be affected by the earthquake preparation process but cannot be monitored with the network of scattered ground stations. In this framework, the study reported in [24] has highlighted that, because of the peak of pre-seismic radon exhalation occurring 4–10 days before the earthquake, the time scale of radon variations and that of the observed air temperature variations on the occasion of the Colima (Mexico) earthquake of 2003 are comparable with results from multi-parametric analysis of the anomalies in surface temperature, latent heat flux, air temperature and relative humidity observed before.

## 1.10 Demeter satellite observations

The French DEMETER (Detection of Electro-Magnetic Emissions Transmitted from Earthquake Regions) satellite was the first one specifically devoted to the investigation of seismo-electromagnetic and volcanic phenomena [25]. It was launched on June 29, 2004, on a quasi-Sun-synchronous circular orbit with an inclination of about 98.23° and an altitude of about 710 km. The altitude was changed to about 660 km in December 2005. The satellite would perform 14 orbits per day and measures continuously between -65° and +65° of invariant latitude.

The DEMETER scientific payloads were: the IMSC detector, composed of three orthogonal magnetic sensors operating on a range from a few Hz up to 18 kHz; the ICE detector, comprising four spherical electrodes with ability to measure signals from DC up to 3.5 $MHz$; the IAP ion analyzer; the ISL Langmuir probe and the IDP, high-energy particle detector.

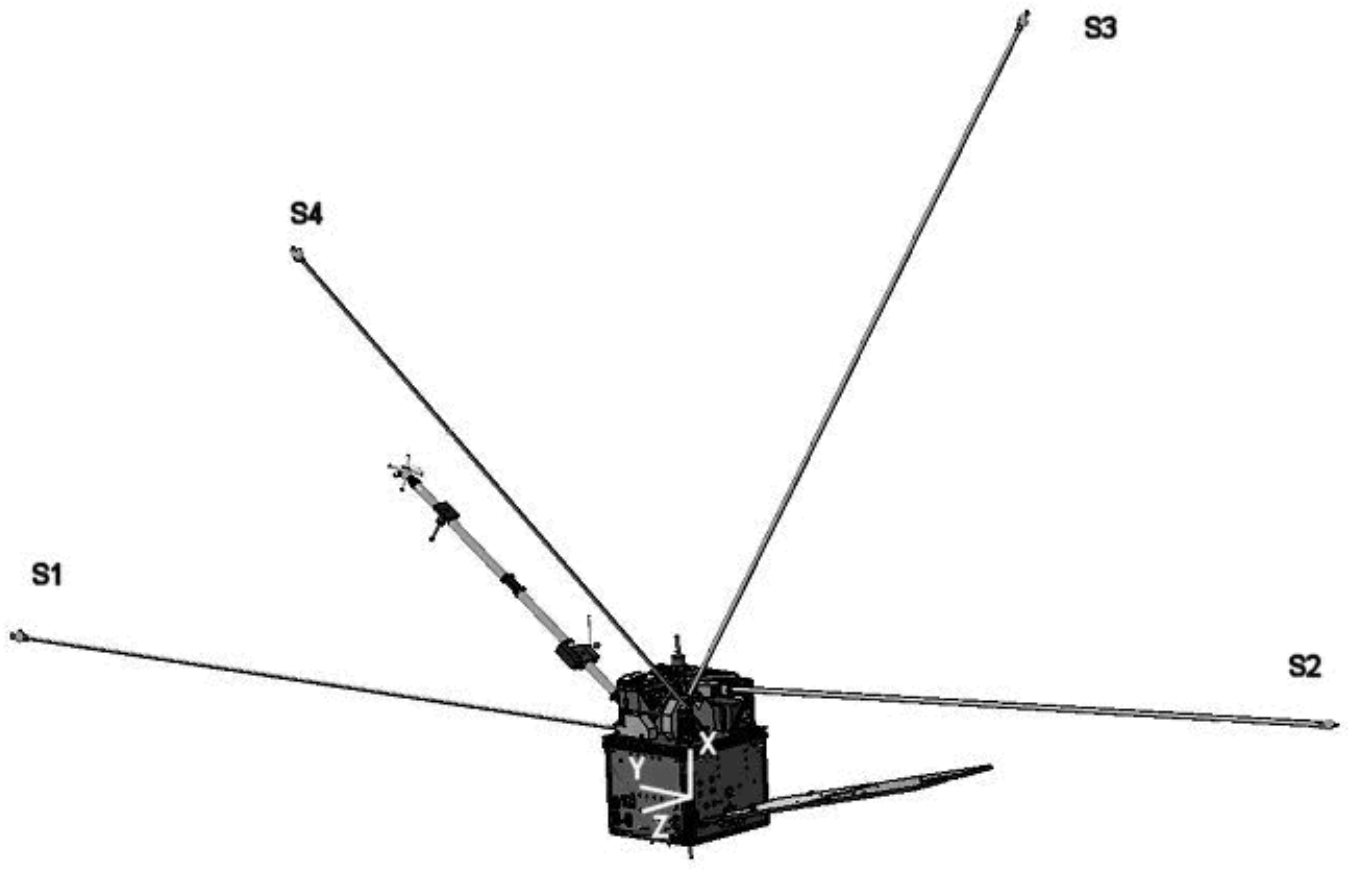

Figure 2 - Position of the ICE sensors on the spacecraft [85]

The analysis of quasi-static electric field data detected by DEMETER during night time for high magnitude earthquakes of Indonesia and Chile regions [26], reported perturbations from 1.5 to 16 mV/m, prior to a set of 27 earthquakes, either over the epicenter or at the end of seismic faults within 2000 km from the epicenter.

The most significant result obtained with the DEMETER data is the statistical analysis - via the superposed epoch and space method - of the disturbances in the electric field power spectrum density (PSD) measured by the ICE experiment - Figure 2 - as a function of the seismic activity [82].

Several analyses of DEMETER satellite data have shown an increase in the number and intensity of the ionospheric perturbations detected before the occurrence of strong earthquakes as well as an increase in the amplitude of perturbations as a function of quake magnitude. In particular, the study in [27] has pointed out a significant increase in the plasma density detected by DEMETER tens of days before the main shock of the Chile earthquake on February 27, 2010.

In [28], an analysis is reported of the ion density (defined as the sum of H+, He+, O+ from IAP) and the electron density data collected over 6.5 years of DEMETER activity and registered during night time. Ion density mainly tends to increase. The fraction of precursor observations increases by a few percent with the earthquake magnitude. The mean number of perturbations per earthquake is larger for stronger events.

From DEMETER data it is possible to draw some conclusions:

- Variability of electromagnetic precursors: the interpretation of the claimed electromagnetic precursors is promising, but still at an early stage. There is a large variability in the detected intensity, frequencies, spatial and temporal distribution, spreading or clustering around the epicenter or along geomagnetically connected areas, etc. After the early sparse observations, even with the most recent devoted missions such as DEMETER, CSES, and FORMOSAT the phenomenology is still barely understood.
- Information about relation between ground and space: based on the observations of an enhancement of VLF fluctuations in the range of acoustic gravity waves measured during some earthquakes, the link between AW/AGW and VLF disturbances seems quite well assessed. On the other hand, the conclusion that the observed phenomenology has a pre-seismic character asks for further confirmations, because the connection is still indirect. Several reports are in favor of a LAIC due to the chain of ionization, changes in conductivity and feedback phenomena originated by radon emissions.
- Correlation between earthquake magnitude, depth, and amplitude of possible precursors: the published analyses about groundwater level variations and radon gas exhalations seem to suggest a correlation between the earthquake magnitude and the amplitude and spatial-temporal distributions of claimed precursors. Reports about electromagnetic precursors seem not univocal, but the variation of the electric field intensity at the ionospheric cutoff is more intense when the magnitude is higher. Nevertheless, it is possible to assert that magnitude should be a key parameter in precursor identification. A similar role is played by the hypo-central depth: most reports of seismic precursors concern shallow earthquakes, although there is no clear or unique threshold for depth, partly because differences in the specific seismic-tectonic conditions of different areas should not be overlooked.

- Extent of the spatial scale of the precursors: the distance of ionospheric precursors could be correlated with magnitude. However, an estimation is still missing of the area that could be involved in the generation mechanism of various (not only mechanical) earthquake-precursors as well as of the extension of detectability region in which the signal to noise ratio would allow a reliable precursor recognition.
- Temporal advance and clustering of anomalous observations: the largest electromagnetic anomalies (measured hours or days before large events) seem to occur more frequently in time and with larger amplitude close to the incoming earthquake.

Moreover, it is worth stressing the recent hypothesis (with first confirmations) that also ionospheric precursors would follow the Rikitake's law according to which larger earthquakes should be associated with longer precursor times. This could be reconciled with the critical nature of the process originating the earthquake in the preparation phase along the fault before the rupture.

The variety of phenomena associated with earthquakes requires the simultaneous observation of many parameters. The need for statistical studies that increase the reliability of results by reducing background effects asks for increasing the number of seismic events analyzed worldwide. Both requirements, for a global coverage system, can only be met from space through satellite remote sensing. Considering this scenario, new multi-spacecraft missions acquire particular scientific value.

## 1.11 Lithosphere - Atmosphere - Ionosphere Coupling (LAIC)

The short-term prediction of earthquakes is an essential issue connected with human life protection and related social and economic matters.

In the first part of this subsection, the mechanisms proposed so far to describe the interaction between the various atmospheric layers and seismic phenomena will be briefly exposed. In the second part, the theories concerning the interactions between plasma and sensors immersed in the plasma will be described, because they will be useful in chapter 4 to better understand the results of tests performed in the plasma chamber, under flight-like conditions.

The study of these processes is of fundamental importance for the understanding of the Earth System, the monitoring of the electromagnetic environment near the Earth, and the study of natural disasters, such as earthquakes. Moreover, it is of interest that the behavior of the transition region between the ionosphere and the magnetosphere is not yet fully known. The coupling mechanisms between the Lithosphere, the Atmosphere and the Ionosphere (called LAIC for short) constitute a complex issue, and very vast, involving many physical phenomena and interactions.

It's out of the reach of current scientific knowledge the deterministic prediction of the time and place of a seismic event. On the other hand, a statistical estimate of the probability of seismic events is possible based on historical series and observations of the geophysical characteristics of the different areas.

The creation of seismic hazard maps of the territory and their use for the adaptation of infrastructures remain to date the only method for reducing the damage caused by earthquakes and volcanic eruptions.

In fact, much remains to be understood about the physics of seismic and volcanic processes. The mechanical and thermal effects of deformation processes associated with earthquakes and volcano eruptions have long been known. More recently, the electromagnetic processes that are believed to accompany the deformation and breaking of the rocks that generate earthquakes have been studied. Electromagnetic emissions have been observed during laboratory experiments on rock samples subjected to strong pressures near the moment of rupture of the materials. [29]
Observations of electromagnetic disturbances, gas emissions (radon) and thermal perturbations observed by ground monitoring networks in relation to seismic events, especially of great magnitude, have been accumulating over time.

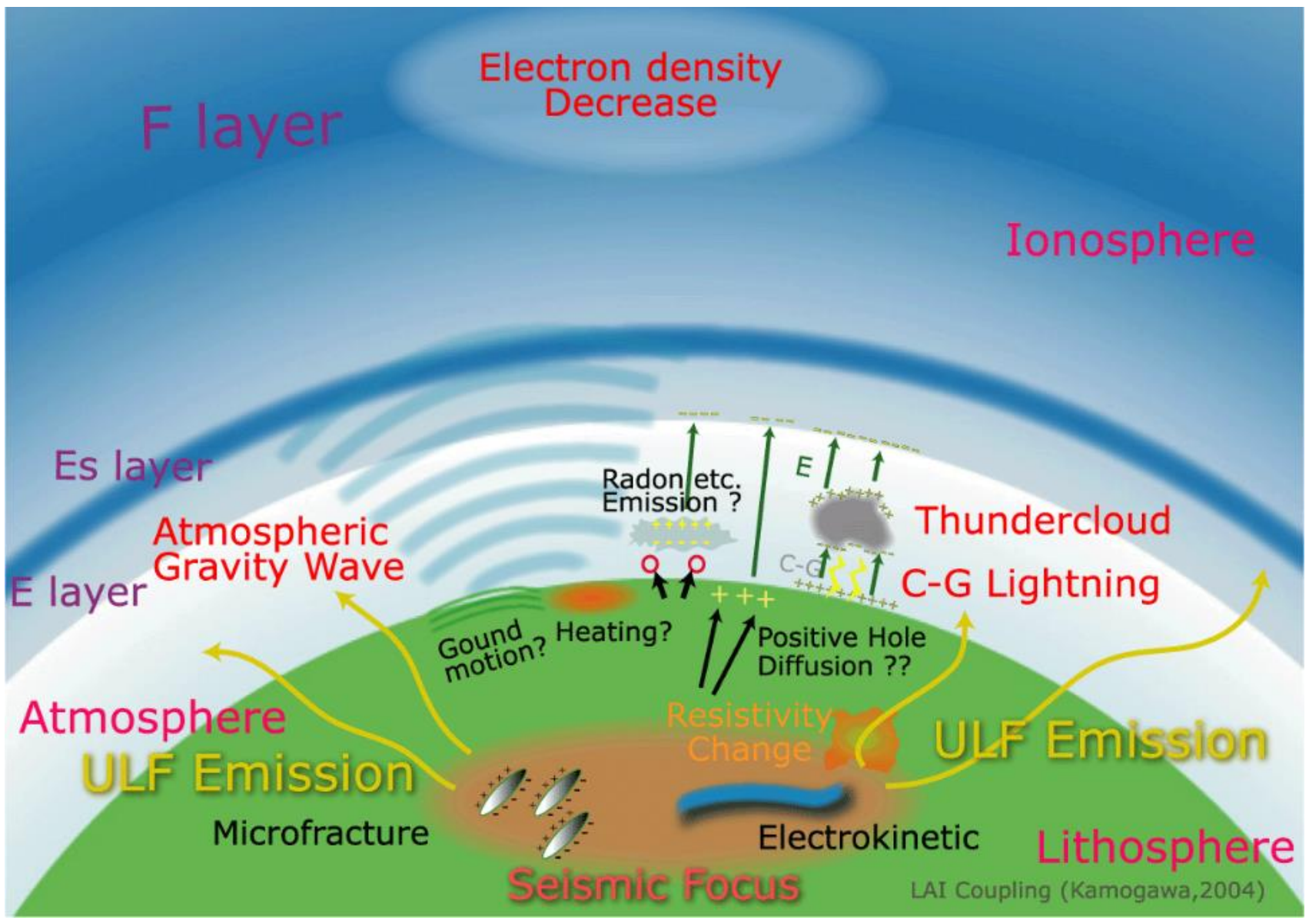

Figure 3 - Drawing describing the fundamental components of the LAIC model [31].

Additional measurements of fluctuations in ionospheric parameters measured on ground and in space are hypothesized to constitute seismic precursors, Figure 3.

In the last decades, many experimental analyses have been carried out, and several theoretical models have been proposed to interpret these ionospheric disturbances, as already explained at the beginning of this chapter. The hypothesis that they are caused by earthquakes and volcanic eruptions, although suggestive, is nevertheless very debated and requires precise verification, also because such disturbances must be carefully distinguished from the much more numerous processes induced by natural non-seismic sources, and from anthropic activities that can generate phenomena of this kind [1, 4, 5].

The intensity of the presumed seismic precursors appears rather difficult to distinguish from the prevailing effects induced by sources external to the geomagnetic cavity and by atmospheric events.

Indeed, the Sun plays a key role in controlling the dynamics of the upper ionosphere and magnetosphere, due to impulsive events such as coronal mass ejections and solar flares. Same story the tropospheric activity (lightning, TLE, etc.) with respect to the ionosphere.

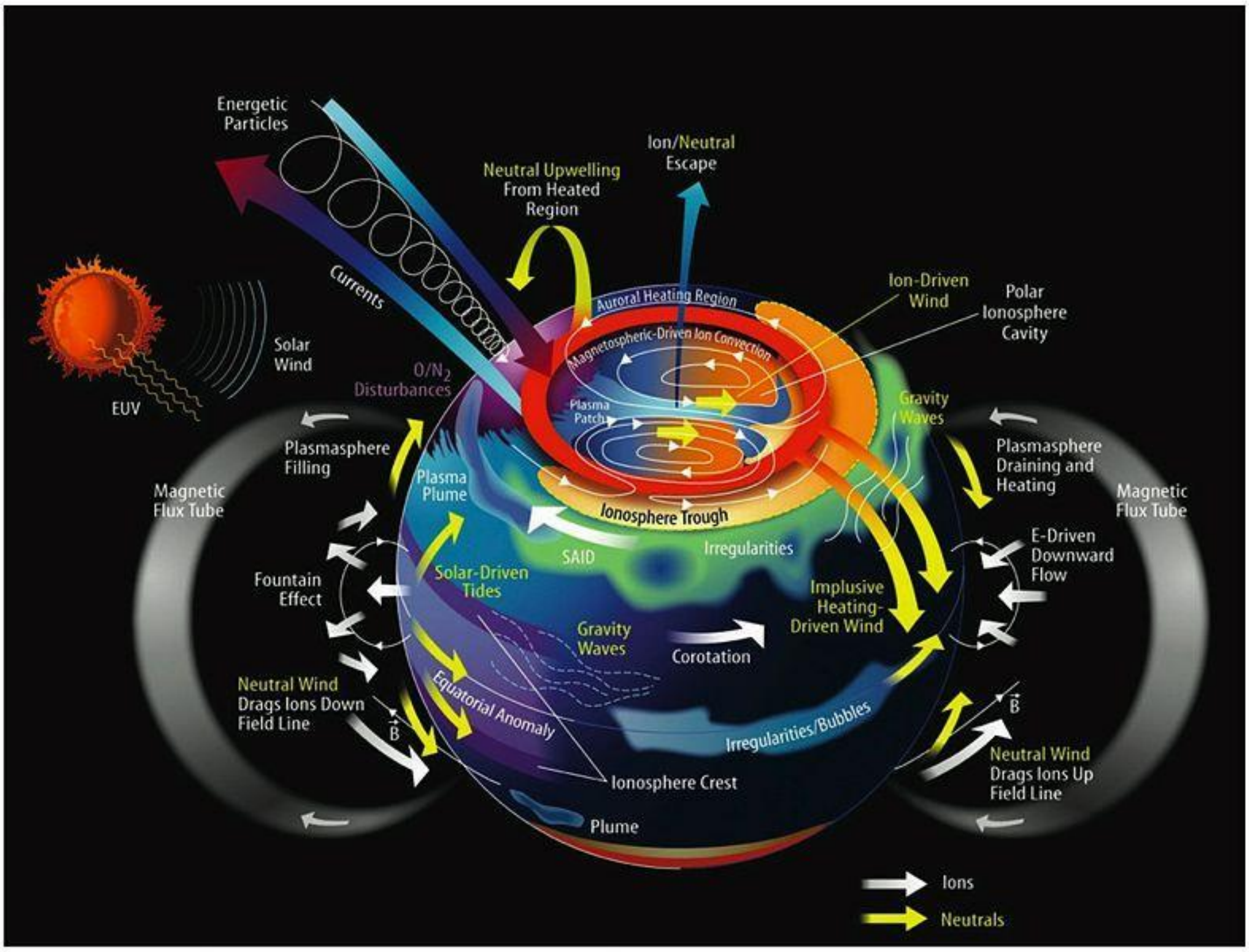

Figure 4 - Representation of the main particle fluxes around the Earth.

The development of low-altitude satellites for Earth observation has made it possible to study these phenomena from space by monitoring various parameters over large areas of the planet, an activity that must be integrated with ground monitoring networks, which alone do not allow to conduct a capillary check of the entire Earth's surface.

In the literature, possible precursor phenomena of earthquakes and volcanic eruptions, include electromagnetic emissions (observed on ground and by LEO satellites), anomalies in ionospheric and atmospheric parameters – such as fluctuations in plasma density and Total Electron Content (TEC), disturbances in VLF transmissions, etc.- and anomalous flows of particles precipitating from the Van Allen belts, see Figure 4. [30]

Variations in the fluxes of trapped particles constitute a relevant piece of research.

Abnormal increases in electron fluxes have been frequently observed several hours before the occurrence of medium and large earthquakes. It has been hypothesized that pre-seismic electromagnetic emissions can modify the trajectory of trapped particles, inducing precipitation.

For example, results from missions like Maria 1 and 2, GAMMA-1, Sampex-PET, ARINA, NOAA etc. show the existence of a statistical correlation between seismic activity and the precipitation of charged particles from the inner part of the Van Allen belts, seemingly preceding the earthquakes by a few hours, see Figure 5.

Particle detectors (HEPD and HEPP) on board the CSES satellite are optimized for studying these effects.

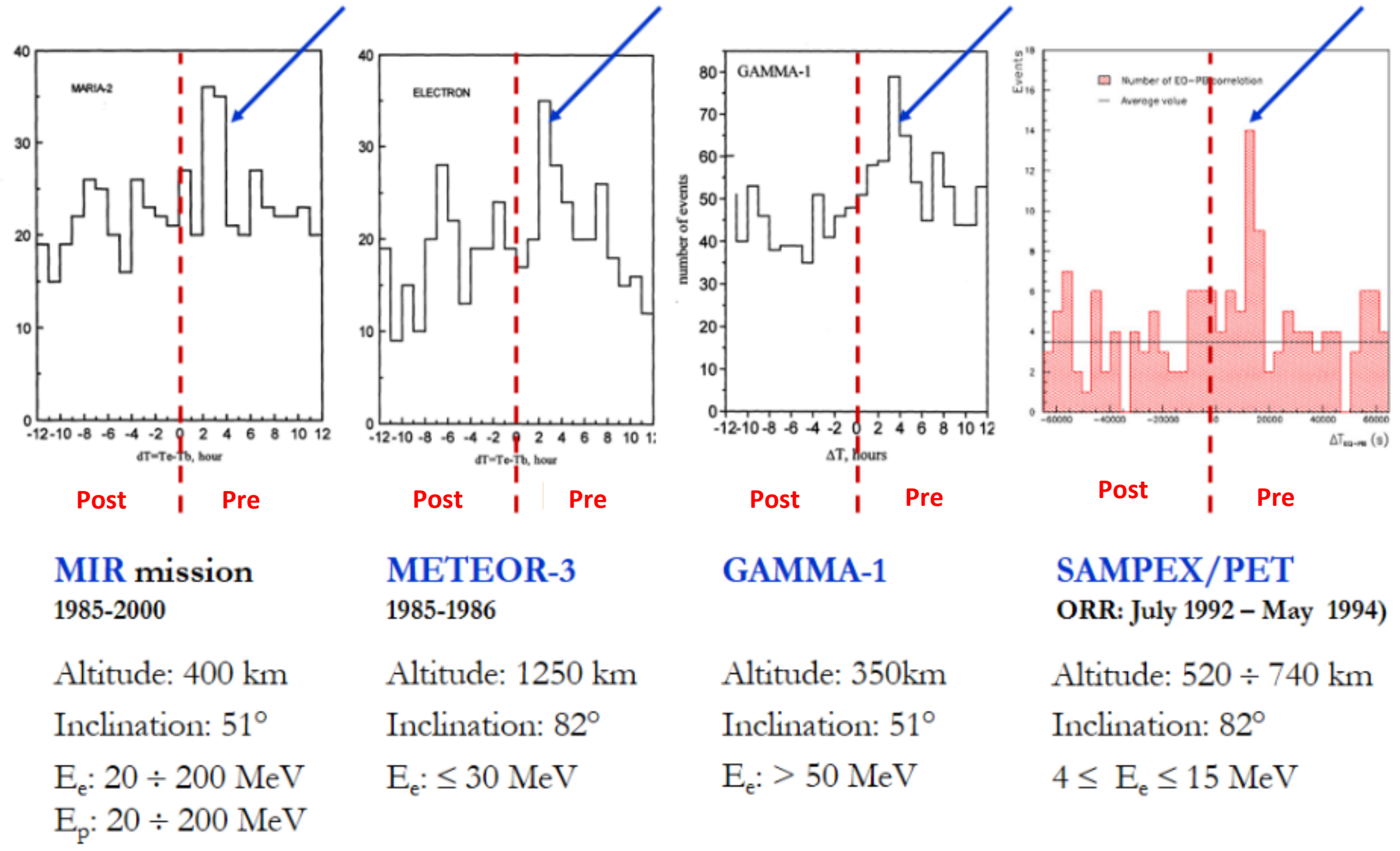

Figure 5 - Time differences between earthquake occurrence and observation of anomalous bursts of falling particles. Same distribution measured from different experiments. The arrows indicate the peak of precipitation which statistically would precede the earthquake by a few hours, pre-seismic [31].

After the first analysis conducted with data from non-dedicated space experiments, in 2000 the ESPERIA satellite project was proposed and developed by an Italian team [32].

The French DEMETER satellite was the first one devoted to investigating precursors from space – was successfully launched in 2004. In particular, the DEMETER mission showed anomalies in VLF emissions and variations in plasma density near the epicenter of some earthquakes in the days preceding the main shock; the intensity of these anomalies has proven greater for earthquakes of high magnitude, lower when the depth of the hypocenter is greater.

The objective of the CSES mission is to drive forward the exploratory study of the DEMETER mission, for the study of ionospheric perturbations induced by seismic activity and by the mechanisms of earthquake preparation, through systematic and detailed measurements over extended periods [1].

## 1.12 Magnetospheric – Ionospheric – Lithospheric Coupling (MILC) Model

Recent papers have provided some evidence of the link between seismic events and couplings between the lithosphere, lower atmosphere, and ionosphere, even though with marginal statistical evidence.

The basic coupling is conjectured as being via Acoustic Gravity Waves (AGW) and Acoustic Waves (AW). In [33], it is analyzed a scenario concerning the low-latitude earthquake (magnitude $M_w = 6.9$) that occurred in Indonesia on 5 August 2018, using a multi-instrumental approach, that relies on ground and satellite high-quality data. As a result, it is derived a new analytical Magnetospheric-Ionospheric-Lithospheric coupling (MILC) model with the aim to provide quantitative indicators to interpret the observations around 6 h before and across the earthquake.

A few hypotheses are introduced to justify the coupling among lithosphere, atmosphere, and ionosphere.

- The first one, which is based on the chemical transmission, assumes that the atmospheric conductivity can be perturbed by radon outflow close to the Earthquake Epicenter (EE), leading to a modification of the atmospheric electric field that drives a variation in the ionospheric plasma density profile [34, 35, 36].
- The second one is based on the emission of acoustic gravity waves (AGWs). Such oscillations, developing around the EE, perturb the atmosphere in terms of changes in temperature, pressure, ground motion, etc. AGWs can propagate upward and, thus, can drive disturbances in the ionosphere [37, 38, 39, 40].
- The third one is electrostatic transmission: an electrostatic effect is produced in the lithosphere and released in the lower atmosphere by "stress-induced positive holes" which alter the ionospheric ionization status [41, 42]. Nonetheless, because of the insufficient amount of experimental evidence supporting these theories, there are many other aspects of the lithosphere–atmosphere–ionosphere coupling process that cannot be explained yet [43, 44, 45].

In this framework, models based on AGWs emission seem to provide the most promising results to correctly understand the coupling processes in concurrence with an earthquake [33].

Anyway, it must be emphasized that all these observations give no information on the lithosphere/atmosphere coupling mechanisms, since only the lower ionosphere has been studied. Therefore, in recent years, many researchers have hunted for a possible link in the lithosphere–atmosphere–ionosphere system during active seismic conditions.

An analysis has been done both relative to the co-seismic and pre-seismic observations, considering atmospheric oscillations, ionospheric plasma, electric field perturbations, and magnetospheric FLR (Field Line Resonance) eigenfrequency variations.

This analysis has led to a scenario that can be explained in terms of a Magnetospheric–Ionospheric–Lithospheric Coupling (MILC) model (see Figure 6), based on three causal steps.

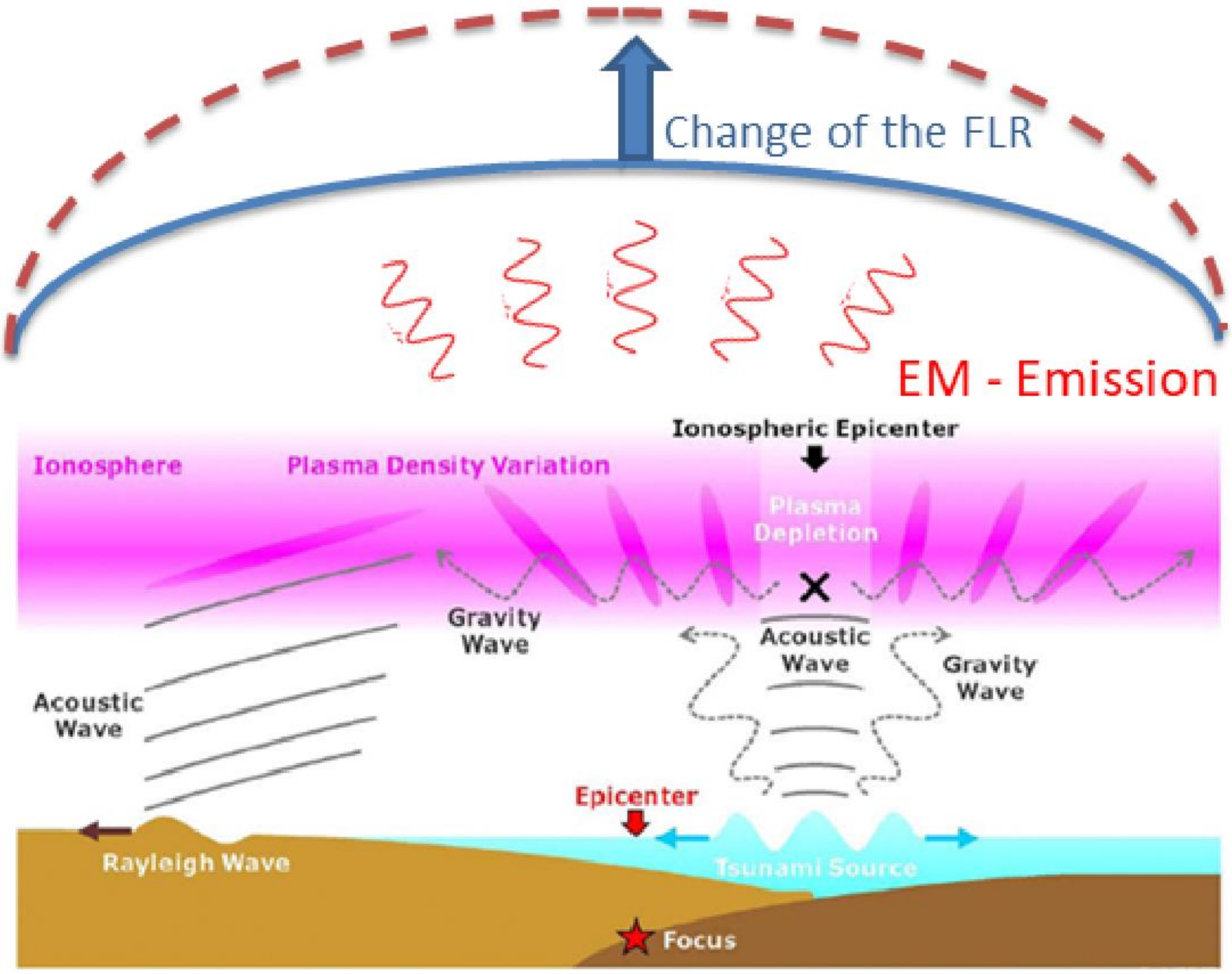

Figure 6 - Representative drawing describing the basic components of the proposed M.I.L.C. model.

1. An AGW is generated around the EE, propagating through the atmosphere.
2. The AGW interacts mechanically with the ionosphere, creating a local instability in the plasma distribution through a pressure gradient. Such a plasma variation puts the ionosphere into a "meta-stable" state, giving rise, in the E-layer, to a local non-stationary electric current. This, in turn, generates an electromagnetic (EM) wave.
3. The effects (especially in terms of local changes in the plasma density) of such EM waves in the ionosphere cause a change in the eigenfrequency of a magnetospheric field line, whose ionospheric footprint is located over the radial projection of the EE.

This picture is supported by the following mathematical description. Starting from the general equations for compressible, inviscid flow under a gravity field, in the absence of external forcing:

$$\begin{cases} \frac{\partial \rho}{\partial t} + \vec{\nabla} \cdot (\rho \vec{v}) = 0 \\ \rho[\frac{\partial \vec{v}}{\partial t} + (\vec{v} \cdot \vec{\nabla})\vec{v}] + \vec{\nabla} p = 0 \\ \vec{\nabla} p_0 = \rho_0 \vec{g} \\ \rho_0 c_s^2 = \gamma p_0 \end{cases}$$

where $\rho, p, v$ are the atmospheric density, pressure, and velocity, $c_s$ is the sound speed and $\gamma$ is the adiabatic index, we can easily evaluate the exact expression for an AGW propagating in a stratified non-isothermal atmosphere in terms of unsteady, non-uniform pressure perturbation $p_0$ related to an adiabatic condition in a convective frame [46].

$$\frac{\partial^2 p}{\partial t^2} - c_s^2 \nabla^2 p + A \cdot p = 0$$

where $A = \gamma g^2 / c_s$. The solution of Equation above is a mechanical wave (Lamb wave) [47, 48]. As expected, the dispersion relation of such an excited waveform mainly depends on the phase speed of the surface waves of the earthquake, $v_s$.

The intensity of the perturbation depends on the height of the atmosphere at the location of the excitation, and on the characteristics of the earthquake, namely the phase speed, $v_S$, the frequency of surface waves, $\omega_S$, the Peak Ground Acceleration and the Strong Motion Duration of the earthquake.

Introducing the $\vec{\nabla} p'$ as the external forcing into the Magneto–Hydro–Dynamic equation of a uniform ionosphere, located at 100 km above the Earth's surface and whose plasma density decrease as $z^{-2}$ with the altitude $z$, an ionospheric plasma density variation is induced [49].

Indeed, if the Magneto–Hydro–Dynamic system is solved for the perturbation of the electric field $\vec{E}$, an EM wave, in the frequency range between 10– 700 $Hz$, is generated, propagating from the top-side ionosphere.

The resonance frequencies of a geomagnetic field line, with both ends fixed in the ionosphere depend on the field line length, the magnetic field intensity, and the plasma mass density $\rho$ along the field line [50 - 56].

A crude estimation of an FLR eigen-frequency $f$ is given by the time-of-flight approximation [57] as:

$$\frac{1}{f} = 2 \int_s \frac{ds}{V_A}$$

$$V_A = \frac{B(s)}{\sqrt{\mu_0 \rho(r)}}$$

where $V_A$ is the Alfvén speed along the field line, $B(s)$ is the magnetospheric field along the field line, and $\rho(r)$ is the density at geocentric distance $r$. Therefore, a change in the field line length and/or the $V_A$, causes a variation in the FLR frequency [58]. Such anomalies can be found in the atmospheric temperature, pressure and/or conductivity [59].

In any case, earthquakes are not the only driver of gravity waves propagating through the atmosphere. In fact, AGW are generally induced by weather systems, by synoptic-scale atmospheric systems and circulations, and they can affect temperature and wind fields in the higher atmosphere. At middle and low latitudes (i.e., where earthquakes are more recurrent), AGWs in general are meteorologically excited by convective activities around the cold air fronts.

There is supported evidence that during an earthquake there is an activation of the lithosphere–atmosphere–ionosphere–magnetosphere chain, which, starting from the fault break, generates an AGW able to mechanically perturb the ionospheric plasma density, which in turn drives the generation of both EM waves and variations in the magnetospheric FLR eigen-frequencies. Interestingly, the observations of the CSES-01 satellite flying over the epicenter around 6 h before the earthquake, confirm both the presence of EM wave activity, coming from the lower ionosphere, and plasma density variation consistent with the TEC anomaly detected ([60] and reference therein).

## 1.13 SENSORS IMMERSED IN PLASMA: BASIC THEORY

Plasma sensors, or rather electric probes, have been used for a long time as a tool to measure the local characteristics of a plasma. These devices are relatively simple, but the theory behind the interaction between a probe and a plasma is quite complicated. Here, the case of an ionospheric plasma will be treated, the type in which the EFD sensors are immersed, while flying on board the CSES satellites.

The ionospheric plasma parameters depend on solar external forcing; plasma electron temperature and density change if exposed to different solar UV irradiation along the satellite orbits.

Plasma density is the parameter which varies more drastically with latitude, showing at the dayside equator an increase of up to 10 times compared with what is observed at polar latitudes. Instead, on the nightside, only a moderate increase at low latitudes (about 2-fold) is observed.

Electron density and temperature can be retrieved from the data-driven International Reference Ionosphere (IRI) model. The IRI dataset shows an expected variability, at the CSES orbit, in the following ranges: Plasma density: $(7 \times 10^{9} - 2 \times 10^{12})\ m^{-3}$; Electron temperature: $(1 \times 10^{3} - 3.2 \times 10^{3})\ K$ [61].

The probes of the EFD behave essentially as floating electrodes immersed within the ionospheric plasma. A conducting probe in contact with a plasma attains a potential (denoted as the floating potential) which can be theoretically estimated by imposing that the net current collected by the probe surface is null.

In the case of the EFD probes, four contributions are relevant:

I. Electron collection.
II. Ion collection.
III. Photoelectron emission.
IV. Current injected to the probe (bias current source).

The floating potential condition can be expressed as:

$$\sum_{k=1,2,3,4} I_k = 0$$

where $k$ denotes the various current contributions listed above. All these terms can be expressed by voltage dependent equations and the procedure required to calculate the floating potential implies the determination of the entire current-voltage characteristic.

This is obtained considering the various currents, expressed as a function of potential, for two cases: $V > V_{pl}$ and $V < V_{pl}$. A qualitative representation of the current-voltage characteristic of an electrode in a plasma is shown in Figure 7. It should be noted that the current generator that is part of the EFD electronics can be set during the flight via telecommand (TC, i.e., remote control), with the aim of modifying the balance of the currents in order to control the probe potential with respect to local plasma potential. Therefore, in EFD, thanks to the tuning of the bias current, the probes float at a specific point of the characteristic curve, as close as possible to the plasma potential.

An important parameter is the contact impedance ($Z_c = R_{pl} \backslash\backslash X_{C_{pl}}$) between probe and plasma.

The dynamical resistance $R_{pl}$ at a point along the current-voltage curve is defined as the reciprocal of the derivative of the current with respect to the potential, according to:

$$R_{pl}^{-1} = \frac{dI_{tot}}{dV} \Big|_{V=V_0}$$

where $V_0$ represents the potential value at which the dynamical resistance is determined.

The contact resistance exhibits its minimum close to the plasma potential point ($V_{pl}$).

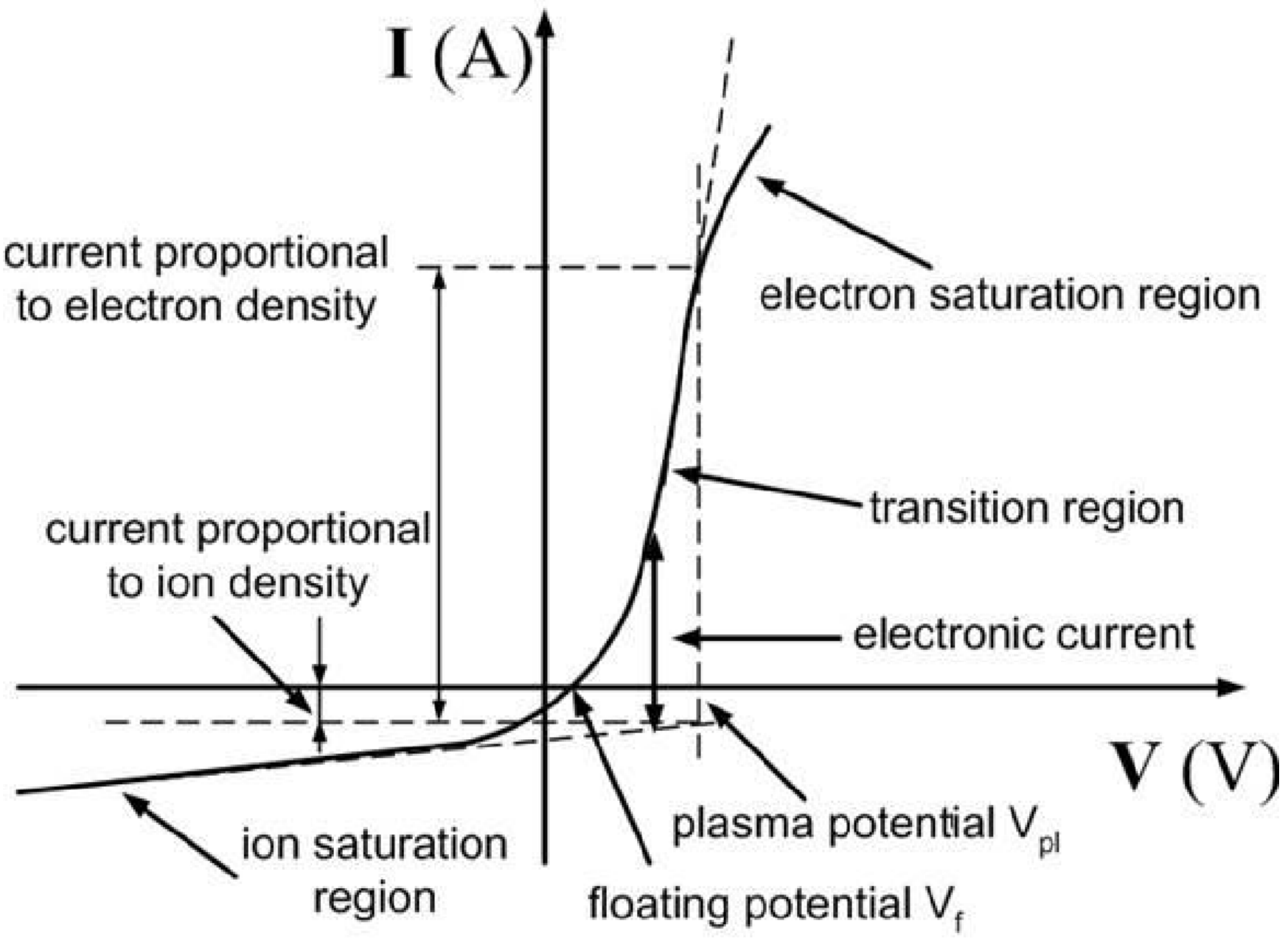

Figure 7 - Characteristic of a conducting electrode in a plasma [68].

The equation above represents only an upper limit for $Z_c$, as a more complete analysis of the equivalent circuit should include the capacitive contribution $C_{pl}$ associated with the plasma sheath which further reduces the impedance at high frequency.

Such a capacitance can be estimated by considering a spherical capacitor with the inner electrode having a radius equal to that of the EFD probe, separated by a Debye length from the outer electrode representing the unperturbed plasma [62]. The capacitance associated with such an element is of the order of 10 pF, varying with the electron temperature and density between about 5 pF and 30 pF [63].

## 1.14 ELECTRON CURRENT COLLECTED FROM PLASMA AT $V < V_{pl}$ (RETARDING POTENTIAL)

In a plasma at thermal equilibrium, the velocities of the electrons are characterized by a Maxwellian distribution function, with thermal speed defined as:

$$v_{thermal} = \sqrt{\frac{8k_eT_e}{\pi m_e}}$$

Where $k = 1.83 \cdot 10^{-23}\ J/K$ , $m_e = 9.1 \cdot 10^{-31} kg$. In the ionosphere, the expected thermal speed varies between $(2 \cdot 10^5 - 3.4 \cdot 10^5\ ) m/s$. In the case of CSES, the satellite velocity is about $7.5 \cdot 10^3\ m/s$, much lower than the thermal speed, so it can be neglected.

The electron current collected by a probe in the ionospheric plasma under retarding potential (i.e., $q(V - V_{pl}) < 0$) is given by:

$$I_e = \frac{1}{4} q\, n \sqrt{\frac{8 k_e T_e}{\pi m_e}}\ S_e e^{\frac{q(V - V_{pl})}{k T_e}}$$

With n the plasma density, $S_e$ the cross-section area of the probe for electron collection, $V$ the probe potential, and $V_{pl}$ the local plasma potential.

If the gyroradius of electrons is larger than the probe radius, the electron speed is assumed to be isotropically distributed in space. This is the case applicable to the EFD probes, thus the cross-section $S_e$ may be assumed equal to the area of a sphere $S_e = 4\pi R_p^2$. Consequently, the electron current is proportional to the plasma density which, at the altitude of the CSES satellite, is expected to vary approximately between $(10^{10} - 10^{12})\ m^{-3}$. It is worth noting that, in the retarding electron collection regime the electron current collected by a probe is completely independent of the size of the sheath [64].

## 1.15 ELECTRON CURRENT COLLECTED FROM PLASMA AT $V > V_{pl}$ (ACCELERATING POTENTIAL)

For an accelerating potential, there are two possible conditions, depending on the relative size of the plasma sheath with respect to probe radius.

- Thin sheath approximation: for plasma sheaths much thinner than the probe radius the collected current, for $(V - V_{pl}) > 0$, tends to flatten at a constant value approximately equal to the random thermal current.
- Thick sheath approximation: for plasma sheaths much thicker than the probe radius, the collected current for $(V - V_{pl}) > 0$ tends to increase linearly with V, maintaining a constant slope equal to that exhibited at the plasma potential (i.e., $V = V_{pl}$) [64].

The thickness of the plasma sheath can be evaluated considering the Debye length $\lambda_D$ as, at a first order of analysis, the sheath can be assumed to extend over several $\lambda_D$'s. The Debye length is:

$$\lambda_D = \sqrt{\frac{kT_e\epsilon_0}{nq^2}}$$

And it can vary between 0.2 cm and 4 cm under ionospheric conditions for $T_e$ and $n$. It is possible to assume that the minimum thickness is of the order of 1 cm, such that “thick sheath approximation” represents a regime that can be reasonably applied to the entire range of ionospheric conditions encountered by the EFD probes during the flight.

The expression that describes the collected electron current $I_e$ as a function of the electrode potential $V$ in the “thick sheath approximation" is [65]:

$$I_e = \frac{1}{4} q\, n \sqrt{\frac{8k_eT_e}{\pi m_e}}\, S_e \left(1 + \frac{q(V - V_{pl})}{kT_e}\right)$$

## 1.16 ION CURRENT COLLECTED FROM PLASMA (ACCELERATING POTENTIAL $V < V_{pl}$ AND RETARDING POTENTIAL $V > V_{pl}$)

Unlike electrons, ions are seen by the satellite as a flux of particles approaching at a velocity equal to that of the satellite itself (i.e., $v_{orb} = 7.5\ 10^3\ m/s$). Therefore, the space distribution of the ion velocity implies that the probe's cross section for ion collection is that of a flux tube, aligned with the satellite velocity vector: for a spherical shape probe $S_e = \pi R_p^2$.

Thus, the value of the ion current collected by the probe can be approximately estimated assuming a flux of mono-energetic ions, as:

$$I_i = \pi R_p^2 q n v_{orb} \left(1 - \frac{q(V - V_{pl})}{K_{ion}}\right)$$

where $K_{ion} = 9.4 \cdot 10^{-19} J$ is the kinetic energy.

Given the positive charge of the ions, the accelerating potential condition is obtained for $V < V_{pl}$, whereas the retarding potential condition occurs when $V > V_{pl}$.

## 1.17 PHOTOELECTRON CURRENT

The photoelectron flux is emitted by the probe surface as an effect of solar irradiation. The emitting surface for the photoelectron current of a spherical probe with radius $R_p$ under solar irradiation is simply $S_{SR} = \pi R_p^2$. Assuming a current density of $J_{ph} \cong 20\mu A/m^2$, the photoelectron current emitted by the EFD probe ($6cm$ in diameter) is [61]:

$$I_{ph} = 20 \cdot 10^{-6} \pi R_p^2 = 0.057\,\mu A$$

Photoelectric emission can be assumed roughly constant for $V \leq V_{pl}$ while it is suppressed for $V > V_{pl}$.

## 1.18 CURRENT INJECTED TO THE PROBE: BIAS CURRENT SOURCE

A current generator is used to modify the balance among the various currents, thus controlling the potential of the probe with respect to that of the local plasma.

The probe potential is modified in order to be placed as close as possible to the local plasma potential, where the contact resistance exhibits its minimum value. Thus, the capability of the probe to follow the fluctuations of the local plasma potential improves significantly.

## 1.19 NON-PLASMA RELATED EFFECTS

A brief parenthesis on the effects of the Earth's magnetic field on the data collected by CSES should be added.

The geomagnetic field has two effects: on the one hand, the motion of CSES through the B magnetic field lines induces an electric field along conducting parts of the satellite body, according to the formula $\vec{E} = \vec{v} \times \vec{B}$, where $\vec{v}$ is the satellite velocity vector (about 7.5 $km/s$).

On the other hand, the local value of the magnetic field represents a natural baseline against which the variable part of the observed magnetic field must emerge.

The value of the Earth's magnetic field is well known and adopting a formal model (such as IGRF or CHAOS), it is possible to estimate its value and orientation along the orbit with respect to satellite booms. This can be used to determine the motion-induced electric field, such that it can be subtracted from the measurements, thus retrieving the true ambient electric field.

# CHAPTER 2

## CSES-02 MISSION

*As in any space mission, it is important to have a deep knowledge of the instrument that will be used for scientific purposes.*

*In this Chapter, general features of the CSES mission will be reported, together with a detailed description of the essential components of EFD-02.*

*After the earliest studies relying on non-dedicated satellites, the DEMETER observations have supported an increasing number of studies claiming the existence of seismo-associated ionospheric and magnetospheric perturbations occurring from two weeks up to a few hours before earthquakes of a large magnitude.*

*In this framework, the CSES-01 satellite is the second mission designed for investigating this kind of phenomena from space, successfully launched in February 2018.*

*Later in this chapter, the innovations that distinguish the EFD-02 instrument (CSES-02 mission) will be highlighted and compared to previous analogous missions.*

## 2.1 THE SATELLITE

Based on the Chinese CAST2000 platform, this 3-axis stabilized satellite has a mass of about 730 kg and a peak of power consumption of about 900 W. The scientific data are transmitted in the X-band at 120 Mbit/s. The orbit is circular and Sun-synchronous (see Figure 8: this is an orbit arranged such that it is always synchronized with the Sun, thus passing over any point always at the same local time [80]), at an altitude of about 500 km, an inclination of about 98°, and descending node at 14:00 LT [71].

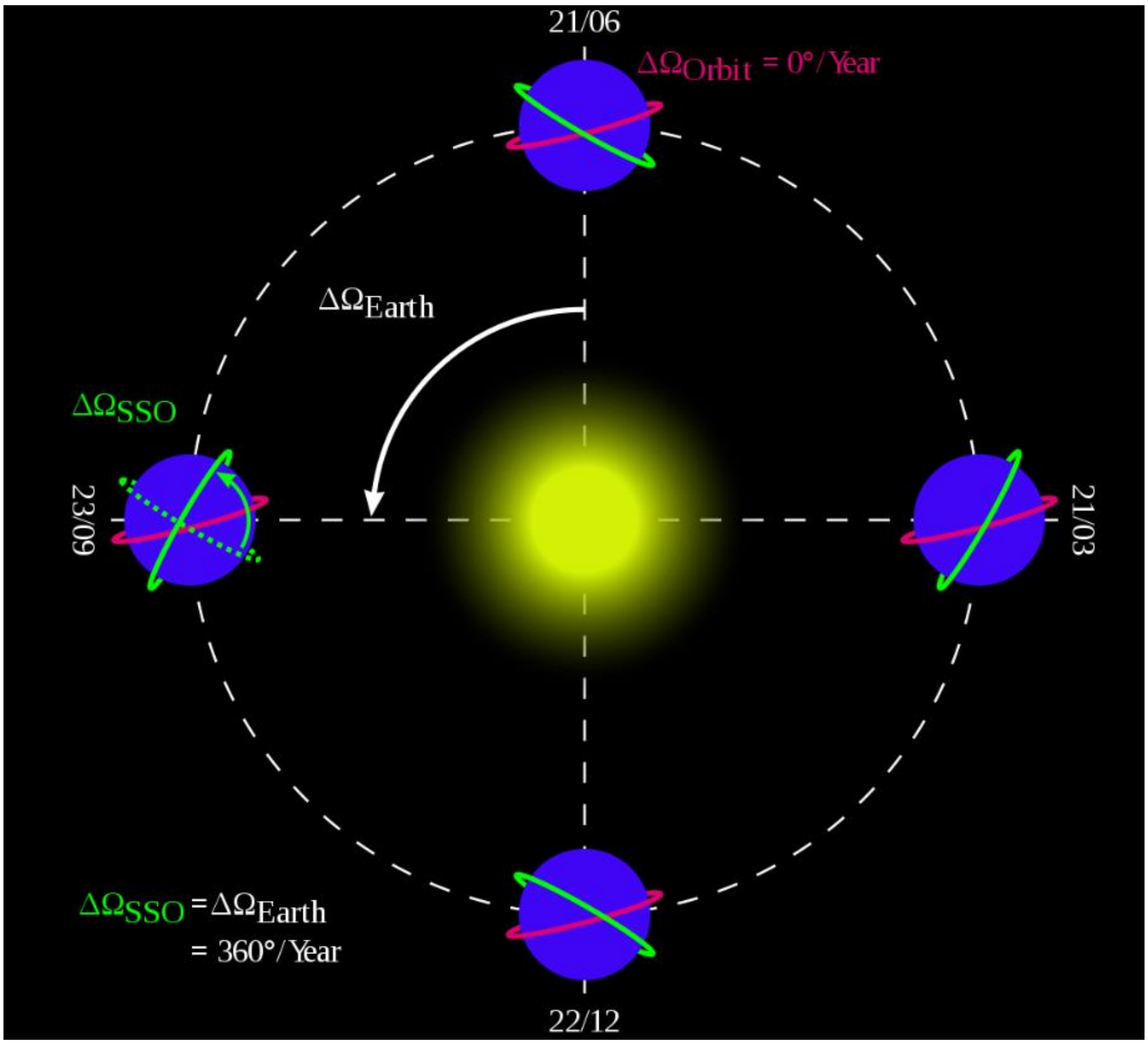

Figure 8 - Diagram showing the orientation of a Sun-synchronous orbit (green) at four points in the year. A non-Sun-synchronous orbit (magenta) is also shown for comparison. Dates are shown in white: day/month.

## 2.2 ITALIAN AND CHINESE PARTICIPATION

The Chinese institutes involved in the project are the China National Space Administration (CNSA), the China Earthquake Administration (CEA), the Lanzhou Institute of Physics (LIP), the Institute of Crustal Dynamics (ICD-CEA), the Institute of High Energy Physics (IHEP), the National Space Science Center (NSSC), the Center for Space Science and Applied Research-Chinese Academy of Science (CSSAR-CAS), the Space Star Technology Co. and the DFH Satellite Co [71].

Italy participates in the CSES mission via the Limadou project, led by Prof. Piergiorgio Picozza (Principal Investigator) and financed by the Italian Space Agency (ASI) and the National Institute for Nuclear Physics (INFN). The Limadou collaboration includes the INFN divisions of Roma Tor Vergata, Bologna, Naples, Perugia, Torino, the TIFPA Center of Trento, and the National Laboratories of Frascati ; together with the Universities of Bologna, Trento, Roma Tor Vergata, and Uninettuno, as well as INAF-IAPS ( National Institute of Astrophysics-Institute for Space Astrophysics and Planetology) and INGV ( National Institute of Geophysics and Volcanology).

The LIMADOU Collaboration has designed, built and tested the High Energy Particle Detector (HEPD) included in the CSES-01 mission, which is conceived for optimized detection of energetic charged particles that precipitate from the Van Allen belts (as a result of seismic and non-seismic electromagnetic perturbations). It has collaborated in developing and testing the Electronic Field Detector (EFD-01) at the INAF-IAPS plasma chamber in Rome, and it currently manages HEPD data analysis, while participating in the data analysis of all remaining payloads [81].

For the upcoming CSES-02 mission, the HEPD-02 payload will be accompanied by the production, testing and qualification of the new EFD-02 instrument, which is endowed with innovative features compared to its CSES-01 counterpart, as described in the next paragraphs.

## 2.3 Payloads on board CSES-02

The goals of the mission will be achieved by the implementation of 10 dedicated instruments aboard the satellite.

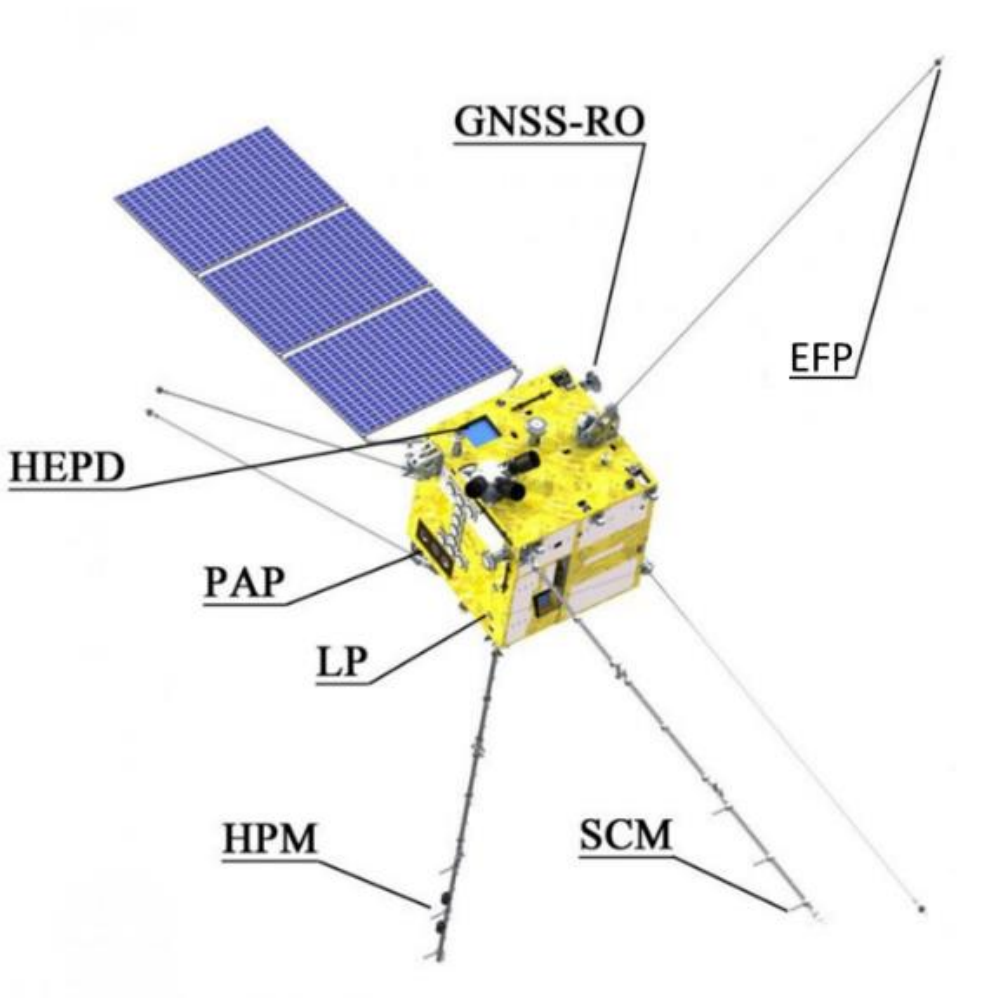

Figure 9 - Some of CSES-02 payloads and their positioning.

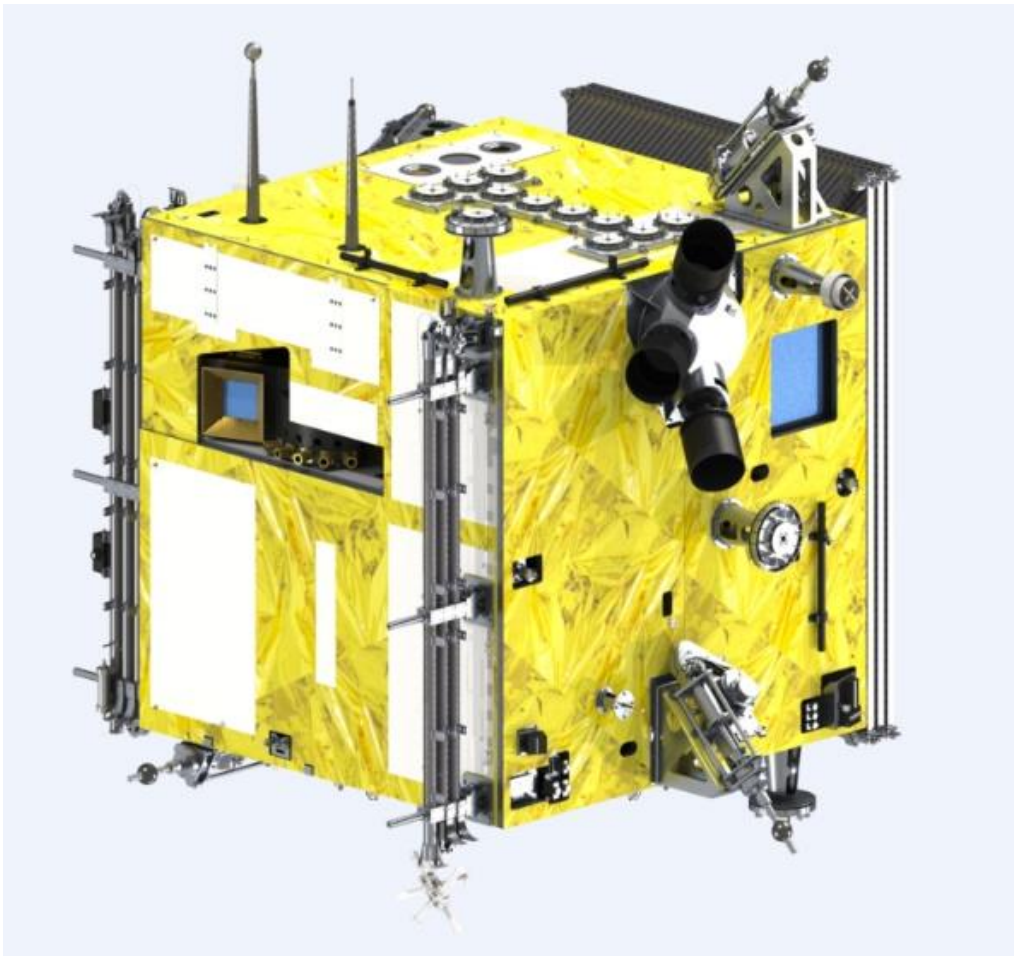

Figure 10 - Rendering of the CSES satellite and related payloads.

Two particle detectors will be mounted, the High-Energy Particle Detector (HEPD) and the Medium Energy Electron Detector (MEED), in order to measure flux, energy spectrum, type, and direction of any impinging particle. HEPD is developed by the Italian Collaboration LIMADOU, and it can detect electrons, protons, and light nuclei. The main objective of the (still operating) HEPD-01 detector has been to measure the variations in the flows of charged particles due to perturbations of the radiation belts caused by solar, terrestrial, and anthropogenic phenomena. Its ranges of energy are 3 - 100 MeV for electrons and 30 - 200 MeV for protons, while it is possible to study nuclei up to oxygen [72].

Also, HEPD-02 identifies the particle type (proton, electron, nucleus), measures particle energy, and determines the angle between the flight line and the local geomagnetic field (pitch angle). Once again, HEPD-02 can detect particle flows coming from the Van Allen belts and determine with great accuracy the magnetospheric region of their origin, with the aim of obtaining the energy spectrum and the particle composition of sudden precipitation to the atmosphere, in case of external disturbances ("Particle burst").

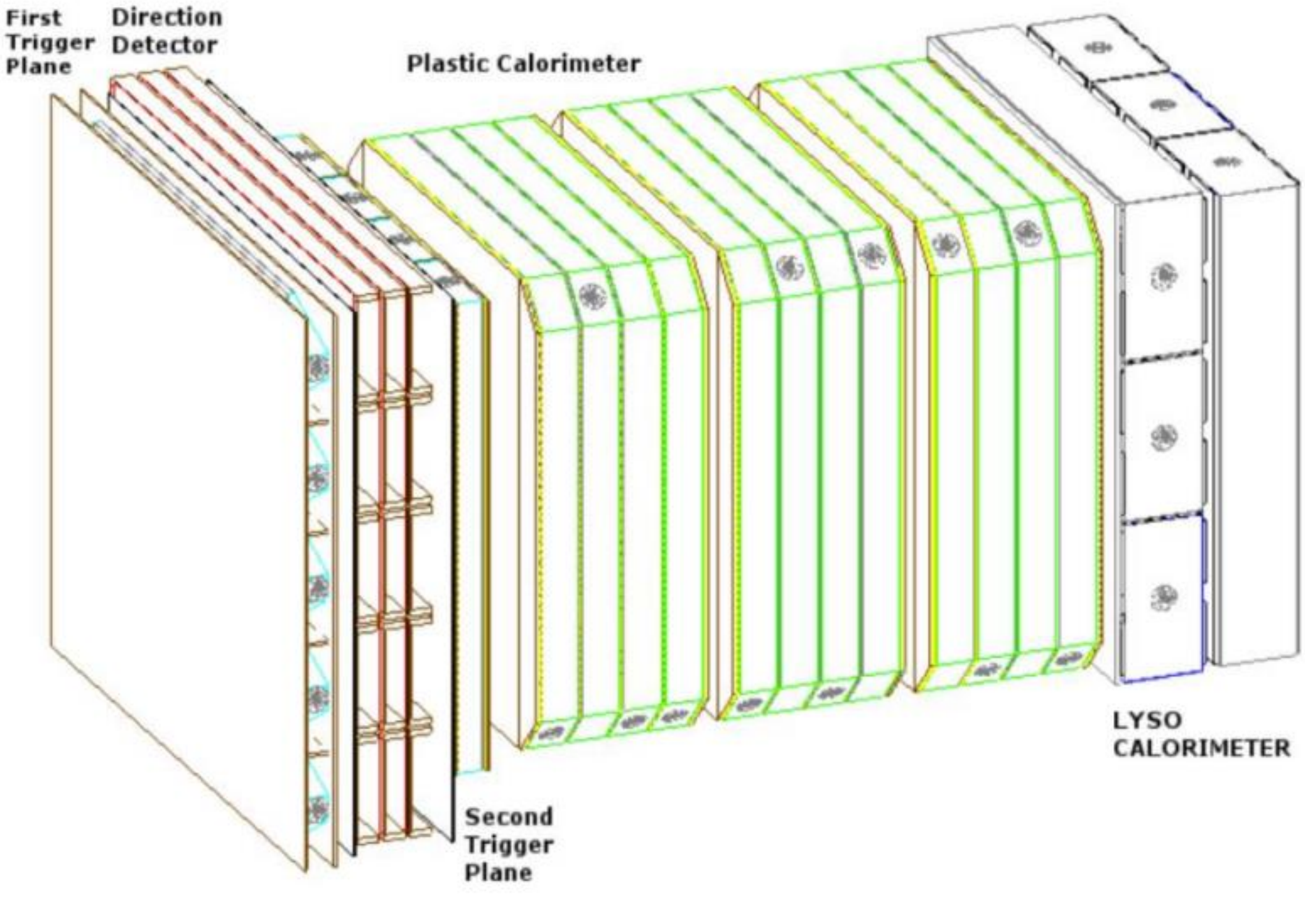

Figure 11 - HEPD-02 scheme.

The instrument, shown in the Figure 11, consists of a first trigger plane TR1 (overall dimensions $200 \times 180\ mm^2$) segmented into 5 plastic scintillator bars ($2\ mm$ thick).

It follows an incident-particle angle detector DD (“tracker”) made of five standalone tracking modules (“turrets”), each composed of three sensitive planes (“staves”).

A second trigger plane TR2 (overall dimensions: $150x150mm^2$) is segmented into 4 plastic scintillator bars (8-mm thick).

Then, an energy detector ED is composed of 12 plastic scintillator planes ($150x150x10mm^3$) and 2 crystal (LYSO) scintillator planes (overall dimensions $150x150mm^2$) segmented into 3 bars ($50mm$ thick).

Finally, a containment detector CD (not shown here) surrounds the calorimeter on 5 sides, and it is made of plastic scintillator planes (4 lateral and 1 bottom plane), each of which 8-mm thick.

The explored energy range is 3 – 100 MeV for electrons and 30 – 200 MeV for protons.

The Qualification Model (QM) was completely assembled and ready for space qualification tests at the time of the writing of this manuscript. The Flight Model (FM) was at its final stage of development for the physics tests scheduled for December 2022.

As in any CSES satellite, a Search-Coil Magnetometer (SCM) and a High Precision Magnetometer (HPM) are included to measure the components and the total intensity of the geomagnetic field, respectively. The HPM is the result of a collaboration between the National Space Science Center (NSSC) of the Chinese Academy of Sciences, the Space Research Institute (IWF) of the Austrian Academy of Sciences (ÖAW), and the Institute of experimental physics (IEP) of the Graz University of Technology.

The instrument includes: two fluxgate magnetometers (to measure the 3 components of the low-frequency magnetic field) and a scalar magnetometer (CDSM). The CDSM is an optically pumped absolute scalar magnetometer for the measurement of the absolute intensity of the magnetic field, used for the calibration of fluxgate magnetometers made by NSSC. The SCM can measure the three components of the magnetic field from about 10 Hz up to about 20 kHz.

Still, the Electric Field Detector (EFD) allows to measure the electric field components in a wide amplitude and frequencies range by the implementation of four spherical probes at the tips of as many booms. It will be copiously described later in this thesis.

A Plasma Analyzer Package (PAP) is intended to measure plasma parameters such as ion density, temperature, drift speed, as well as the composition and fluctuation of the ion density, while a Langmuir probe (LP) unit allows to measure plasma density and electronic temperature. The LP unit consists of a pair of spherical Langmuir probes with diameters of 5 cm and 1 cm respectively, installed at the end of arms about 50 cm long.

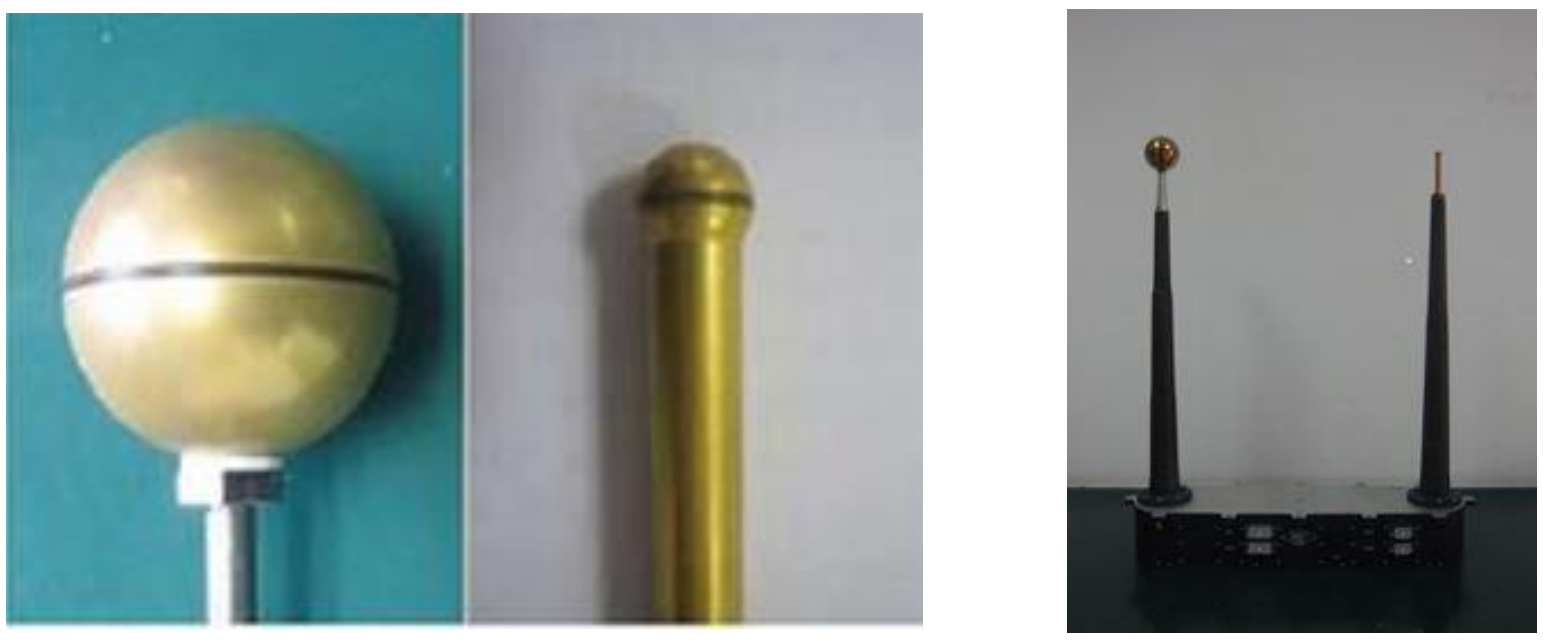

Figure 12 - Langmuir probes.

A GNSS Occultation Receiver is used for the vertical sounding of the ionosphere. It allows to measure the total electron content (TEC) and to obtain electron density profiles.

A Tri-Band Beacon is a three-frequency beacon developed to transmit in the VHF / UHF and L bands (150/400/1067 MHz). The primary objective of the instrument is to study the electron density in the ionosphere and produce 2D maps and 1D profiles of electron density, respectively. The tool also allows to study the influence of ionospheric irregularities on VHF, UHF and L band transmissions from space to ground [73].

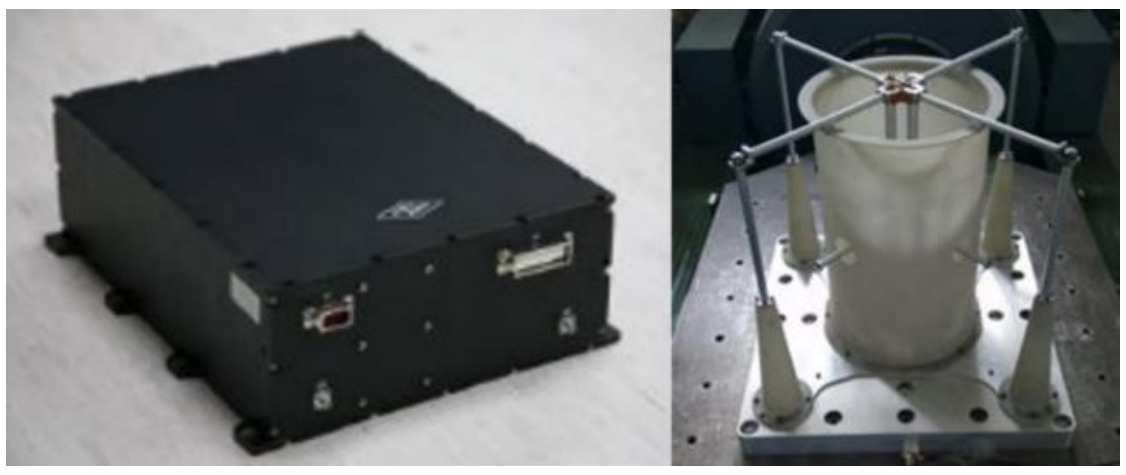

Figure 13 - Tri-Band Beacon.

Each instrument is arranged to collect data in two different operating modes: 'the 'Burst mode", activated when the satellite passes over China and the more seismic regions of the Earth, and the "Survey mode" for other areas of the planet (see Figure 14) [71].

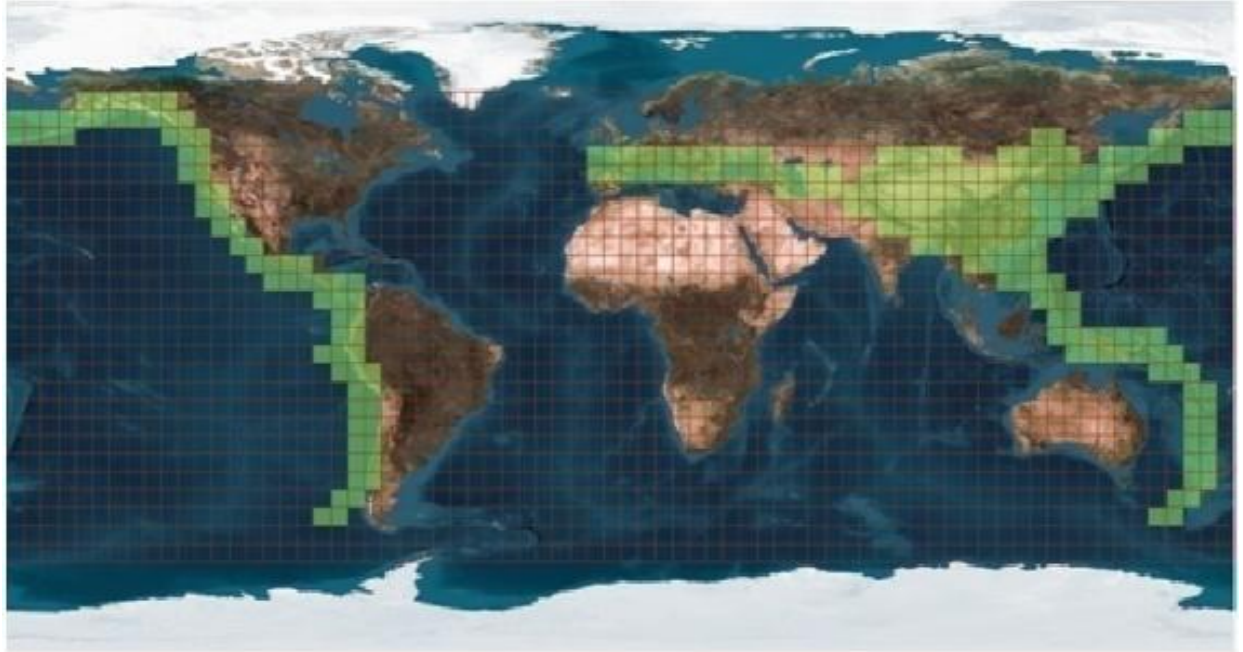

Figure 14 - Orbital operating modes, the burst mode is activated in the areas highlighted in green.

In addition, there are two different orbital working zones: the "payload operating zone", for geomagnetic latitudes between -65° and +65° (where the instruments collect data), and the "platform adjustment zone", at higher latitudes (where all detectors are switched off to perform the satellite altitude control and the orbital maintenance activities, AOCS). Table 1 shows the specifications of some CSES payloads.

| ***Category*** | ***Payload Name*** | ***Observation Target*** |
|---|---|---|
| ***Energy Particle*** | *High Energy Particle Detector (HEPD* | *Proton: 30 MeV~200 MeV*<br>*Electron: 3~100 MeV* |
| | *Medium Energetic Electron Detector (MEED)* | *Electron: 25 KeV~3.2MeV* |
| ***Electro-Magnetic Field*** | *Electric Field Detectors (EFD)* | *Electric Field: DC~3.5 MHz* |
| | *High Precision Magnetometer (HPM).*<br>- *Vector Magnetic Field: FGM1, FMG2.*<br>- *Scalar Magnetic Field: CDSM, CPT.* | *Magnetic Field 10 Hz~20 kHz* |
| | *Search Coil Magnetometer (SCM)* | *Magnetic Field 10 Hz~20 kHz* |
| ***In Situ Plasma*** | *Plasma Analyzer Package (PAP)* | *Composition: H+, He+, O+*<br>*$N_i$: $5x10^2$~$1x10^7$ $cm^{-3}$*<br>*$T_i$: 500K~10000 $cm^{-3}$* |
| | *Langmuir Probe (LP)* | *$N_i$: $5x10^2$~$1x10^7$ $cm^{-3}$*<br>*$T_i$: 500K~10000 $cm^{-3}$* |
| ***Plasma Profile Construction*** | *GNSS Occultation Receiver* | *TEC by transmit VH/U/L Signal* |
| | *Tri-Band Beacon* | *TEC by transmit VH/U/L Signal* |
| | *Ionospheric* | *$O_2$ 135.6 nm and $N_2$ LBH airglow* |

Table 1- Specification table of some CSES payloads, [71].

## 2.4 THE ELECTRIC FIELD DETECTOR (EFD)

Although it is a very advanced and electromechanically complex instrument, the operating principle of the EFD is very simple: it measures the differences in electric potential between pairs of sensors, called Electric Field Probes (EFPs), installed at the tips of 4 booms deployed at about 4 m from the satellite.

Each of the three electric field components is obtained as the difference between two probe voltages divided by their relative distance (8.3 m on average), $d_{1,2}$:

$$E_{1,2} = \frac{V_1 - V_2}{d_{1,2}}$$

It is possible to determine the expected theoretical values of any EFD sensor potential thanks to key parameters such as [74]:

- the knowledge of the geomagnetically induced electric field $v_s x B$ ("V cross B"), where $v_s$ is the satellite speed and B the local magnetic field.
- the plasma variation effect, described by the Orbit Motion Limited (OML) theory.

Thus, EFD is intended to measure:

- Electric field variations.
- Plasma density depletions.
- Electromagnetic signals from natural and artificial sources (e.g., Schumann resonances and signals from VLF antennas).

The measurements carried out by the EFD are considered essential for the entire mission, as they will allow, in association with the magnetic field data, the monitoring of EM waves along the orbit.

The low frequency (LF) band will allow the fine monitoring of plasma structures modulated by solar perturbations in transit in the interplanetary medium. Oscillations in the medium frequencies (MF) will give information on the dynamics of the ionosphere also in association with the precipitation of particles from the radiation belts.

Finally, high frequencies (HF) may provide local information on the violation of plasma neutrality induced by pulses from the lithosphere [75].

## 2.5 THE CSES-01 MISSION AND HEPD-01 MAIN RESULTS

The satellite is in orbit since Feb 2nd, 2018. The payload on board consists of: one High-Energy Particle Detector and one High Energy Particle Package (HEPD and HEPP, respectively) designed to measure particle flux and energy spectrum [86] ; a Search-Coil Magnetometer (SCM) and a High Precision Magnetometer (HPM) to measure the components and the total intensity of the magnetic field, respectively ; an Electric Field Detector (EFD) to measure the electric field ; a Plasma analyzer and a Langmuir probe to measure the disturbance of plasma in ionosphere ; a GNSS Occultation Receiver and a Tri-Band Beacon to measure the density of electrons.

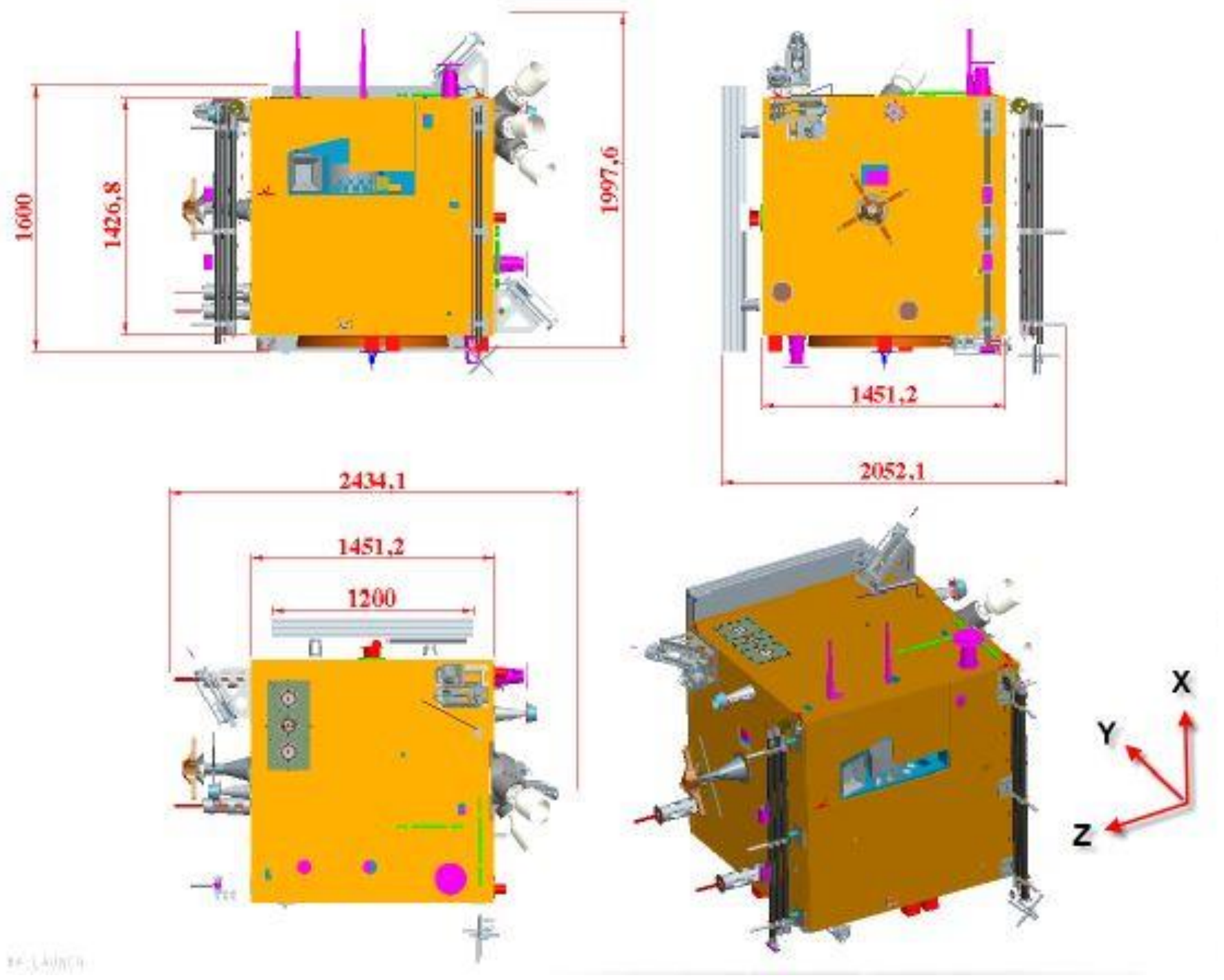

Figure 15 - 3D representation of the CSES-01 satellite.

Below, some of the main HEPD-01 results.

**The G3 geomagnetic storm of August 26, 2018, detected by HEPD-01** [89].

Here, the disturbance of the electron population is detected by HEPD-01 at the storm onset (Figure 16).

HEPD trigger rate variations observed for electrons in the MeV energy range during the Aug 2018 storm have clearly shown a depletion during storm's main phase, followed by a robust enhancement in the outer belt (at $L > 3$ for energies above 3 $MeV$ and, to a lesser extent, at $L > 4$ for energies above 4.5 $MeV$) during recovery, with support by large (AL index $> 1000\ nT$) and prolonged (> 2 days) substorm activity downstream of the main phase.

The temporal and spatial distributions of ELF/VLF wave activities and 0.1-3 MeV electron fluxes during the same storm have been investigated in [94].

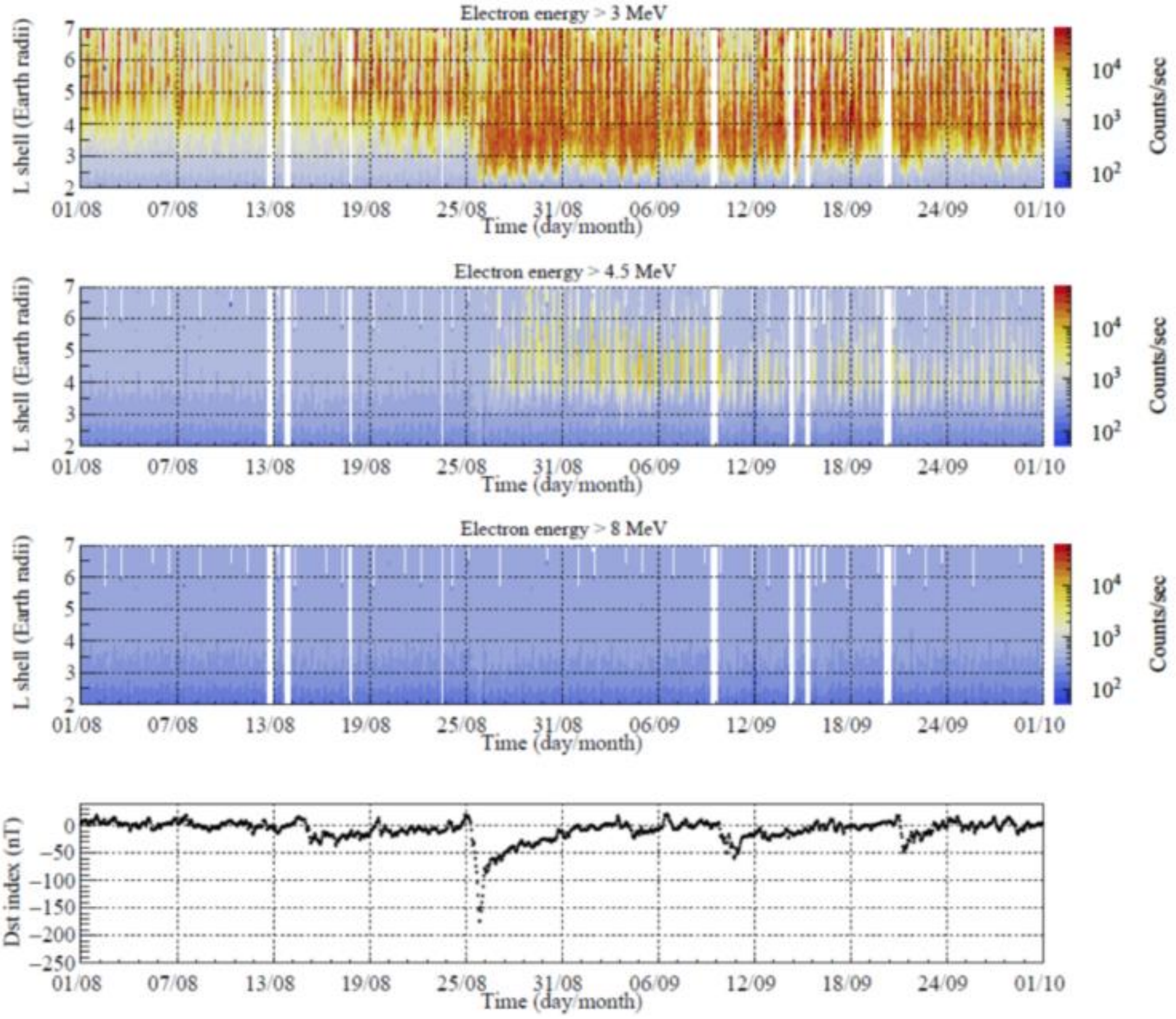

Figure 16 - a) 3-MeV electrons particularly disturbed by the arrival storm
b) and c) this effect gradually weakening and completely disappearing at about 8 MeV
d) Disturbance index, describing the intensity of the storm on Earth.

**Galactic Cosmic Rays (GCR) Solar modulation** [87].

In the large panel of Figure 17: galactic proton spectra as a function of energy measured by HEPD in the three intervals (from 2018 August 6 to 2019January 15, from 2019 January 16 to 2019 June 28, and from 2019 June 29 to 2020 January 5, respectively). Systematic uncertainties are also present as a yellow shaded area. The continuous curves represent, respectively, the HelMod theoretical spectrum averaged over the period under study (blue solid line), the maximum (dashed line) and minimum (dotted line) expected deviation from the model itself. The red square represents data obtained from the SOHO/EPHIN spacecraft. In the narrow panel shows: the ratio between HEPD data and HelMod model, as a function of energy; the errors on HEPD data are a sum of statistical and systematic uncertainties.

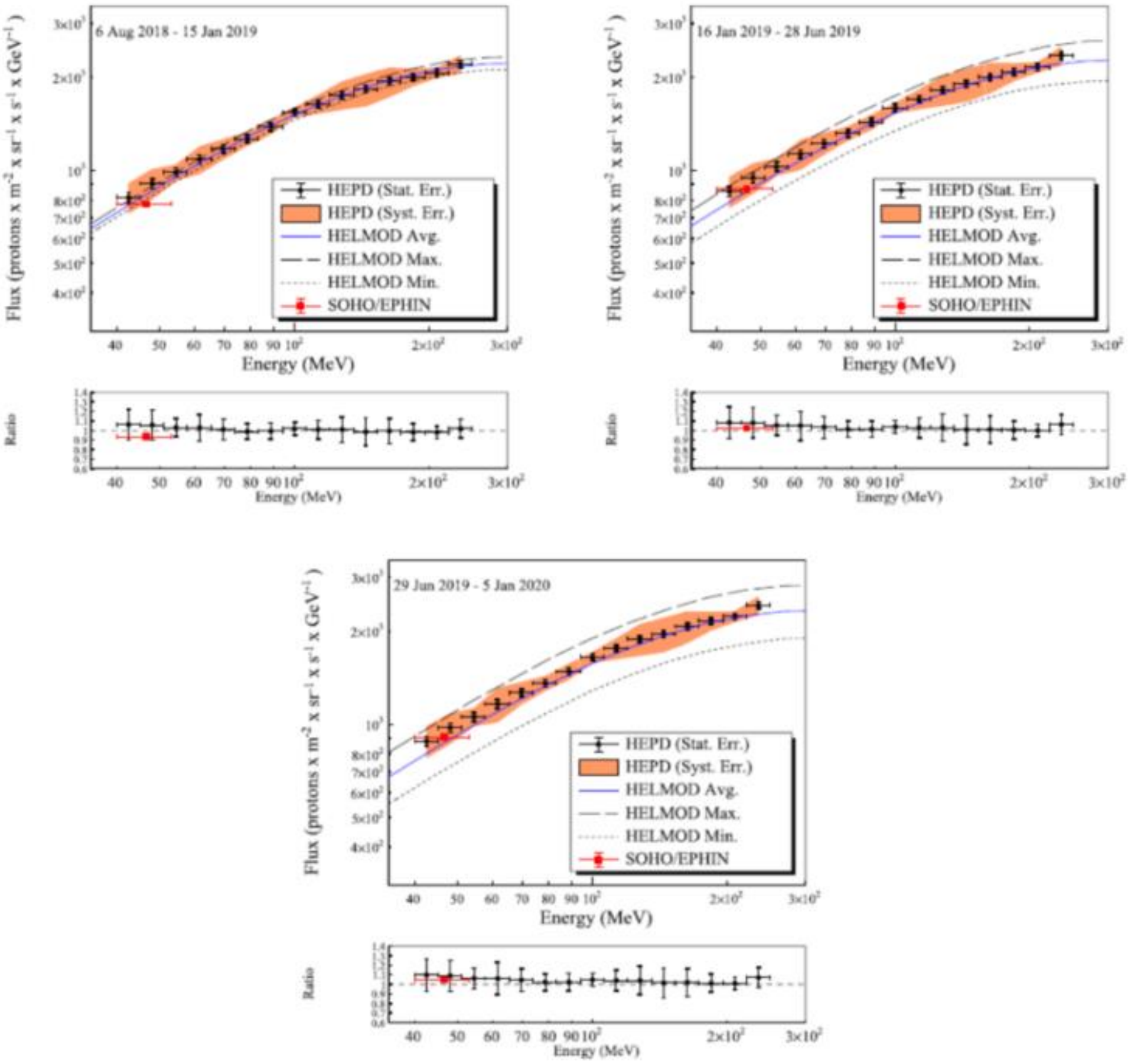

Figure 17 - GCR protons between 40 and 250 MeV, in 3 different periods of solar cycle 24th (from August 2018 to January 2020) measured by HEPD-01. Experimental data are compared with theoretical prediction by the HelMod model (solid curves). The agreement between data and model is very good [95].

Other important results can be found in the following bibliographical references [ 88, 90, 91, 92, 93].

## 2.6 SOME MAJOR RESULTS FROM EFD-01 DATA ANALYSIS

At a first approximation, the Earth can be considered as a conducting sphere, covered by the neutral atmosphere. In the ionosphere, where the conductivity is significant, any atmospheric electric discharge (such as Transient Luminous Event or TLE, lightning, elves, and so on) can produce broadband electromagnetic waves propagating between ground and the ionosphere.

This produce standing waves whose wavelengths are directly linked to the radius of the cavity, and their occurrence probability peaks over the continents.

Such a phenomenon is collectively named the set of Schumann resonances, whose lowest four eigenmodes are approximately placed at $7.8, 14.3, 20.8$ and $33.3 Hz$, respectively ([69] and reference therein, Figure 18).

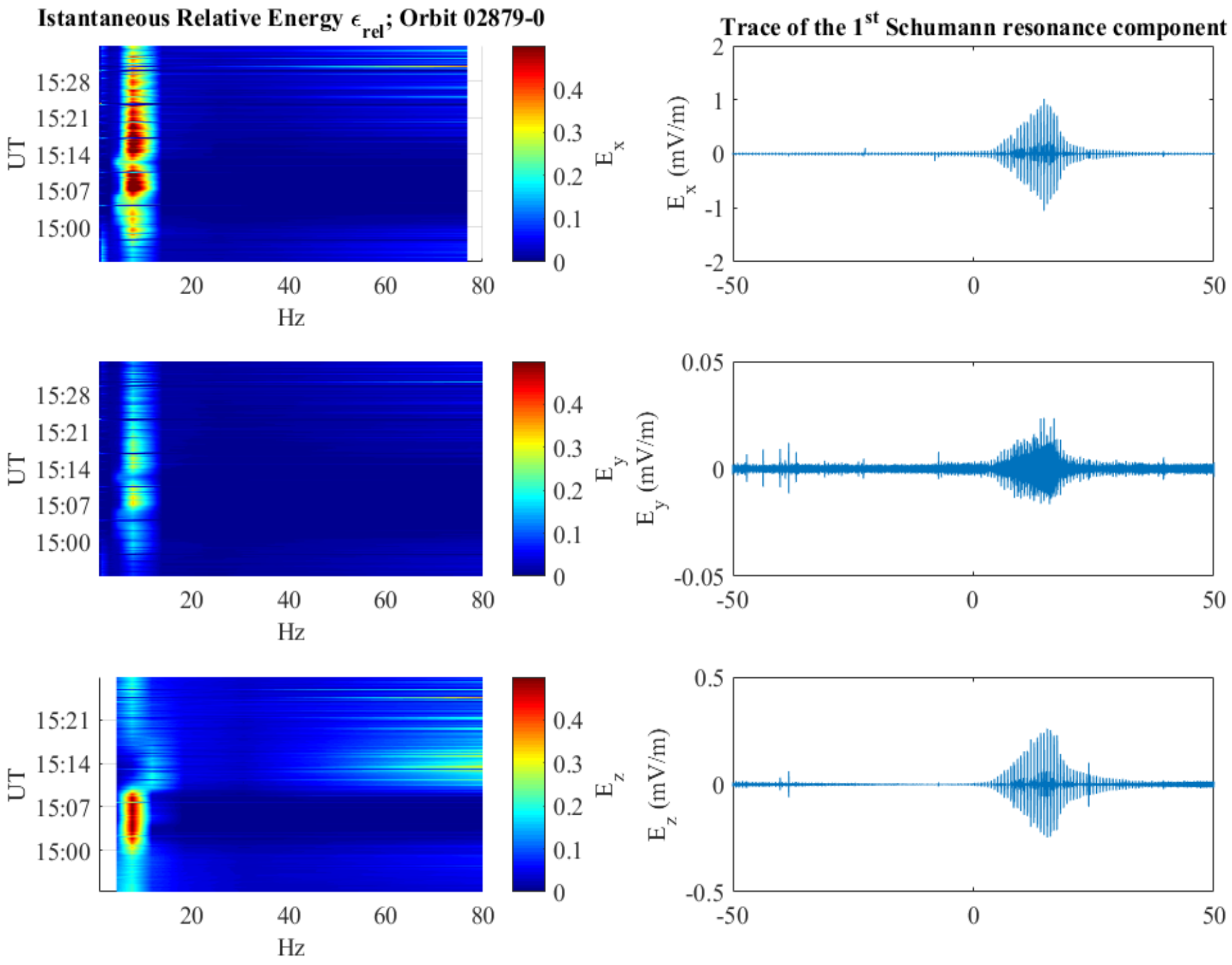

Figure 18 - - Schumann resonance signals in spectrogram in the left panel and trace of $1^{st}$ Schumann resonance component in time domain in the right panel [69].

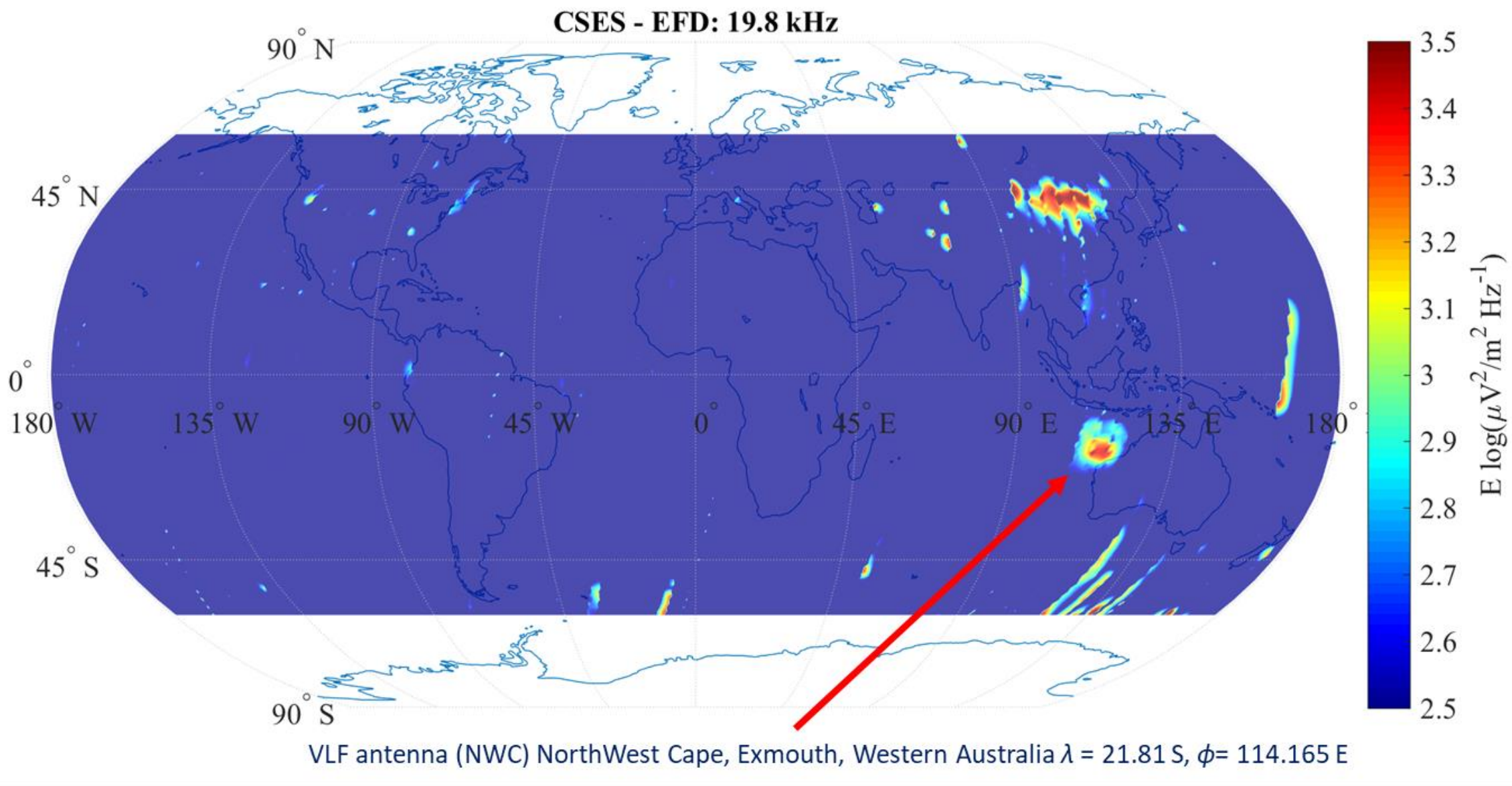

Figure 19 -- Geographical map of the total electric field, power integrated along the entire 2019 set of night-side semi orbits, and filtered at 19.8 kHz, as recorded by the EFD detector on board CSES-01. The intensity is color coded according to the scale on the right [69].

An example of CSES-01 detection of VLF ground transmitter signals is reported in Figure *19*, which shows the spectrogram of the electric field components recorded on 2 February 2019 between 18:27 UT and 19:02 UT close to the 19.8 kHz NWC transmitter (Northwest Cape, Exmouth, Western Australia $\lambda = 21.81° S$, $\phi = 114.165° E$).

The transmitter frequency clearly appears at 19.8 kHz in the spectrogram for all the electric field components. A huge increase in the power is recorded above the emitter, as expected for ionospheric heating induced by the VLF transmitter, and quite well observed by the EFD.

The intense electromagnetic field excited by NWC transmitters can be clearly observed in both the areas above the transmitter and the one in its geomagnetic conjugate hemisphere. The electric field response in the conjugate hemisphere is mainly evident at $L = 1.3–2.5$ (where L is the McIlwain parameter).

This feature can be explained in terms of VLF waves in both ducted and non-ducted way. Indeed, Kulkarni et al.  and Zhao et al. [21] suggest that non-ducted VLF transmitter signals can reach the opposite hemisphere very close to where a ducted signal could be reached, which means that ducted and non-ducted propagation modes cannot be separated at some L-shell.

The amplitude of the VLF wave is relatively smaller at the conjugate region if compared to the space region over the transmitters, due to the Landau damping effect occurring when it approaches the high wave normal angle [69 and reference therein].

During the commissioning phase of the satellite, these VLF antennas can be used as in-flight calibration sources, additional to the necessary calibration already performed on ground.

## 2.7 ESSENTIAL COMPONENTS OF EFD-02

1. **The Electric Field Probes** (EFPs) are four, identical sensors, each of which is housed inside a spherical shell placed at the end of a satellite boom (see Figure *20*). Each sensor has the task of detecting the electric potential with high precision. A voltage adapter with a high input impedance is the core of the sensor's Front-End electronics.

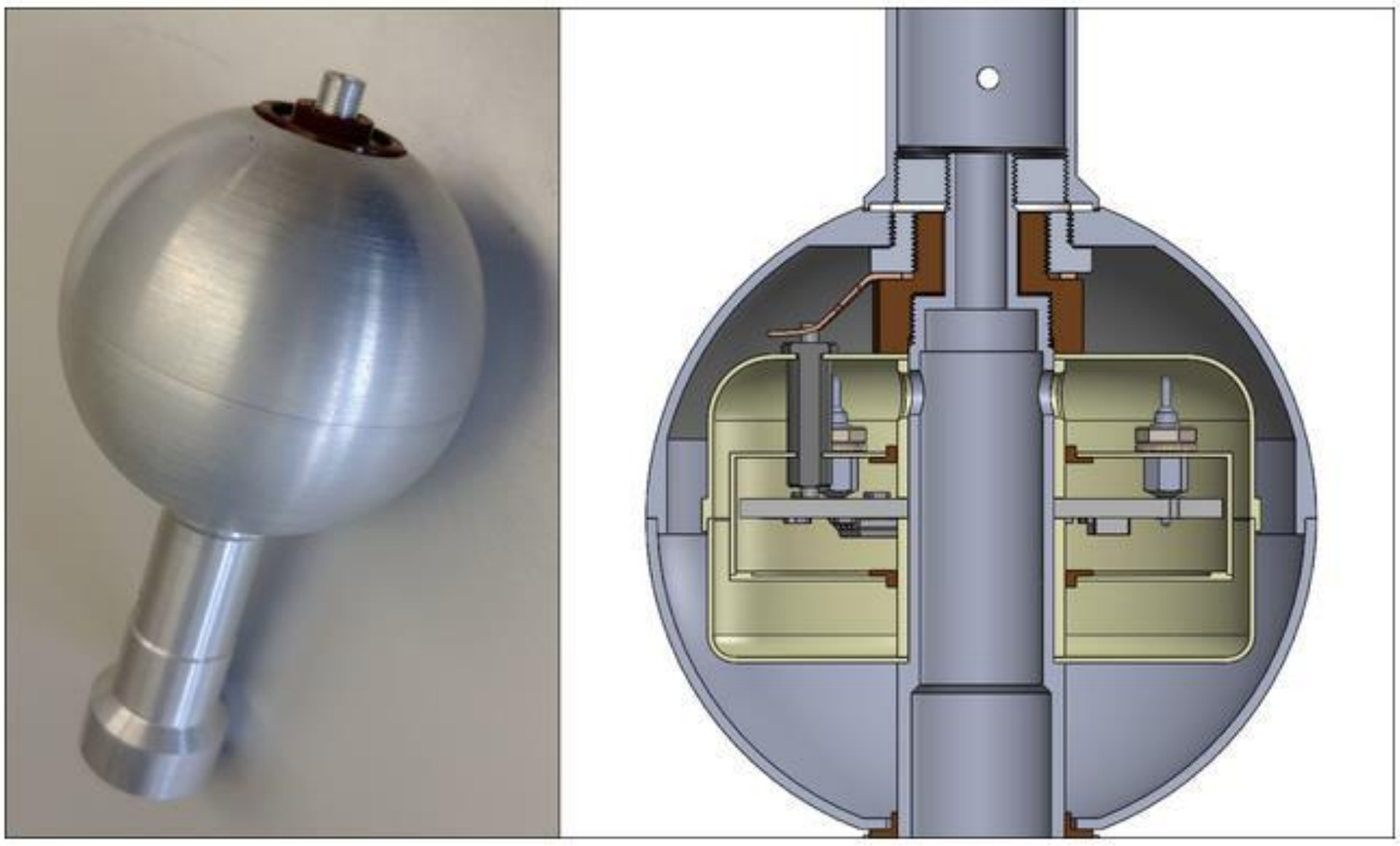

Figure 20 - Photo of the probe and representation of its internal sections.

2. The **5 electronic boards**, placed inside a metal rack (the box in Figure *21*) at a specific position are:
   - **The Low Voltage Power Supply** and **Control** (LVPS & CTRL): power supply, housekeeping and $TM/TC$ interface towards the satellite.
   - **The Analog Processing Unit** (APU): analog signal processing board.
   - **The Digital Processing Unit** (DPU): digital processing board, On-Board Data Handling (OBDH), command and control of the payload.
   - **The Splitter,** which enables the switch between HOT and COLD electronics in case of failure.
   - **The Backplane,** designed for the interconnection between electronic boards.

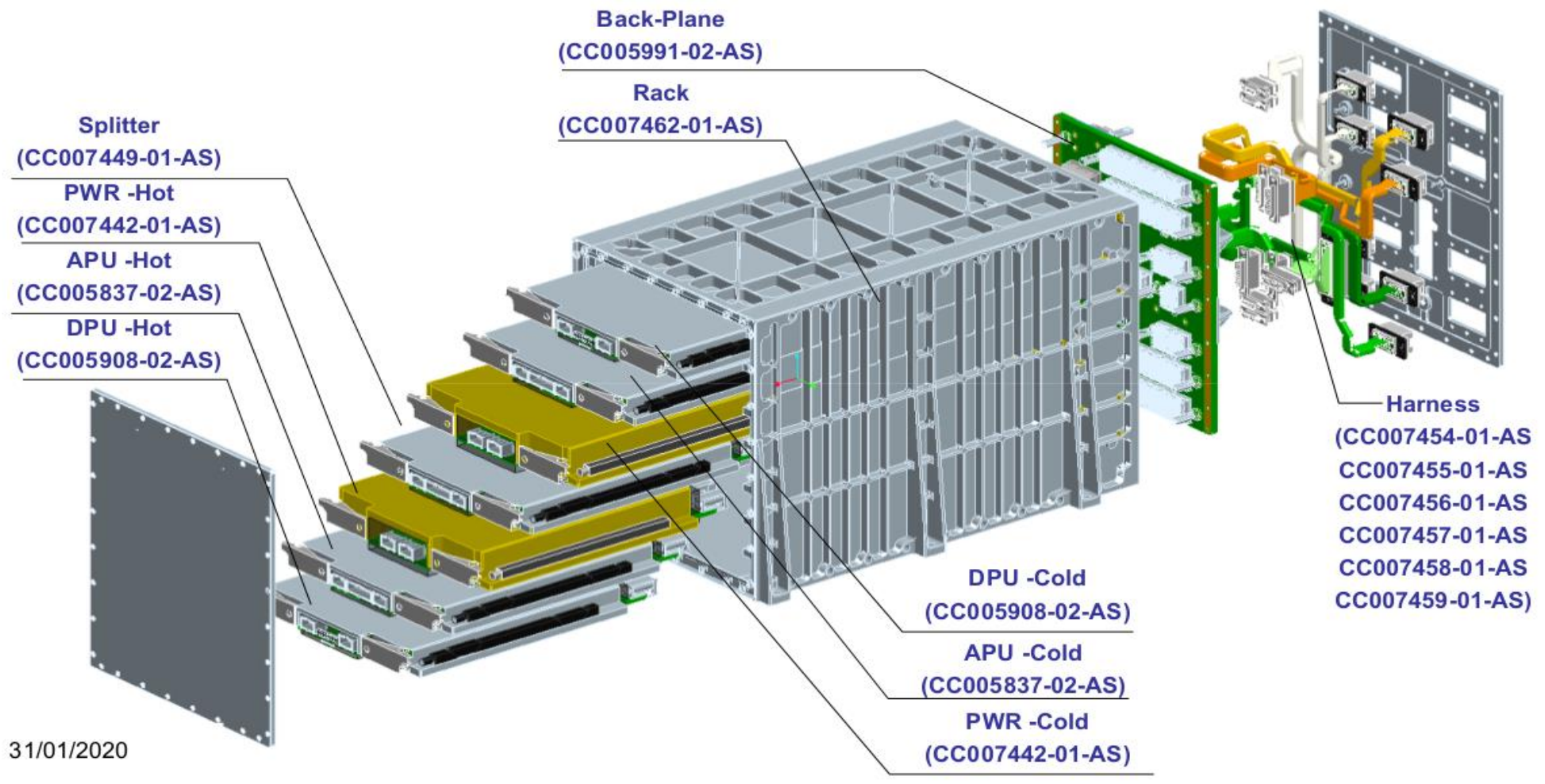

Figure 21 - Arrangement of the various EFD's electronic boards inside the satellite rack.

All the boards, except the Backplane and the Splitter, are duplicated into a "Hot" and "Cold" version for the sake of redundancy. In Figure 22 it is shown a scheme of the connections between the EFPs and EFD-02 boards.

The signals picked up by the probes, through the backplane, are routed to the Splitter board. The Splitter directs the signals and activation of the power supplies to the selected hot or cold part, a choice that can be modified from ground via telecommand. Once the signals reach the APU board, they are processed and digitized, and finally sent to the DPU that packages them in an appropriate way before delivery to Earth.

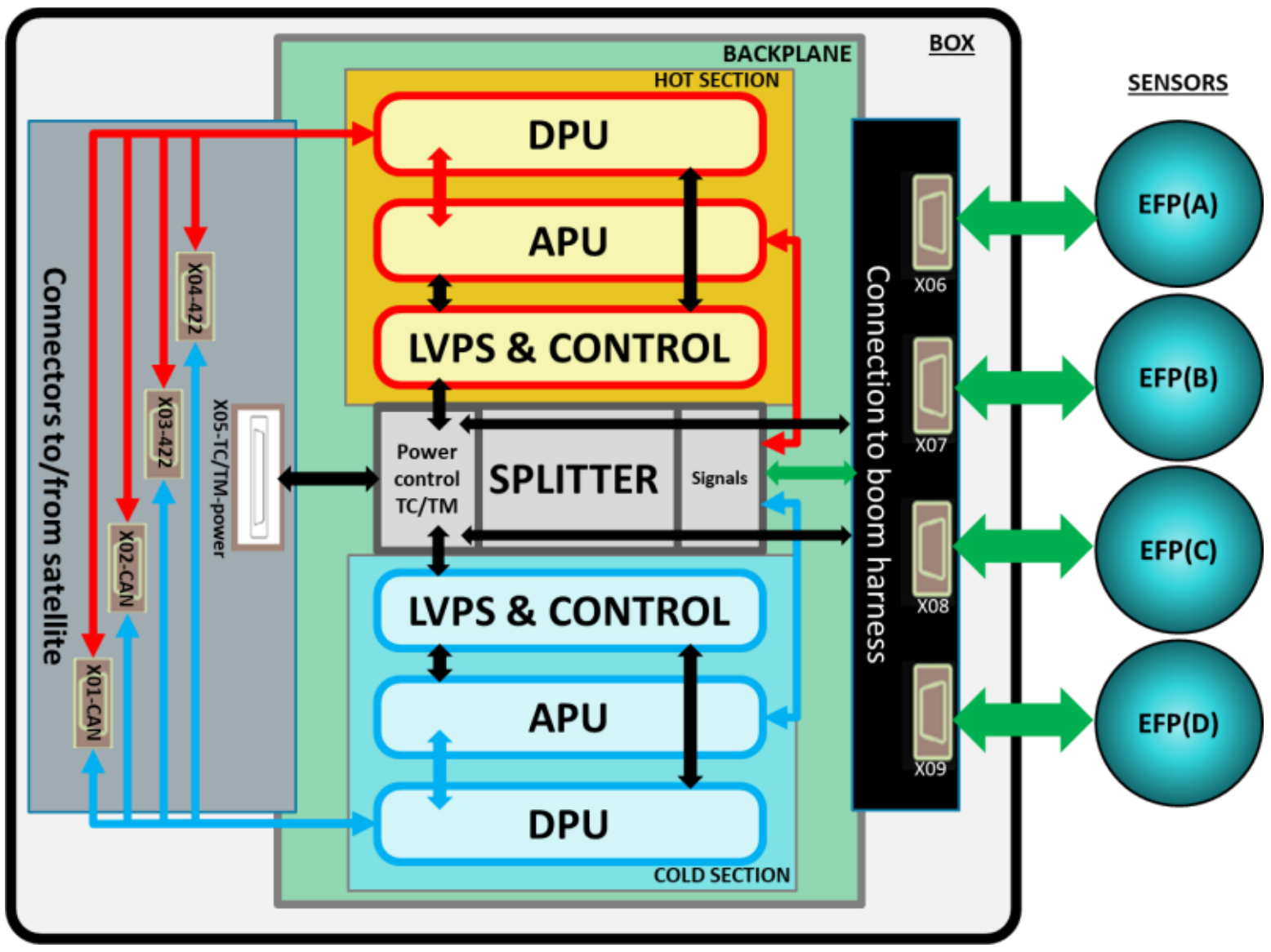

Figure 22 - Scheme of the connections between sensors and EFD boards.

## 2.8 TECHNOLOGICAL INNOVATIONS OF EFD-02

Compared to previous detectors, exploited for the same purpose, (EFD-01, ICE) on board CSES-01 and DEMETER missions respectively, EFD-02 provides:

- a better organized subdivision of acquired signals into 5 bands (ULF, ELF, VLF, VLFe, HF), as shown in the Table 3.
- a higher sampling frequency, allowing to observe more thoroughly the variation in the electric field due to perturbations from solar, seismic, and anthropic phenomena.
- An increased number of acquired channels and a higher bit depth of data (both for scientific purposes and redundancy).
- The possibility to choose, via remote control from ground, the pair of probes for the computation of the 3 components of the electric field (Switch Matrix).
- The possibility to turn off and isolate any of the probes, both for test reasons and in case of failure.
- A new bias current control algorithm, to avoid saturation phenomena in perturbed plasma conditions.

| Band | Type | Frequencies band | #channels | Sampling Frequency | Resolution (bit) |
|---|---|---|---|---|---|
| ULF | Wave | 0 – 100 Hz | 4 | 244 Hz | 24 |
| ELF | Wave | 19 Hz – 2 kHz | 3 | 4000 Hz | 20 |
| VLF | Wave+FFT | 1 kHz – 30 kHz | 3 | 60 kHz | 16 |
| VLFe | Wave+FFT | 21 kHz – 100 kHz | 3 | 200 kHz | 16 |
| HF | Wave+FFT | 21 kHz – 3,7 MHz | 1 | 8 MHz | 12 |

Table 2 - EFD band specifications.

One of the most important features is that EFD-02 will be able to measure the electric field components over a wide-band frequency range ($DC - 3.5MHz$), and with high sensitivity (about $1\,\mu V/m$) in the ULF band, which is the band of greatest scientific interest. All these features will be discussed in detail later.

It has been possible to test the instrument functionality using a Plasma Chamber in an INAF facility at Tor Vergata, as described in detail in Chapter. The Plasma Chamber makes it possible to create a flight-like environment suitable for performance tests to be compared with modeling.

All the choices are the result of a trade-off between the "highest performance" ensured by the updated electronic technologies and the constraints imposed by the satellite spacecraft in terms of power consumption and downlink data rate.

In general, resolution and bandwidth have been increased to optimize data quality and maximize the records in each band; for all the bands, the ADC sampling frequency is suitable for ensuring correct sampling, in order to recover clearer signals and optimize the S/N ratio.
The digital data processing unit applies the necessary and unavoidable reduction of the dataflow requested to be compliant with the downlink data rate.

One of the novelties is the Switch Matrix, i.e., the possibility to choose the pair of probes for the computation of the three components of the electric field thanks to an array of switches that carry out any possible signal difference from the four EFPs, unlike the fixed default pair differences in EFD-01.
An important feature is the possibility to turn off and isolate any of the probes, as required for testing and in case of failure.

In the following, some details of EFD-02 specifications:
• The Ultra-Low Frequency (ULF) band has been extended up to 100 Hz. This broadening allows the monitoring (at very high resolution) of a larger variety of low-frequency waves and perturbations that can propagate inside the ionospheric cavity and through the ionosphere up to the satellite orbit, thus providing insight in the lithosphere-atmosphere-ionosphere coupling mechanisms.

• The Extremely Low Frequency (ELF) band (having bit and frequency resolution higher than in the VLF band) ranges from 19 Hz to 2 kHz. The upper frequency limit set to 2 kHz is to ensure a significant overlap with the VLF band (which – as in all space missions – has a coarser sampling). In this way, in the overlapping range, the higher quality of ELF waveform can complement poorer information from FFTs in the VLF band. The ELF band is useful for the detection of secondary emissions due to gravito-acoustic waves (order of hundreds Hz).
For the medium frequencies (MF), 15 Hz up to 17.4 kHz in the case of the DEMETER/ICE instrument, the band has been extended up to 100 kHz and split into two sub-bands: 1 kHz – 50 kHz (VLF) and 21 kHz – 100 kHz (VLFe, or VLF extended). The lower portion of VLF frequencies - significantly overlapping the ELF band - allows an optimal detection of whistlers and seismo-associated phenomena.

I Indeed, according to literature, experimental DEMETER data, and modeling, the medium frequency range should be the most promising for detecting electromagnetic precursors.

To this purpose, it must be emphasized that observations and analyses of seismo-associated electromagnetic precursors have been mainly reported (both on ground and in space) in the ELF/VLF frequency band, whereas no significant detection in space has been reported in the HF band. At the same time, the widening of the VLFe band ensures a better frequency resolution for FFT data up to 100 kHz (w.r.t. the HF band). The FFT average values, together with standard deviation (SD) and Kurtosis, provide signal information in higher VLF frequencies without significant loss of information (FFTs are computed every 40ms with a spatial resolution of about 300 m).

• At the highest attainable frequencies (HF band, from 21 kHz to 3.5 MHz), the instrument sequentially monitors one of the three electric fields components (x, y, z) at a time with a high switching speed, a mode necessary to meet the budget constraints of consumed power. HF is useful for the monitoring of fast plasma oscillations (FFTs are computed every $16ms$ with a spatial resolution of about $120m$). It is worth to note that observations of seismo-associated electromagnetic perturbations have been rarely reported in the HF range and essentially only by ground-based measurements, the reason being that the ducted propagation along magnetic field lines is more efficient in the ELF/VLF range. Typically, the radio wave propagation, through the ionosphere up to satellite altitude, is limited below the cut off at the plasma frequency. Over daytime crossings, the plasma frequency is of several MHz (higher than the maximum frequency detectable by ICE, EFD-01 and EFD-02), whereas, at night, at middle latitudes, its value can reduce down to a few MHz, thus allowing some HF emissions to propagate beyond the ionosphere.

On the other hand, local plasma instabilities may generate secondary HF disturbances. Therefore, the monitoring electric phenomena in the HF range is needed, even though HF is not the privileged band, candidate for detecting seismo-electromagnetic phenomena. For these reasons, a trade-off (between scientific requirements, data rate, and power consumption) has been applied to limit to one the number of HF components measured at a time. Also, the French TARANIS mission (that is, the second generation of the Myriad satellite series after the Demeter mission) has adopted the same approach [76].

The normal operations in the HF band involve acquisition in sequence of all the three components of the electric field by including a minimum delay time, needed for shifting (sequentially) the input signal from the HF ADC between the three components (x, y, z).

Since the earthquake preparation zone (e.g., Dobrowolski et al. ,1979), where the seismo- associated deformation can be detected, is of the order of hundreds of kilometers (for medium and strong events), the satellite crossing time is many orders of magnitude larger than the delay time needed to sequentially acquire all three electric field components.

Therefore, the acquisition in fast sequence of the three HF components does not introduce a significant limitation.

Finally, in case of need to acquire statically one of the three HF components, it is possible to change the EFD-02 data acquisition system to an appropriate TEST mode.

EFD-02 data output and weight are shown in **Errore. L'origine riferimento non è stata trovata.**.

| Band | ULF | ELF | VLF FFT | VLF extended FFT | HF FFT | Total |
|---|---|---|---|---|---|---|
| Frequency | DC-100 Hz | 15Hz- 2kHz | 1kHz-30kHz | 21kHz-100kHz | 21kHz-4MHz | |
| Sampling (sps) | 2.50E+02 | 4.00E+03 | 6.00E+04 | 2.00E+05 | 8.00E+06 | |
| Bit depth | 24 | 20 | 16 | 16 | 12 | |
| Channels | 4 | 3 | 3 | 3 | 3 | |
| Weight per second (bits) | 2.40E+04 | 2.40E+05 | 4.32E+06 | 1.44E+07 | 4.32E+08 | |
| Weight per day (bits) | 2.07E+09 | 2.07E+10 | 3.73E+11 | 1.24E+12 | 3.73E+13 | |
| %age | 100 | 100 | 1 | 0.5 | 0.032 | |
| Survey weight | 2.07E+09 | 2.07E+10 | 3.73E+09 | 6.22E+09 | 1.19E+10 | **4.47E+10** |

The FFT is provided in the form of average values (over 50 FFTs), with relative SD and Kurtosis. In Burst Mode (Survey + VLF WF + VLFe WF for 2h/d), the total data weight is 82 Gbit/day. The Time flow of data is shown in Figure 23.

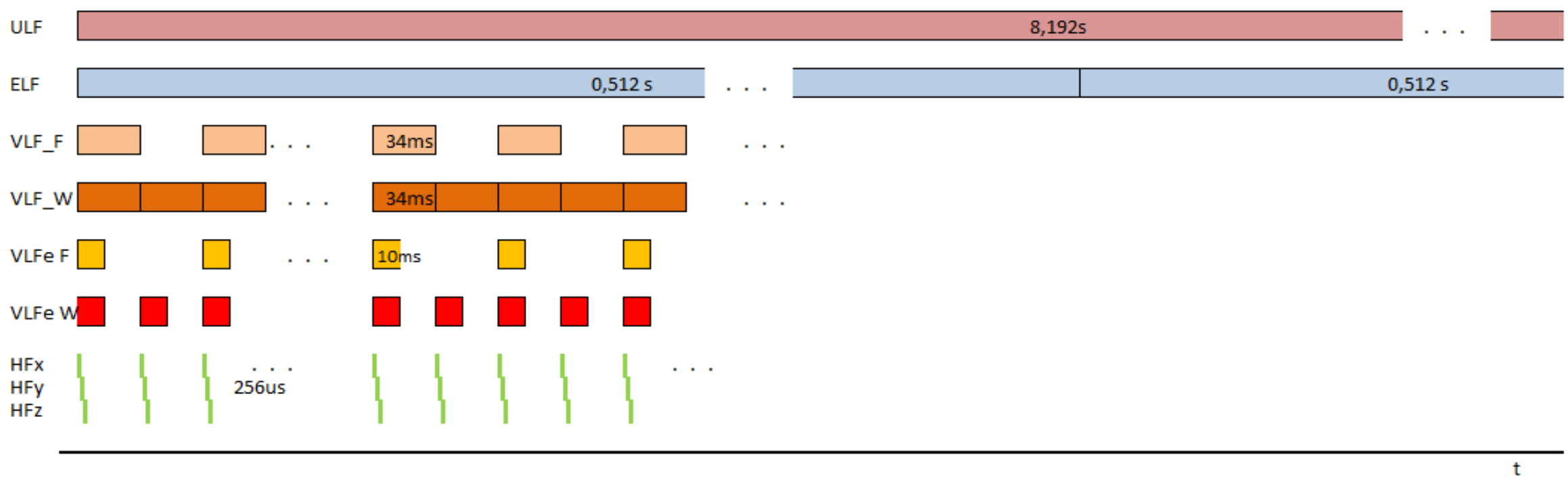

Figure 23 - Time flow of data.

## 2.9 THE ELECTRIC FIELD PROBE (EFP)

The EFP consists of a spherical sensor located at the tip of a conductive boom. It is equipped with cylindrical conducting stubs bootstrapped at the electrode potential, (see Figure 24).

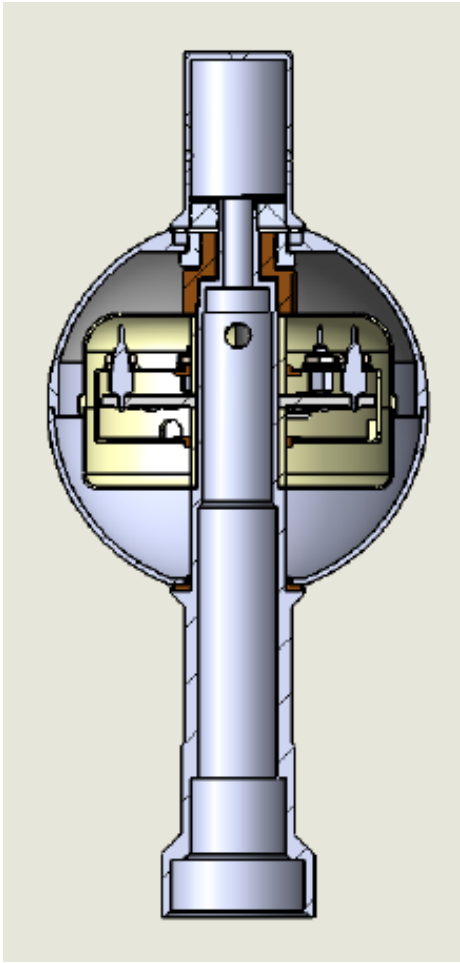

Figure 24 - Electric Field Probe shape and internal sections.

The inner stubs are needed to minimize possible perturbations due to the boom potential (which is that of satellite ground). Outer stubs, formerly introduced to obtain an average symmetry for spinning satellites, introduce a spurious electric field in DC measurements. It is possible to eliminate this spurious field through the removal of outer stubs and the orientation of any sensor toward the Ram direction (so called because it is the side impacting/ramming into the “fluid” the satellite moves in). Nonetheless, since these mechanical modifications would deeply impact to the satellite design, outer stubs were saved in the final sensor design.

The diameter of the probe is $60mm$. The EFP is made of an aluminum alloy covered by a $DAG213$ black graphite coating to ensure thermal stability of the probe electronics along the orbit.

The EFP Front-End (FE) electronics is included in the sphere and located around the upper part of the stub. Its electrical connections are provided by eight spacer screws and two pass-through, Figure 25.

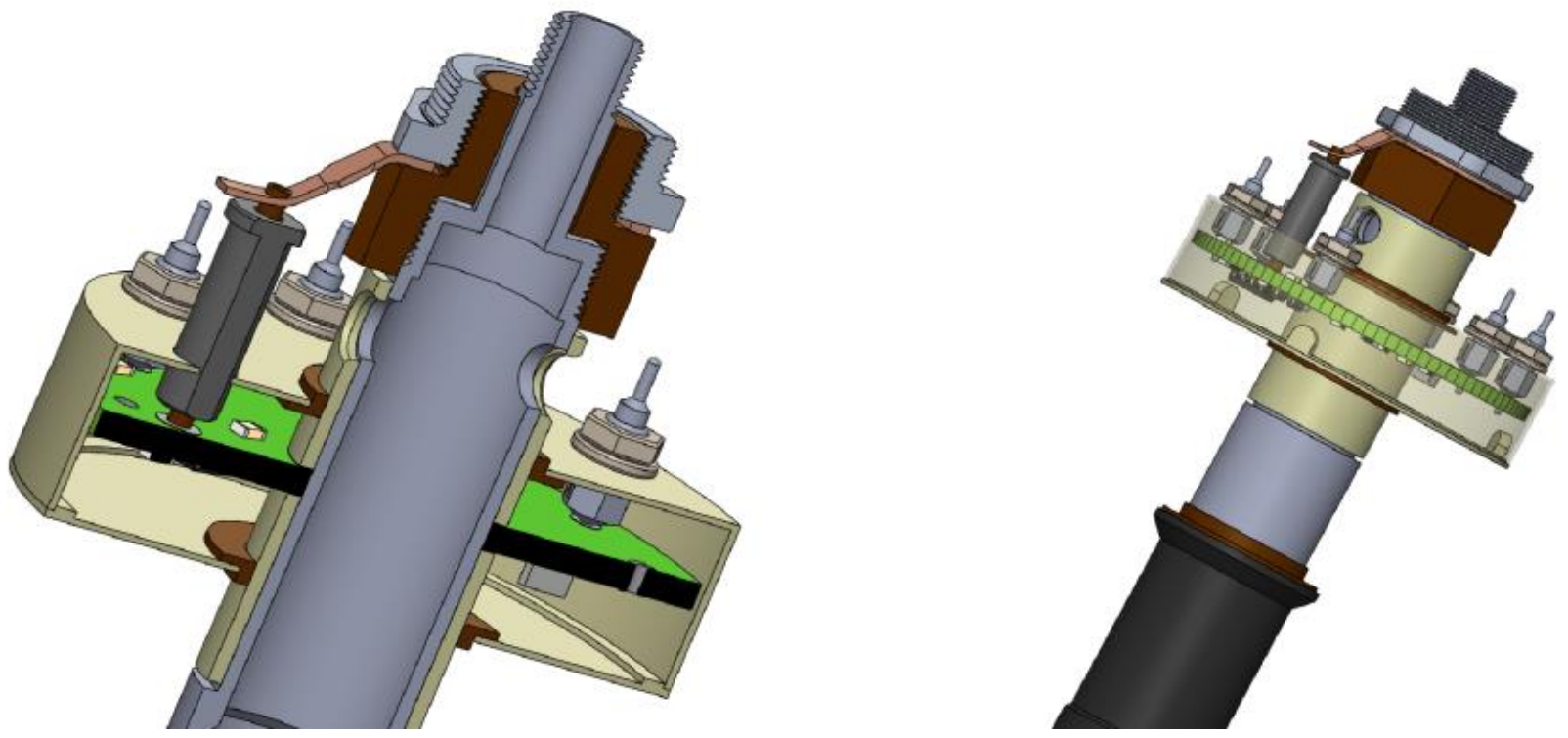

Figure 25 - Displacement of electric connectors of the PCB inside the sphere; a sight-trough of the box shows the PCB and its connectors.

The PCB (printed circuit board) is surrounded by a shielding box. The contact with the sphere and the electronics is provided by an electric connector locked between the insulator (Vespel) and the Aluminum screw. The electrical connection between the sphere and the front end is provided by a copper connector ring held tightly between the insulator (brown barrel in Figure *25* and Figure *26*) and the aluminum screw. This solution allows to separate the locking sequence, thus, improving the overall mechanical rigidity.

The upper hemisphere is then screwed onto the upper part of the aluminum screw ensuring the electrical connection. Finally, the upper stubs can be screwed to the inner stubs assuming the same potential.

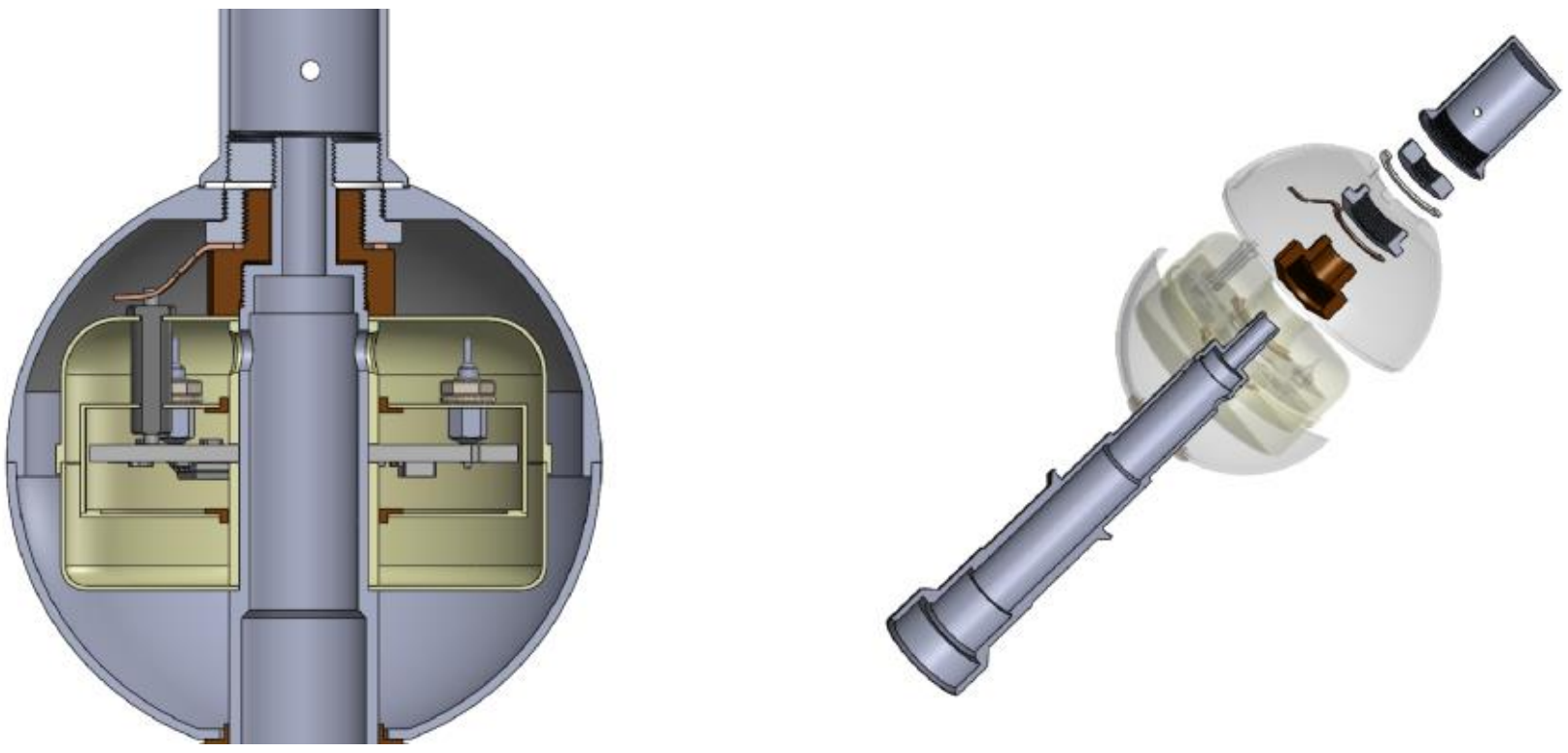

Figure 26 - Left panel shows inner parts assembly. The locking system of the EFP provided by separate screw on insulator, Aluminum screw, and the inner stub. The upper hemisphere is locked by acting with a special key on the top side aperture.

The architecture of the EFP sensor consists of 3 main blocks: the simplified block diagram of the probe electronics is shown in Figure *27*.

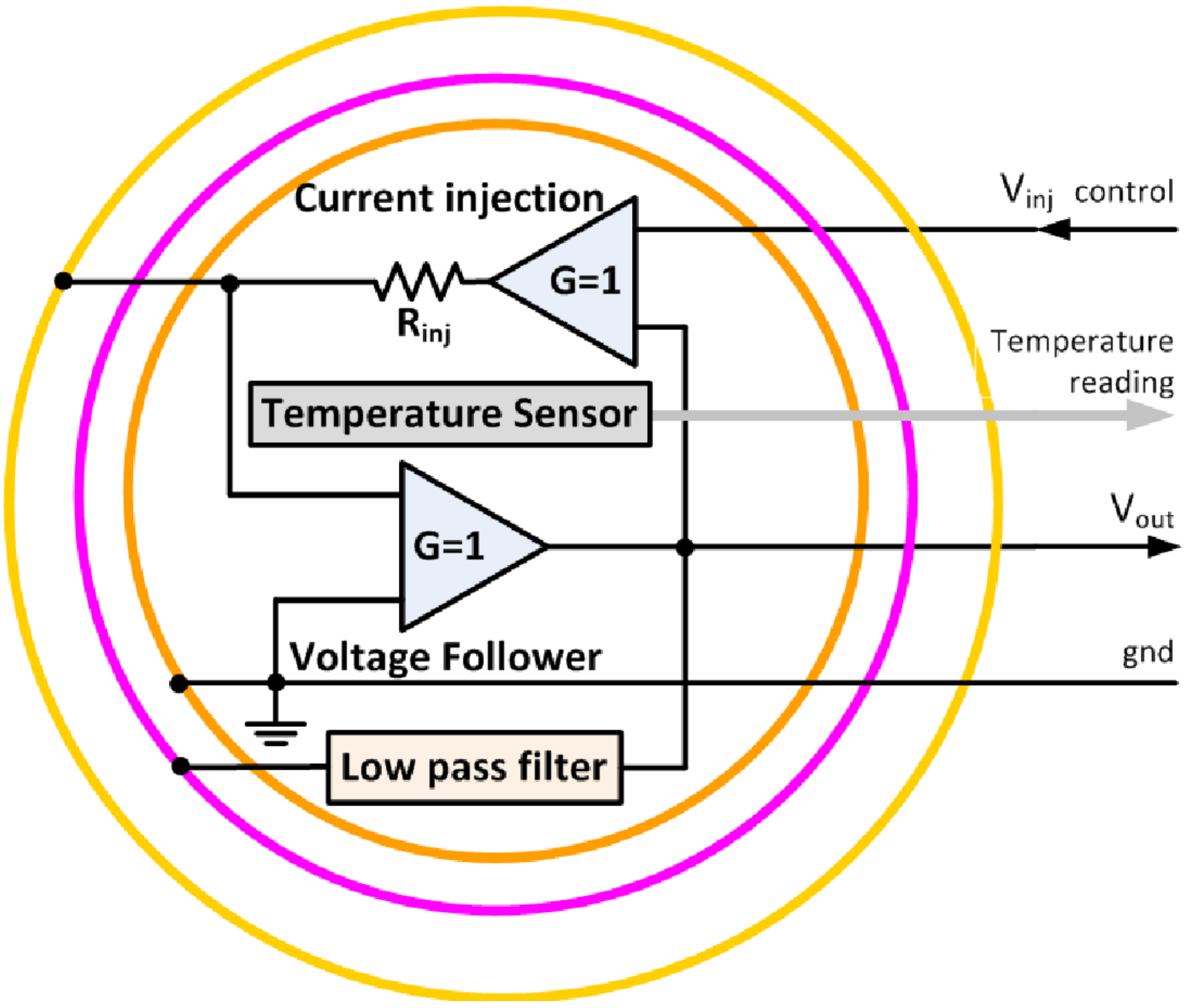

Figure 27 - Probe internal circuit diagram.

- Voltage follower buffer.
- Current injection circuit.
- Bootstrap circuit.

A temperature sensor (PT-1000) allows the monitoring of the temperature inside the probe.

The outermost sphere ($3cm$ in radius) serves as the measuring electrode; the inner one is a screen for the electronics, and it is grounded; while the intermediate shield has the aim to minimize the parasitic couplings between the outer measurement sphere and the ground.

The voltage-follower is a buffer stage obtained by implementation of a low-noise operational amplifier with high input impedance and non-inverting unity gain. It transforms, over a wide range of frequencies, the high impedance exhibited by the floating electrode (in contact with the local plasma) to a value sufficiently low to allow, with good accuracy, the voltage signal transmission through the cables along the booms, down to the processing and control electronics box located within the spacecraft.

The intermediate sphere is bootstrapped to the output voltage of the preamplifier to minimize the capacitance to ground in parallel with the input preamplifier capacitance and to improve the frequency response. The low-pass filter, present in the bootstrap feedback, removes the potential high-frequency instability due to the capacitive positive feedback of the operational amplifier.

The current injection circuit (Figure 28), used to minimize the plasma sheath impedance, consists in a low noise operational amplifier used as voltage controlled current source (voltage to current converter). The non-inverting input is connected to the output voltage of the probe, while the inverting input is driven by the voltage of the processing and control electronics box.

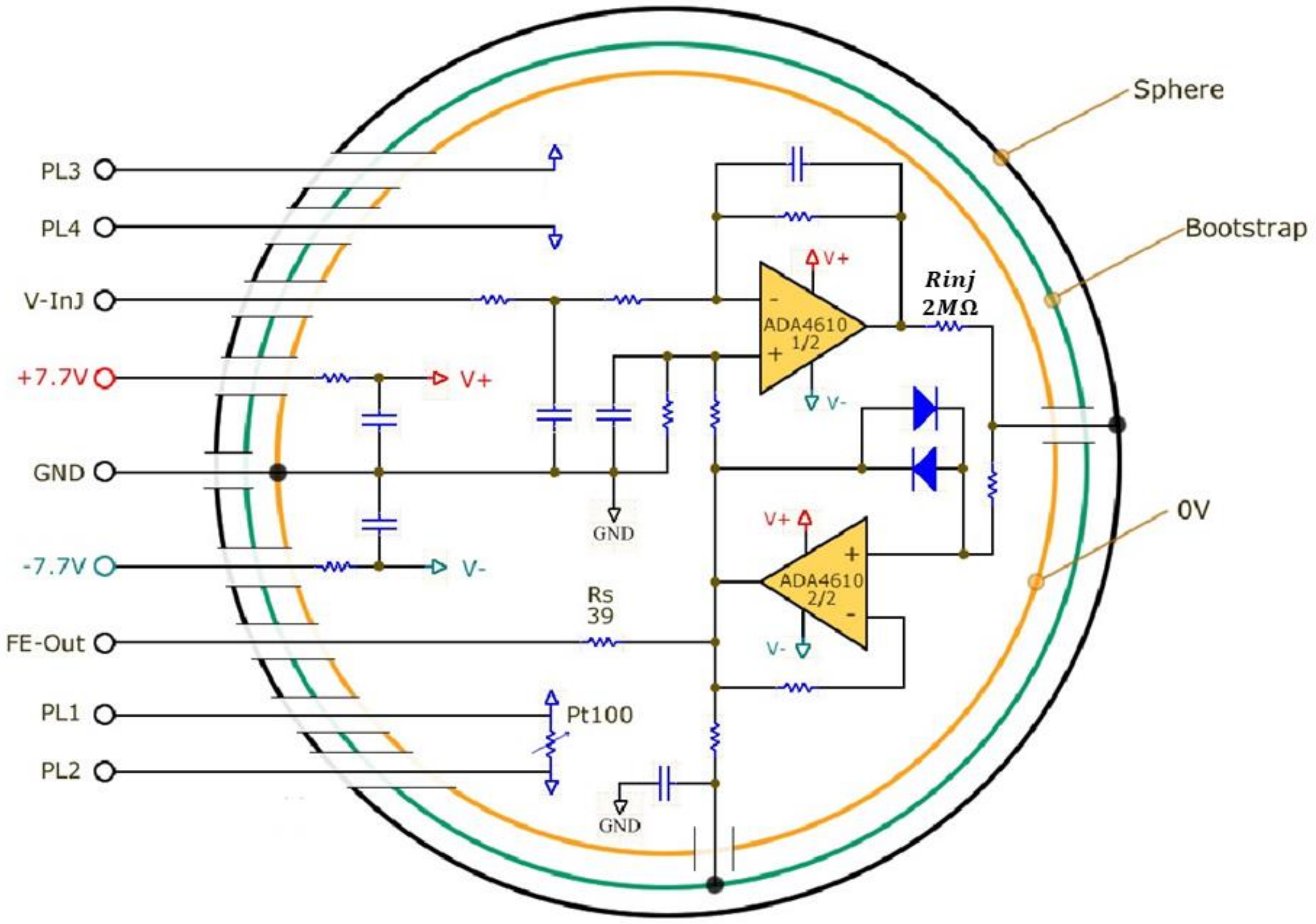

Figure 28 – Schematic of the probe.

Thus, the output current is nominally equal to the ratio between the control voltage and the value of the resistance $Rinj = 2M\Omega$ (see Figure 28). The range for the control injection is consistent with the expected range of the plasma sheath impedance.

## 2.10 Analog Processing Unit

The Analog Processing Unit (APU) board represents the core of the detector and performs a preliminary subdivision into 3 frequency bands (LF, MF, HF) and has the important role of providing analog to digital conversion of signals. See Figure *29* for an overall look at the blocks that make up the APU.

| LF | MF | HF |
|---|---|---|
| DC – 100 Hz | 21 Hz – 100 kHz | 21 kHz – 3.5 MHz |

Table 3 - APU frequency bands.

The ELF, VLF and VLFe bands are produced by the DPU from the MF band. The LF band is further filtered by the DPU to obtain the ULF band.

It is important to note that, due to the constraints on data-budget and power consumption, it has been necessary to contain the throughput for the higher frequency bands. For this reason, for signals in the highest frequency band (HF), the three components are multiplexed and sent to a single ADC, while for Ultra-low frequency signals (ULF) the four potentials are always acquired at the same time, thus maximizing data quality (as to signal-to-noise ratio) in this specific band.
Regarding the Medium Frequency (MF) bands (specifically ELF, VLF and VLFe), the three electric field components are generated via the difference of signals from three pairs of Probes.

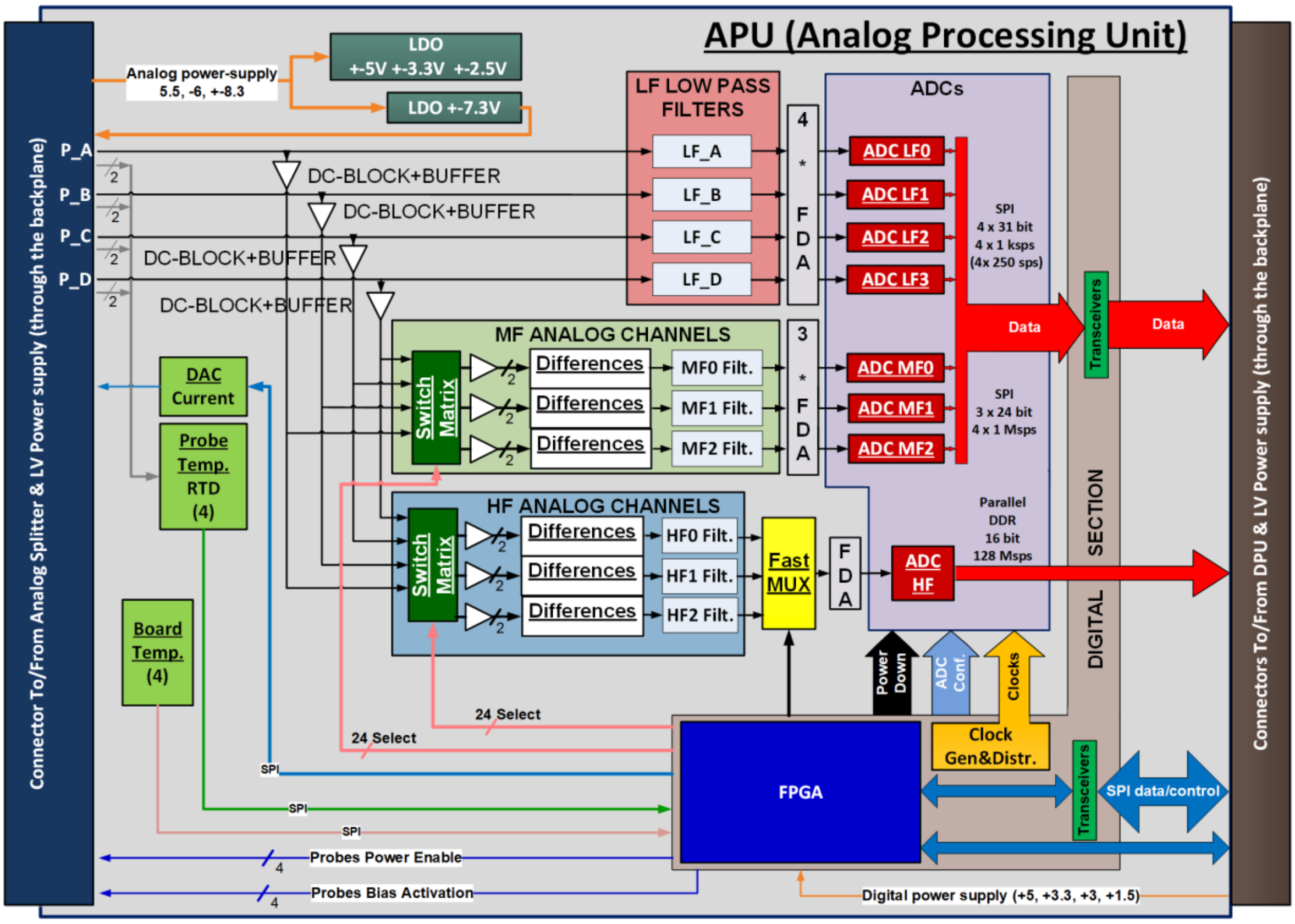

Figure 29 - APU block scheme.

In the LF band, the acquired signals are the four potentials directly measured between each probe and the ground reference.

The four LF channels have nominal voltages in the ± 7.3 V range at their inputs, and each low-pass is a third-order Butterworth filter implemented with a Sallen-Key architecture plus a single pole filter, (Figure 30).

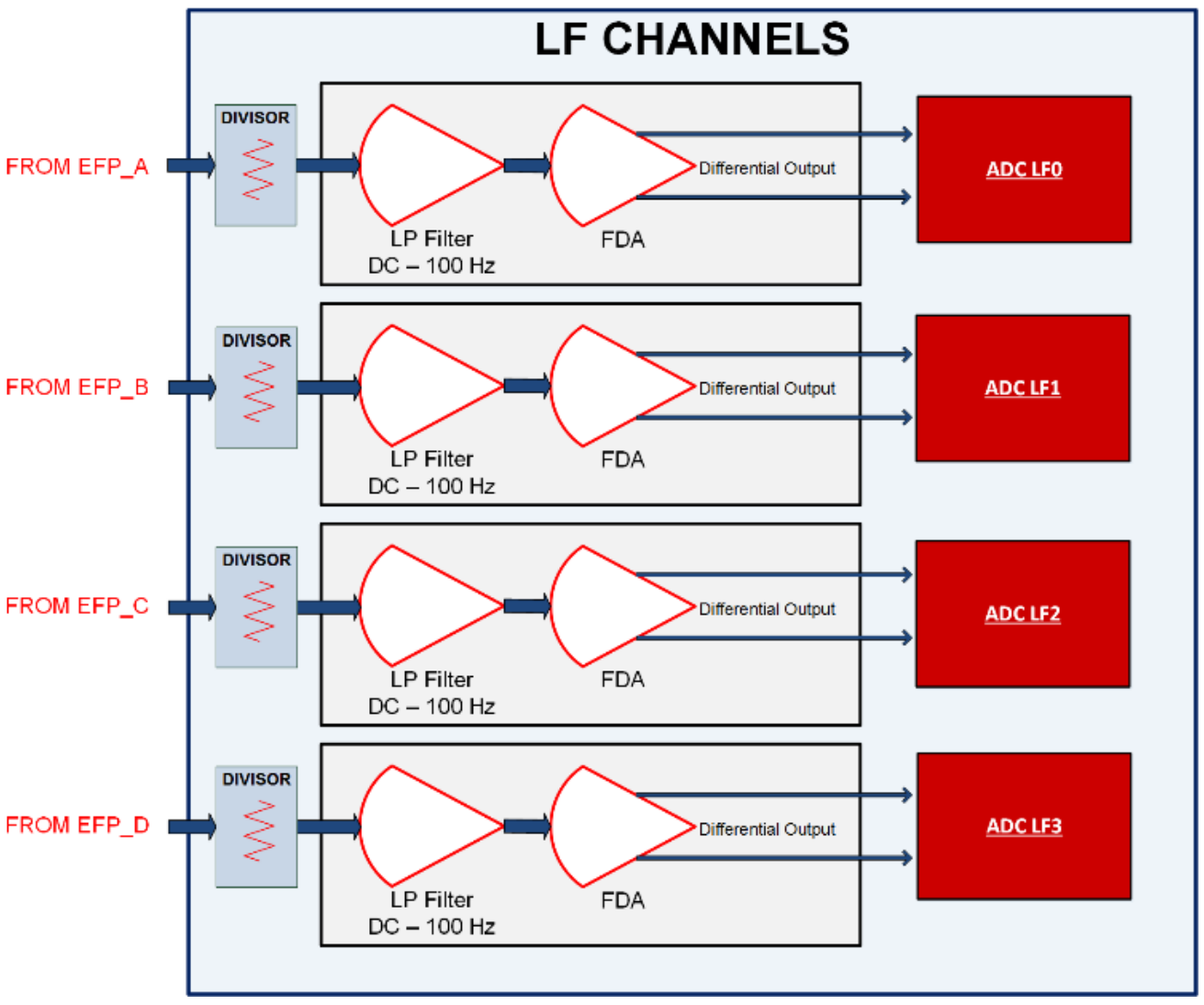

Figure 30 – LF channel filters.

The 3 MF channel input range is $\pm 3\,V$: here, each band-pass is a second- order Butterworth filter implemented with a Multiple-Feedback (MFB) architecture, (Figure 31).

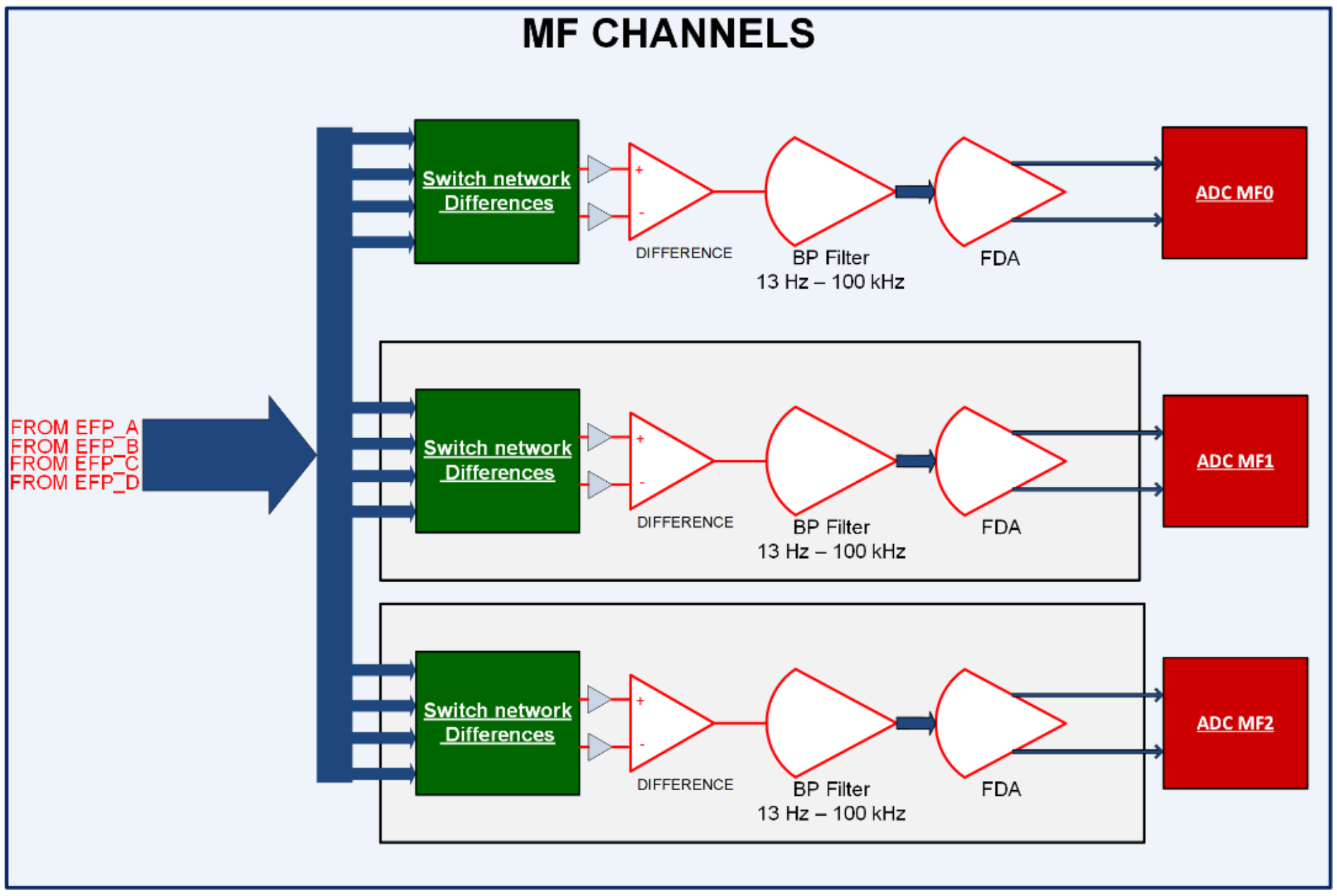

Figure 31 – MF channel filters.

The 3 HF channels are in the range $\pm 1.5V$ : in this case, the band pass filter comprises a cascade of fourth-order Butterworth high-pass elements and a fourth-order Butterworth low-pass filter with Sallen-Key architectures. Low noise op-amps are used for buffers and analog chains, (Figure 32).

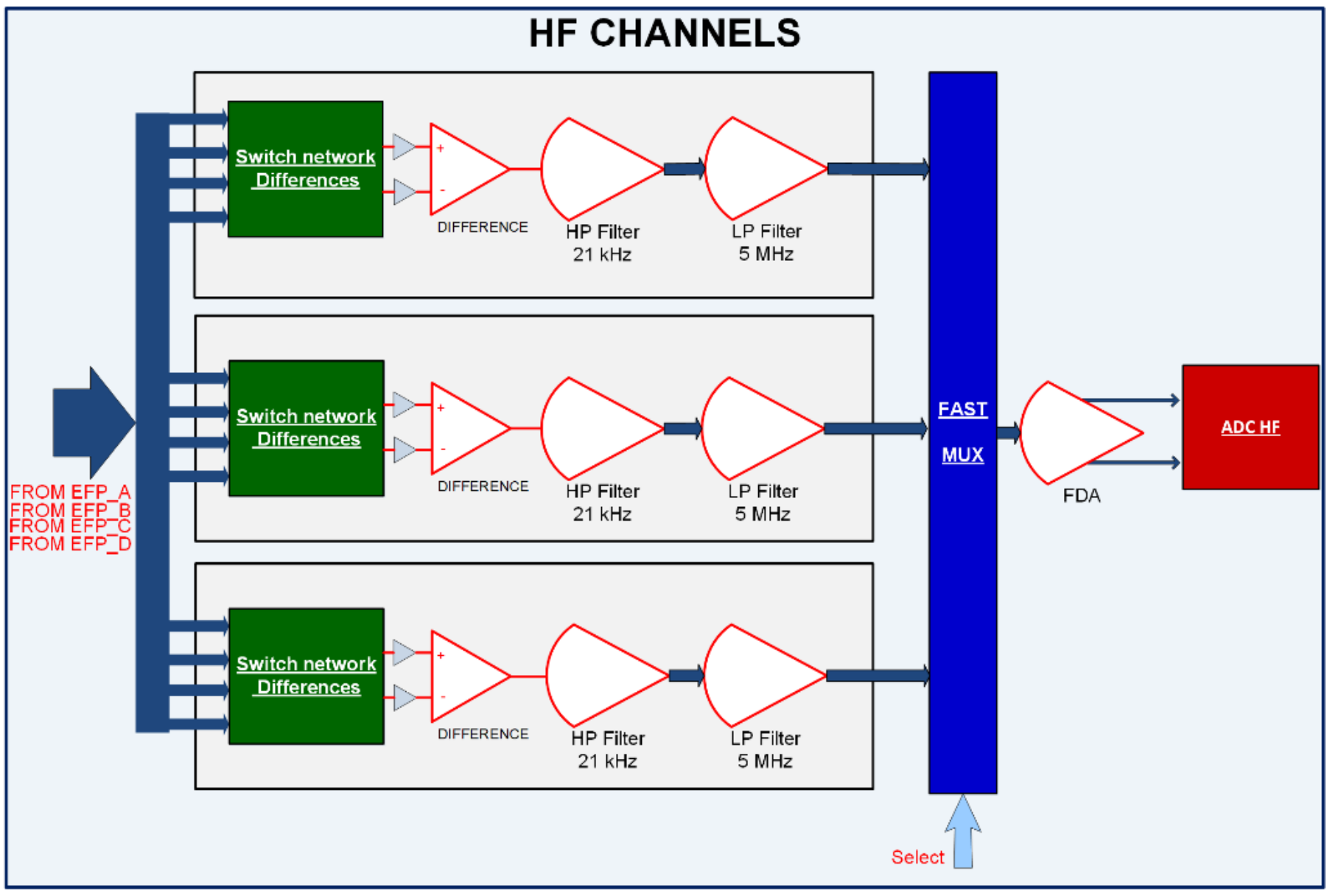

Figure 32– HF channel filters.

The ADCs are directly configured by the DPU, with the exception of the one for HF. Generated data are sent directly to the DPU, in serial mode for the LF and MF channels, and in parallel mode for the HF channel. An FPGA (Field Programmable Gate Array) communicates via SPI (Serial Peripheral Interface) with the DPU in order to generate all switch-matrix block control signals, drive the ADC power-down mode, produce the bias current control voltage, read the Probes and Board temperatures via SPI, and configure the clocks of the ADCs.

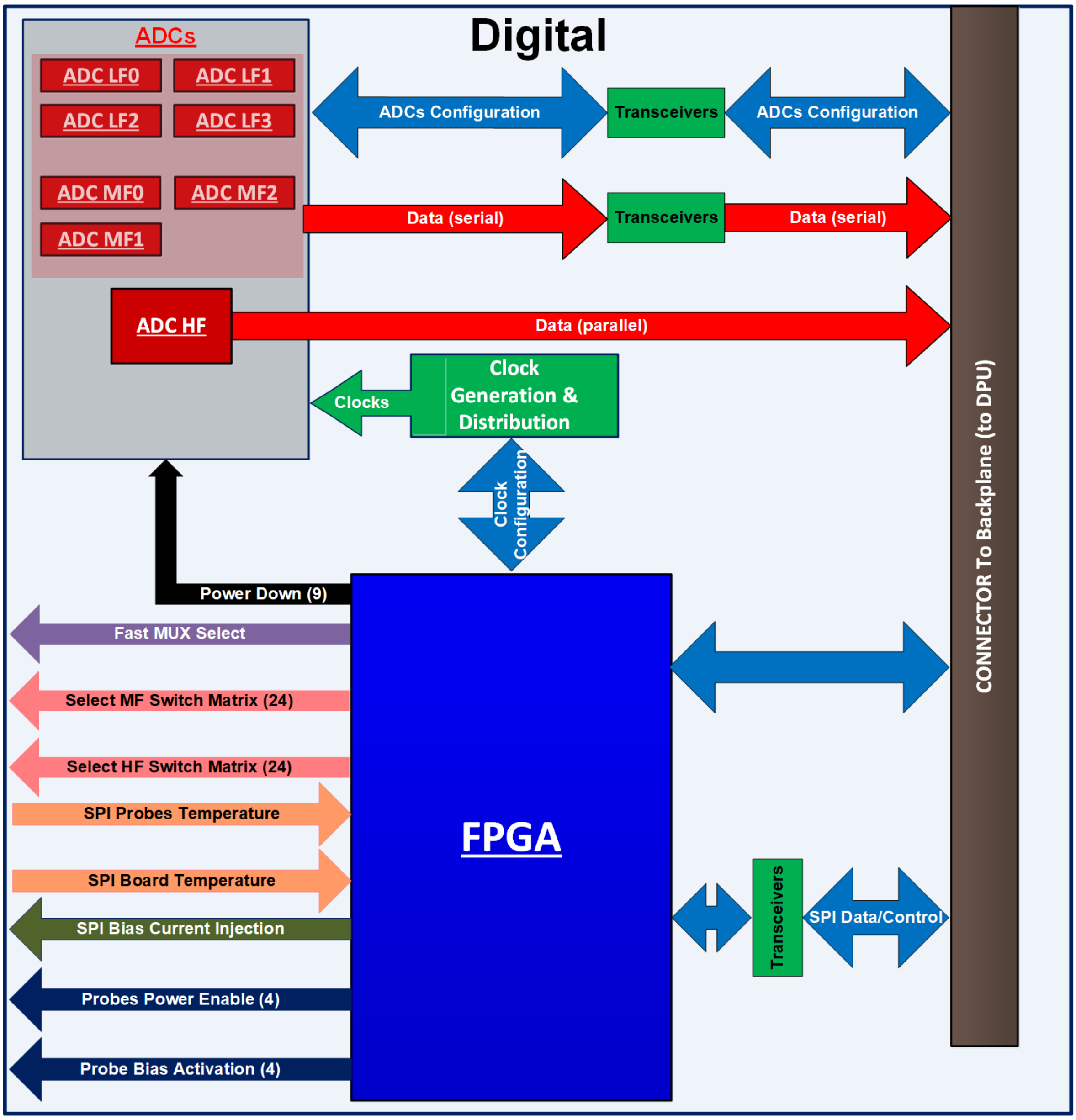

Figure 33 - APU digital section.

All ADCs have differential inputs; the red boxes in the Figures represent the ADCs, each driven by an FDA (Fully Differential Amplifier) for single-ended to differential mode conversion. A 16-bit DAC generates voltages in the ±5 V range, which are sent simultaneously to the four probes and then converted, through a voltage-current converter, into the bias current. Then, data are sent to the DAC from the DPU board. The AC current signal, with amplitude ΔI, produces a corresponding variation in the potential $\Delta V$ of the sensors, and the coupling resistance is thus determined from the ratio $\Delta V/\Delta I$. Such a signal is superimposed on the DC bias current.

The potential is sent to all EFPs through the SPLITTER board. The DAC can operate both in static configuration and in continuous updating, with clock speed and settling time high enough to generate the desired periodic waveform.

Four RTD-to-Digital Converter devices convert the temperature read from a PTC-1000 housed in each Probe. The temperature sensors are placed in properly chosen areas of the APU board to monitor board temperature at various locations.

The contact impedance can be estimated in a special satellite TEST MODE using a sine current waveform with an amplitude that can be adjusted from a few tens to a few hundreds of $nA$ at about $22Hz$ of frequency in the ULF band.

## 2.11 DIGITAL PROCESSING UNIT

The Digital Processing Unit (DPU) board deals with the management and packaging of data on board the satellite, and it is based on a Zynq xc7z7045 system on chip (SOC) from Xilinx. The block diagram is shown in Figure 34.

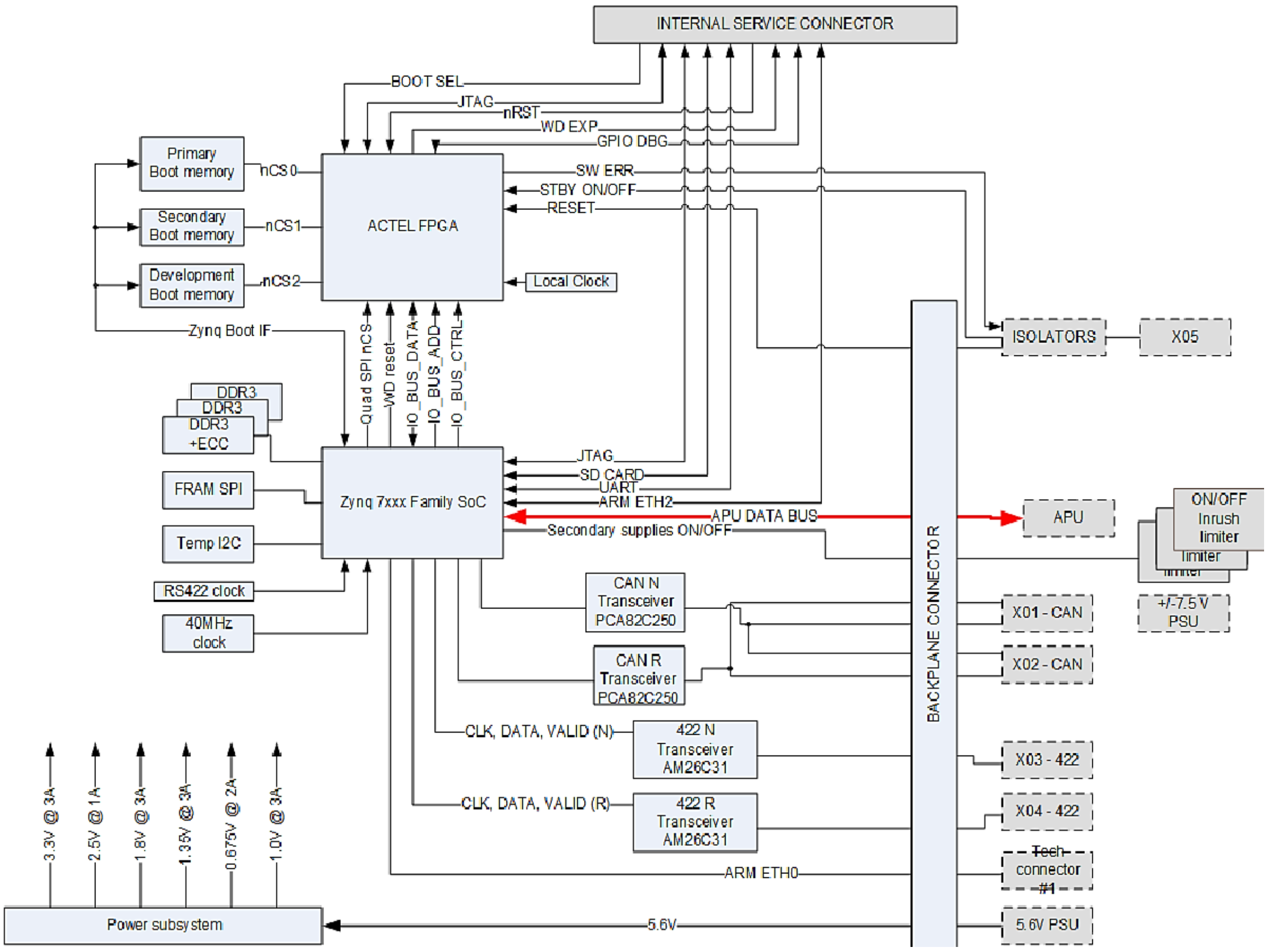

Figure 34– DPU block diagram

The SoC boot sequence is managed by an ACTEL FPGA that select the boot memory using a window watch-dog approach in TMR logic. As visible in Figure 34, the SoC directly interfaces the Analog to digital converters (ADC) on the APU board. For each ADC there is an interface that performs the functions of: configuring, controlling, and input data buffering. The acquired data are processed into the FPGA DSP section of the SOC and stored in the Scientific data memory.

The entire system is controlled by the Soc ARM processor. The processor has the task of managing the communication links (CANBUS and RS422) to / from the satellite, monitoring the health status of the EFD and managing the operating modes of the detector.

The processor changes the configuration tasks and management apparatus through the configuration and status registers. These registers are the point of exchange of information between the processors and the finite state machines (FSM) that handle the parts of the subsystem. The main state machines of the subsystem are:

- **APU TC/TM FSM**: it has the task of configuring and managing the logic on the APU (switch configuration, Current injection configuration) and monitoring the APU's health status (temperatures, voltages, etc.).
- **APU Power up sequence FSM**: it has the task of turning on the power supplies of the digital section of the APU in the correct sequence and reporting any power-on problems in the status register.
- **DPU housekeeping FSM**: it has the task of monitoring DPU power supplies, temperatures, status records, and of generating alarms for the ARM processor.
- **Acquisition control FSM**: It has the task of managing acquisition start and stop for each type of operating mode of the apparatus. This machine directly communicates with interfaces to the ADCs, and with the signal processing section.

Communication between the APU and DPU is handled by the APU Interface (APU IF). This block has the task to implement the communication protocol for conveying the data of configuration, control, and telemetry for the APU.

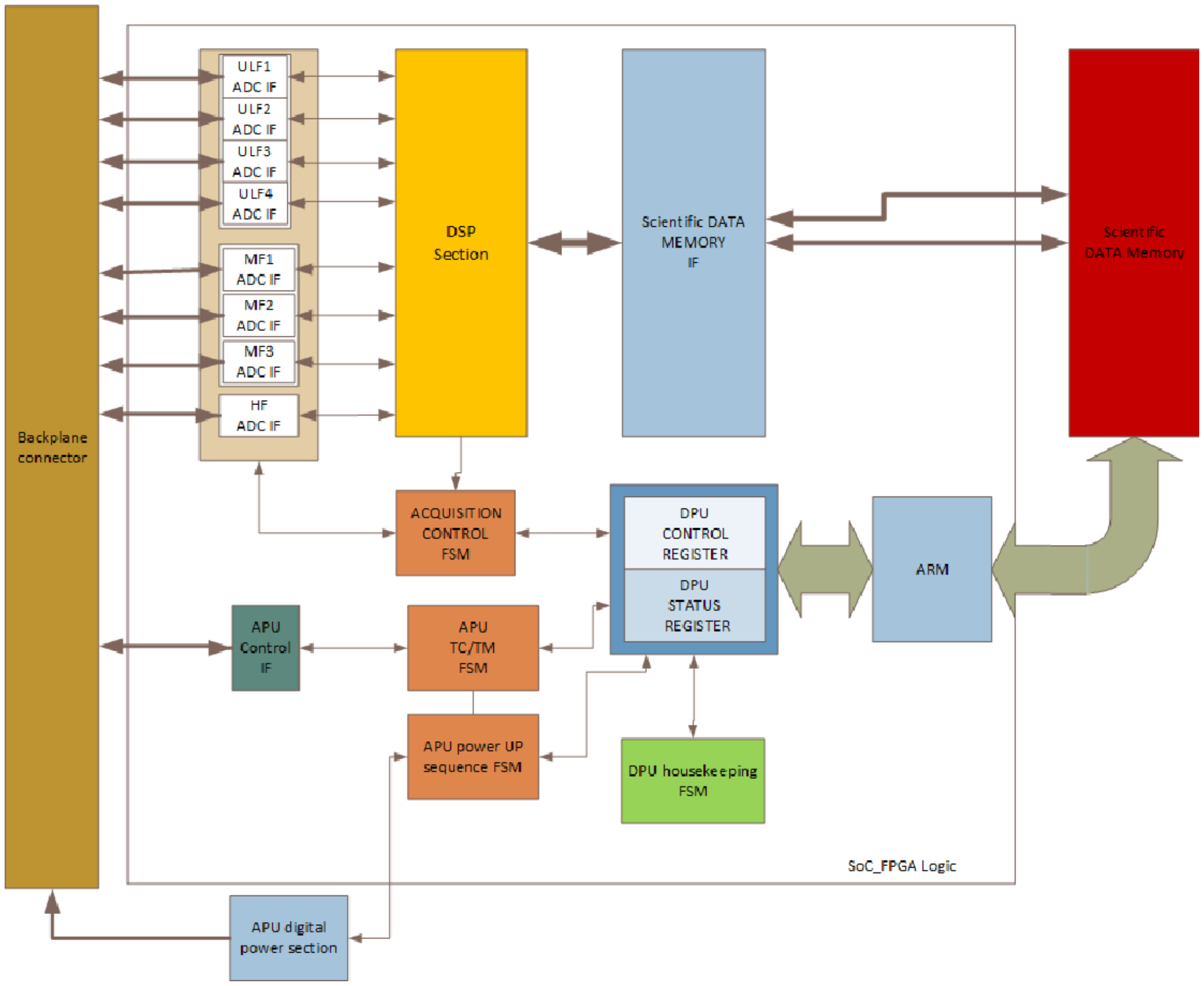

Figure 35 - SoC logic block diagram.

Figure 35 sets out the block diagram of the DATA processing section in the Zynq SOC. Since the entire electric field detector has been designed using oversampling advantages, the acquired data in each frequency band are processed by multi-rate filters.

These filters perform the task of sample rate reduction and dropping of outbound noise. Also, in some frequency bands such as the MF, the filter chains play the further role of subdivision into frequency sub-bands (ELF, VLF and VLFe).

The data for the VLF, VLFe and HF bands are processed through an FFT core to obtain spectra. All processed data are collected by a state machine (MEM storage IF FSM) and stored in the DDR scientific data memory. Once available on the main DDR memory, scientific data are encapsulated into a data packet and transmitted to the satellite platform using the dedicated data channel on the RS422 physical layer.

## 2.12 ANALOG SPLITTER

The analog splitter enables the switch of all signals/controls, and of the power supply lines of the Probes, between the hot and cold electronics, (see Figure 36).

The "power-supply mux" provides the power supply to the probes from hot or cold lines depending on which is activated. All the switches go in high impedance when their power supply is missing, automatically providing a switch between the APU hot and APU cold for the bias current control voltage, probe signals, and wires of the temperature sensors on probe connections.

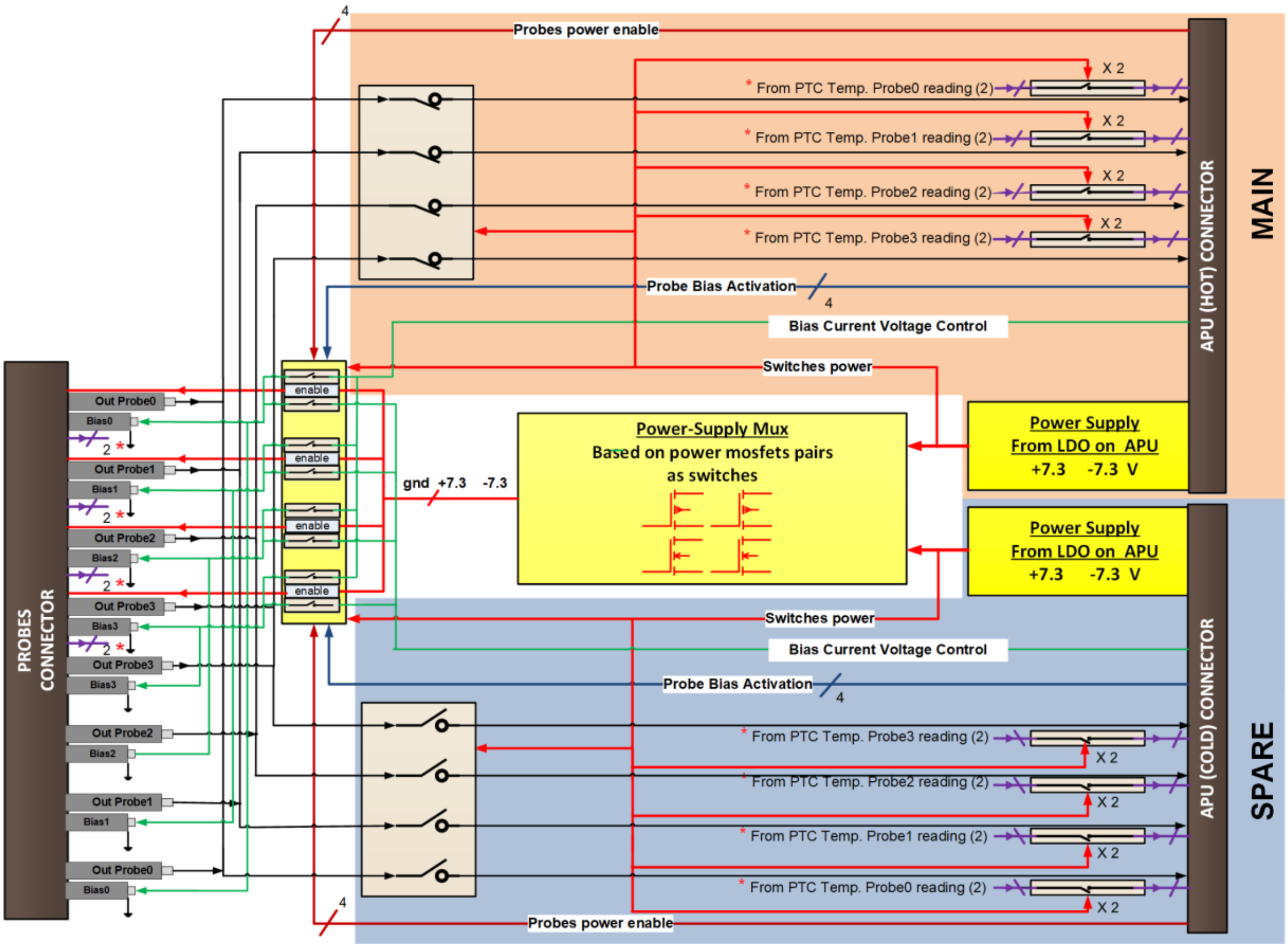

Figure 36 – Splitter connections diagram.

## 2.13 BIAS CURRENT ADJUSTMENT

As already explained in chapter 1, a DC bias current is directly injected in the probes to minimize the contact impedance between probe and plasma (from Plasma Sheath).

Electric field payloads in previous experiments, such as EFD-01 and ICE, show how plasma variations can affect the probe floating potential ($V_f$): in normal plasma condition, the $plasma\ density$ is about $n = 10^{10} m^{-3}$ and the electron temperature, $Te$, is between $2 \times 10^3 - 4 \times 10^3 K$, such that $V_f$ does not exceed the spacecraft ground (S/C GND) for more than about $2V$ A conductive body in contact with an ionospheric plasma attains a more negative potential than the plasma potential, $V_{pl}$. This is due to the larger electron collecting surface (almost isotropic) with respect to ion collection. In general, the spacecraft (S/C) GND is a few Volts more negative than the unknown plasma potential, but the EFP surface is less negative. As a result, the EFPs show a positive $V_f$ with respect to the S/C GND.

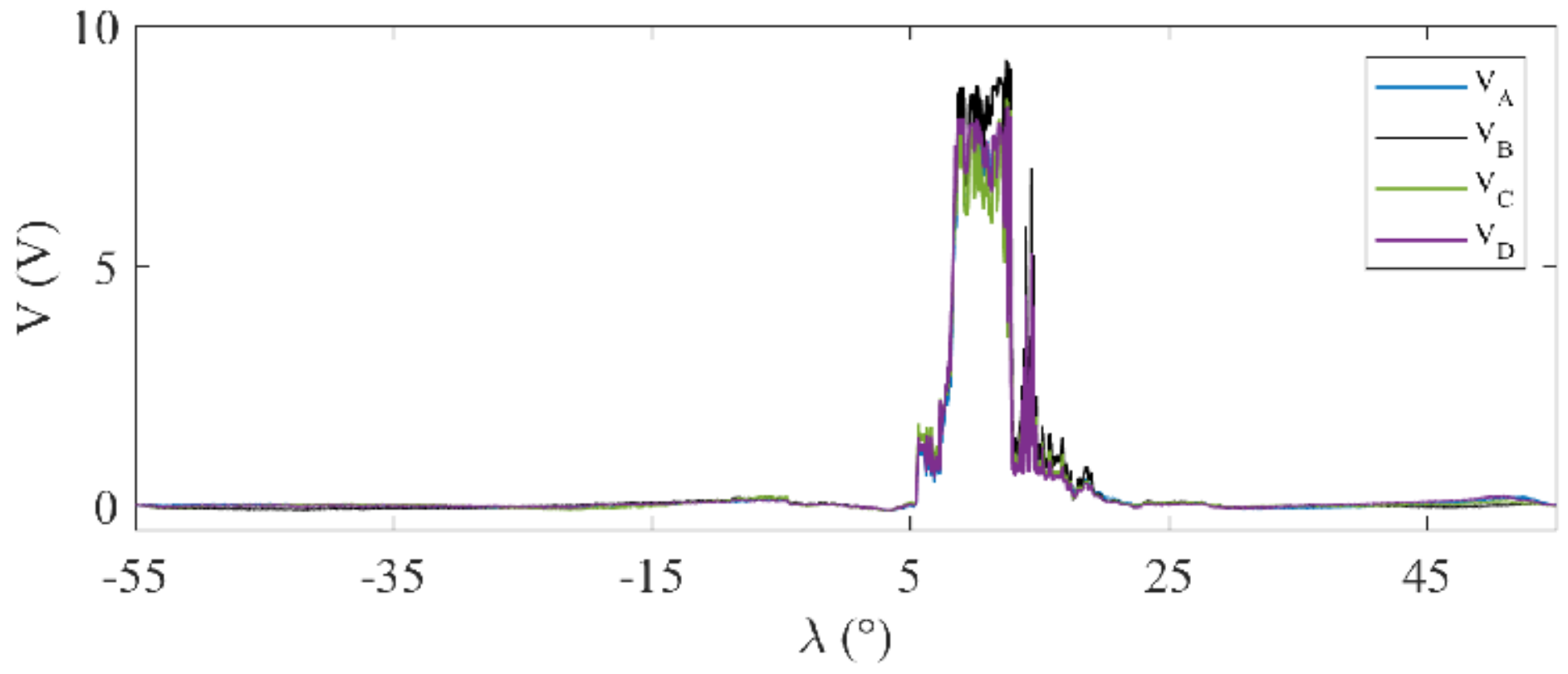

Figure 37 - This plot shows the potential variations in all the four probes, from which the $v_s \times B$ geomagnetic field effect has already been removed [69].

When a plasma depletion is encountered at low latitude, as shown in Figure 37, it is usually detected by both the Langmuir probe (LP) and EFD instruments [69]. The plasma density from LP measurements is abruptly reduced by about two orders of magnitude, that is, from about $10^{10} m^{-3}$ down to $10^{8} m^{-3}$ (or lower) in the central part of the bubble.

The EFD sensors reach high positive values due to the increasing ratio between the injected bias (fixed at $500nA$ in CSES/EFD-01 and DEMETER/ICE) and the collected plasma currents. In this condition, the measured $V_f$ is, on average, about 3 times higher than $V_{theo}$. This effect depends on the induced unbalancing in ion and electron collection on the probes, due to the electric field effect on the electron flow direction.

Indeed, according to the Orbital Motion Limited (OML) theory, the electron flow should be isotropic, such that any sensor collect electrons on the whole of its surface (i.e., $4\pi r^2$ for a spherical probe).

On the other hand, when a strong electric field is present, the OML is no longer valid in describing electron collection, because particle flow becomes more directional. In the case of very strong fields $\left(E > 100\frac{mV}{m}\right)$, the flow is completely dominated by the electric field direction, and the collecting surface on the probes is reduced up to a factor 4 as for the case of an ion collecting surface, which, in OML, is simply the cross section of the sphere under a unidirectional particle flow (i.e.,$\pi r^2$) [70].

So, if a dramatic plasma density decrease occurs (as under extreme plasma conditions or in a Plasma Bubble), when a bias current is injected, the floating potential can easily increase over a saturation level. This produces a persistent peak in the probe potential that is due to the injected bias current (even a very small one of the size of tens of nA). In order to avoid any saturation, the bias injection should be promptly decreased. So, if the bias current is properly reduced (rather than nullified), it is possible to compute the plasma density needed to obtain the current balance on the EFPs.

Here it comes a great novelty introduced with the EFD-02 instrument, that is, an algorithm that automatically adjusts the bias current injected into the probes to keep the floating potential below the saturation level and at an adequate value to minimize the contact impedance.

This algorithm has already been tested on ground, both in a laboratory (using a signal generator) and in a plasma chamber, demonstrating its effectiveness in acting automatically and quickly.

# CHAPTER 3

## CHARACTERIZATION AND QUALIFICATION OF THE ANALOG SECTION

*Many functionality and performance tests have been performed on the analog sections of EFD-02 (EFPs and APU). In addition, robustness test has been designed to guarantee the correct and desired electronic operations under different conditions of temperature, humidity and gamma irradiation. All these tests will be exposed in this chapter.*

### 3.1 ELECTRONIC STABILITY IN A CLIMATIC CHAMBER

The Analog Processing Unit, APU, must deal with different temperature and humidity conditions. It is commonly known that the gain of operational amplifiers, as well as the value of the parameters of various passive electronic components (resistors, capacitors, etc.), can vary with temperature, even in a non-linear way. Therefore, already in the design phase of the filters for the analog chains of the three different frequency bands (LF, MF and HF), it must be ensured that electronic stability is preserved in the temperature range encountered in orbit, i.e., between -35 and +85 degrees Celsius.

Thermal tests have shown that, thanks to the chosen circuit configuration and the selected electronic components, this electronic stability is guaranteed throughout the whole temperature range examined.

## Technical insight into the LF chain: diode protection test in the Climatic Chamber measurements

In the case of the low frequency filter (LF), operational amplifiers (ADA4528) have been used. The relative datasheet shows a maximum value of $300mV$ as the voltage on the input node over the power supply voltage, beyond which the protection of the operational is not ensured.

The setup used for the low-pass filter consisting of two operational amplifiers, establishes the addition of two protection diodes to ensure that the voltage on the input node of the first operational amplifier never exceeds the value of breakdown, as previously indicated.

In a simulation, using the Texas Instruments simulation software, it was observed that the addition of BAT54SW diodes (chosen as a trade-off between minimum diode capacity and maximum protection offered) effectively protects the inputs operating up to the required value.

Since the forward voltage of the used Schottky diodes, as known from the literature, increases as the temperature decreases, a temperature simulation was carried out in which it was seen that this protection is valid in a temperature range from -35$°C + 85°C$, up to the maximum voltages reached by probe signals, i.e., $\pm 7.3V$.

To physically verify the functioning of this protection, the same apparatus was used for measurements at different temperatures in a climatic chamber. A $36.9\Omega$ resistor was added to the output (chosen because it was experimentally useful for not making the circuit oscillate, due to the large capacitance seen in parallel by the operator on the feedback loop because of the length of the LEMO cables used). LEMO cables, about $4m$ long, were connected (for signal, supply and voltage reading) to simulate a flight-like wiring scenario.

In summary, in the entire temperature range studied, the correct protection due to the diodes has been guaranteed even for the maximum voltage value of the signals that can come from the EFPs.

## 3.2 APU BOARD VERIFICATION

For each APU board produced (for the EM, QM, and FM Models) accurate tests were carried out to assess electrical functions and general performance, both very important for the scientific purposes of the mission.

First, on every APU board, it was performed a defect test procedure, which aims to detect manufacturing defects that can invalidate the APU board functionality from the mechanical point of view. This is a standard procedure based on visual inspection of the board: mechanical interface visual inspection, appearance inspection and dimension procedure (232.25 x 154 mm).

## 3.3 APU FUNCTIONAL & ELECTRICAL VERIFICATION TESTS

These tests are necessary to verify the electrical functionality of the APU as follows.

The functional and electrical verification was conducted using the “APU_ADAPTER” test setup. This setup consists of a mechanical structure comprising a Xilinx (EZC706XILINXEV) evaluation board and an APU_ADAPTER electronic board, the latter of which connects all the control signals (splitter circuit) and data from the APU board with the evaluation board. It also allows the application of all APU board power supplies through external power-supplies.

The APU Test Equipment (ATE for short) includes two waveform generators to inject signals into a pair of inputs of the Probe Emulation Box, a precision oscilloscope and multimeter, the analog benchtop power supplies, a digital power supply which provides power to four DC-DC converters housed in the APU_ADAPTER, and a PC equipped with test code installed under the Xilinx Vivado application (using the "ILA" and "VIO" tools) for the APU configuration operations and digitized signal acquisition. Figure 38 shows the ATE instruments/PC and APU_ADAPTER connections.

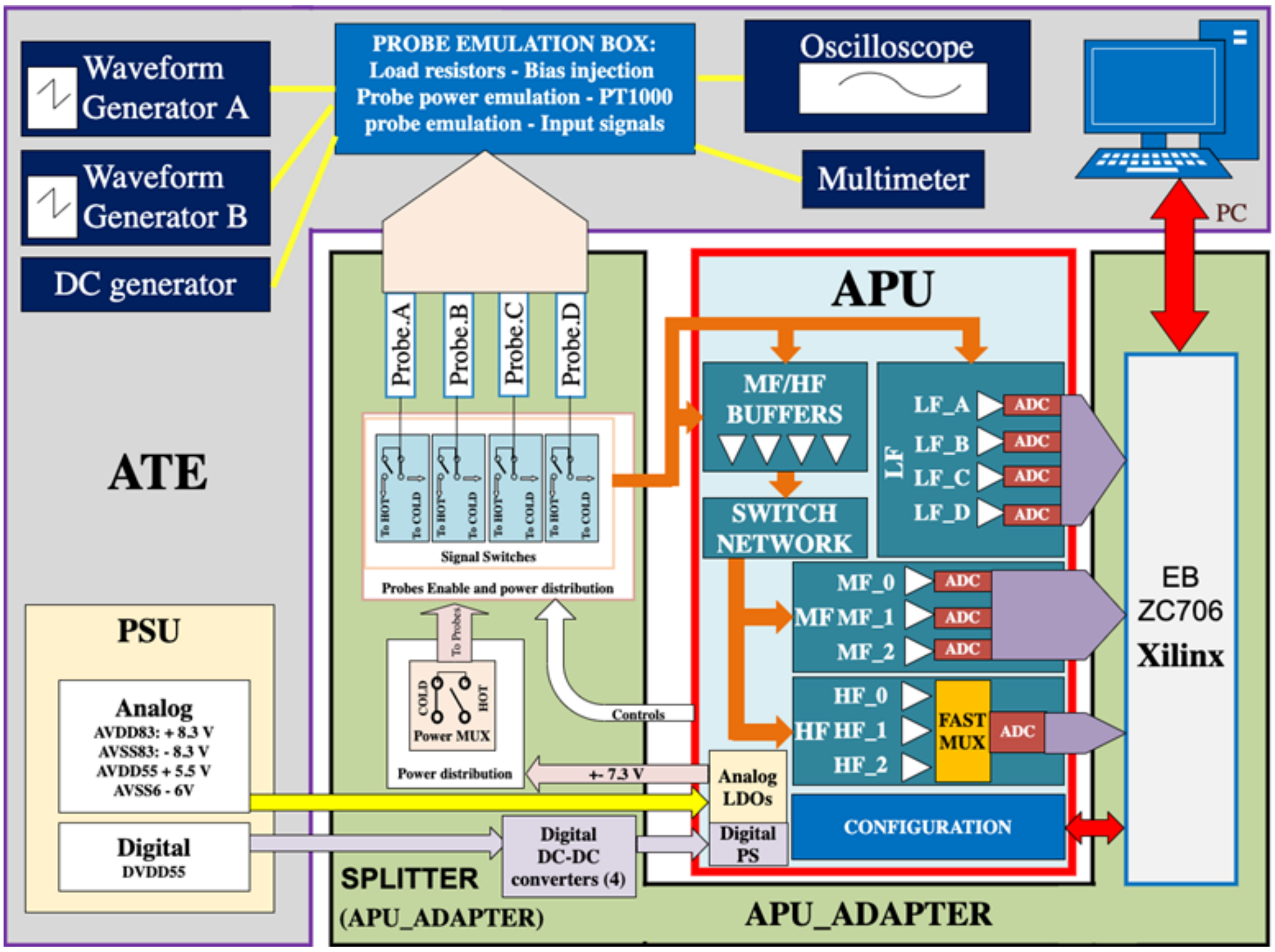

Figure 38 - APU_ADAPTER and ATE for the APU board.

All the tests performed on this set-up for the verification of the APU board are listed below.

1. **Nominal voltages generated by internal LDO verification**

First, the correct value of the voltages generated by the internal LDOs has been verified.

2. **Nominal power consumption verification**

This test is necessary to assess the power consumption of the board, which must meet the following defined requirements as to APU Power budget:

$$15\ W \pm 5\%$$

3. **Probe power enable verification**

The verification is intended to verify the correct functionality of enabling/disabling the probe power supplies.

This is to offer the possibility of interrupting the power to each single probe, a useful function especially in the event of a fault.

### 4. Probe temperature acquisition

The purpose of the check is to verify the correct reading of the PT1000 sensors housed in each probe.

The devices read the resistance value from an internal ADC, which is converted into temperature with an appropriate algorithm.

### 5. Board temperature sensors verification

The check has the purpose of verifying the reading correctness of the four board temperature sensors.

The verification is obtained by comparison with a thermo-camera measurement directed to each sensor.

### 6. Switch Matrix verification

The verification is intended to verify the correct functioning of all possible switch configurations in the switch matrix, for all the possible probe-pair differences.

This test consists of a sequence of 144 subtests in total. A waveform generator sends a sine wave with 1Vpp amplitude, at 50 kHz. The difference signals of each channel (MF0, MF1, MF2, HF0, HF1, and HF2) are read at the test connectors placed at the final analog stage of each channel before digital conversion by an oscilloscope together with the reference signal sent.

### 7. Fast mux verification

The verification is intended to test the fast mux functionality for the HF band. The check is done on both the static configuration of the mux, with the possibility of displaying the output of one of the HF channels at a time, and the output in the "rotation mode", in which the signals of the three HF channels are acquired in sequence, (*Figure 39*).

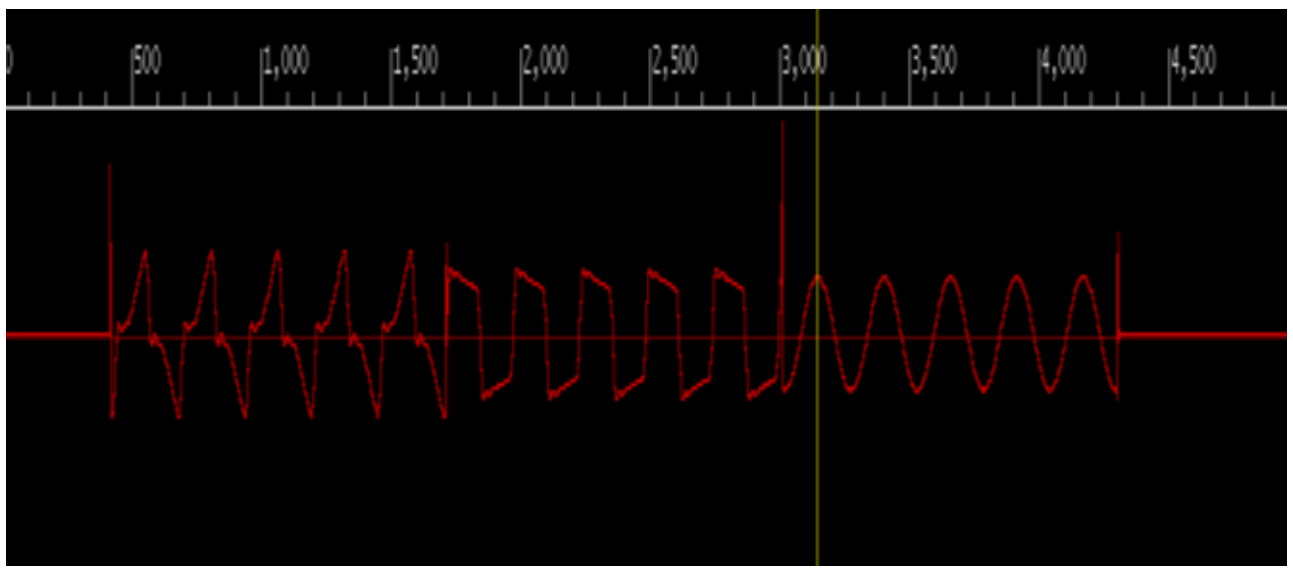

Figure 39 - Display of the mux output signal in rotation mode.

## 8. Bias current verification

In this test, the DAC is tested for correct generation of the desired voltage values, in the voltage range $\pm 5V$, in the case of either a static voltage or a sine wave with fixed frequency and variable amplitude.

## 3.4 APU PERFORMANCE VERIFICATION TESTS

The performance test plan was performed as follows:

- Transfer Function estimation (magnitude)
- Dynamic range and linearity
- Intrinsic noise estimation

All measurements given in the tables of this section are raw level digitized values measured at the output from each channel. The characteristics of the relative input signal are given in the "setup" column.

### 1) Offset tests for all LF, MF and HF channels and variability

These tests have the purpose of verifying the digitized offset at the end of each channel for each of the tested APU boards, when no input signal is injected. The measured values proved acceptable and compliant with what expected.

### 2) Transfer Function estimation and variability

The test was performed with a sine waveform of $1V_{pp}$ at the connector input for all bands, and a frequency around the center for each band: $MF: 10kHz; HF: 500kHz; LF: 40Hz.$

This test is usually intended to verify the cut-off frequencies of the acquired digital signal referred to the center band amplitude for each channel and for each input of the pair.

The same test was performed for each LF channel (A, B, C and D).

I took part in all acceptance tests for the APU boards, and in the study of the transfer function as reported in Table 4. These tests are meant to point out how the variability, defined as the ratio of the standard deviation to the mean value of the measured peak-to-peak output from the channel (over the entire set of five boards produced), is always less than 5%, except in extreme cases at the frequency bandwidth limit. where it never exceeds 10% anyway. The values of center-band frequencies used for board acceptance are highlighted in green.

| TRANSFER FUNCTION ESTIMATION | | | | | | | | | |
|---|---|---|---|---|---|---|---|---|---|
| set-up | | QM1 | QM2 | QM3 | QM4 | QM5 | MEAN | STD | % STD/Mean |
| LF A | 40 Hz | 6,57E+07 | 6,62E+07 | 6,62E+07 | 6,63E+07 | 6,63E+07 | 6,614E+07 | 2,782E+05 | 0,42% |
| | 105 Hz | 4,73E+07 | 4,68E+07 | 4,70E+07 | 4,68E+07 | 4,65E+07 | 4,688E+07 | 2,851E+05 | 0,61% |
| LF B | 40 Hz | 6,58E+07 | 6,65E+07 | 6,61E+07 | 6,63E+07 | 6,66E+07 | 6,621E+07 | 3,251E+05 | 0,49% |
| | 105 Hz | 4,80E+07 | 4,71E+06 | 4,77E+07 | 4,65E+07 | 4,70E+07 | 4,730E+07 | 6,444E+05 | 1,36% |
| LF C | 40 Hz | 6,56E+07 | 6,61E+07 | 6,52E+07 | 6,63E+07 | 6,71E+07 | 6,606E+07 | 7,073E+05 | 1,07% |
| | 105 Hz | 4,64E+07 | 4,64E+07 | 4,46E+07 | 4,61E+07 | 4,65E+07 | 4,597E+07 | 7,872E+05 | 1,71% |
| LF D | 40 Hz | 6,57E+07 | 6,64E+07 | 6,64E+07 | 6,66E+07 | 6,56E+07 | 6,613E+07 | 4,544E+05 | 0,69% |
| | 105 Hz | 4,70E+07 | 4,66E+07 | 4,68E+07 | 4,64E+07 | 4,60E+07 | 4,655E+07 | 3,793E+05 | 0,81% |
| MF 0 | 10 kHz B | 4,88E+05 | 5,07E+05 | 5,05E+05 | 5,05E+05 | 5,13E+05 | 5,036E+05 | 9,317E+03 | 1,85% |
| | 10 kHz A | 4,86E+05 | 5,05E+05 | 5,04E+05 | 5,04E+05 | 5,10E+05 | 5,018E+05 | 9,176E+03 | 1,83% |
| | 21 Hz B | 3,46E+05 | 3,54E+05 | 3,50E+05 | 3,52E+05 | 3,74E+05 | 3,552E+05 | 1,092E+04 | 3,07% |
| | 21 Hz A | 3,47E+05 | 3,58E+05 | 3,51E+05 | 3,48E+05 | 3,72E+05 | 3,552E+05 | 1,033E+04 | 2,91% |
| | 100 kHz B | 3,39E+05 | 3,52E+05 | 3,34E+05 | 3,41E+05 | 3,90E+05 | 3,512E+05 | 2,266E+04 | 6,45% |
| | 100 kHz A | 3,38E+05 | 3,48E+05 | 3,29E+05 | 3,49E+05 | 3,87E+05 | 3,502E+05 | 2,213E+04 | 6,32% |
| MF 1 | 10 kHz B | 4,97E+05 | 5,12E+05 | 5,12E+05 | 4,99E+05 | 5,18E+05 | 5,076E+05 | 9,127E+03 | 1,80% |
| | 21 Hz B | 3,48E+05 | 3,72E+05 | 3,44E+05 | 3,69E+05 | 3,82E+05 | 3,630E+05 | 1,631E+04 | 4,49% |
| | 100 kHz B | 3,47E+05 | 3,62E+05 | 3,55E+05 | 3,41E+05 | 3,45E+05 | 3,500E+05 | 8,426E+03 | 2,41% |
| MF 2 | 10 kHz C | 5,15E+05 | 5,25E+05 | 5,27E+05 | 5,19E+05 | 5,15E+05 | 5,202E+05 | 5,586E+03 | 1,07% |
| | 10 kHz D | 5,18E+05 | 5,29E+05 | 5,29E+05 | 5,22E+05 | 5,17E+05 | 5,230E+05 | 5,788E+03 | 1,11% |
| | 21 Hz C | 3,66E+05 | 3,84E+05 | 3,82E+05 | 3,71E+05 | 3,78E+05 | 3,762E+05 | 7,563E+03 | 2,01% |
| | 21 Hz D | 3,68E+05 | 3,87E+05 | 3,81E+05 | 3,76E+05 | 3,83E+05 | 3,790E+05 | 7,314E+03 | 1,93% |
| | 100 kHz C | 3,57E+05 | 3,59E+05 | 3,71E+05 | 3,50E+05 | 3,63E+05 | 3,600E+05 | 7,746E+03 | 2,15% |
| | 100 kHz D | 3,58E+05 | 3,62E+05 | 3,79E+05 | 3,63E+05 | 3,66E+05 | 3,656E+05 | 8,019E+03 | 2,19% |
| HF 0 | 500 kHz B | 3931,00 | 3904,00 | 3874,00 | 3935,00 | 3844,00 | 3897,60 | 38,6820372 | 0,99% |
| | 500 kHz A | 3894,00 | 3885,00 | 3859,00 | 3911,00 | 3821,00 | 3874,00 | 35,0856096 | 0,91% |
| | 21 kHz B | 2722,00 | 2811,00 | 2674,00 | 2735,00 | 2724,00 | 2733,20 | 49,4236785 | 1,81% |
| | 21 kHz A | 2700,00 | 2800,00 | 2682,00 | 2730,00 | 2719,00 | 2726,20 | 45,1464284 | 1,66% |
| | 4,3 MHz B | 2625,00 | 2682,00 | 2937,00 | 2989,00 | 2742,00 | 2795,00 | 159,904659 | 5,72% |
| | 4,3 MHz A | 2648,00 | 2670,00 | 2955,00 | 2998,00 | 2731,00 | 2800,40 | 164,311594 | 5,87% |
| HF 1 | 500 kHz D | 3860,00 | 3929,00 | 3899,00 | 3867,00 | 3853,00 | 3881,60 | 31,8088038 | 0,82% |
| | 500 kHz C | 3807,00 | 3904,00 | 3864,00 | 3870,00 | 3834,00 | 3855,80 | 36,9079937 | 0,96% |
| | 21 kHz D | 2662,00 | 2741,00 | 2730,00 | 2726,00 | 2707,00 | 2713,20 | 31,1400064 | 1,15% |
| | 21 kHz C | 2655,00 | 2721,00 | 2721,00 | 2719,00 | 2707,00 | 2704,60 | 28,3337255 | 1,05% |
| | 4,3 MHz D | 3633,00 | 3257,00 | 3393,00 | 3489,00 | 3134,00 | 3381,20 | 194,741367 | 5,76% |
| | 4,3 MHz C | 3623,00 | 3268,00 | 3391,00 | 3456,00 | 3128,00 | 3373,20 | 187,591844 | 5,56% |
| HF 2 | 500 kHz D | 3844,00 | 3931,00 | 3890,00 | 3882,00 | 3855,00 | 3880,40 | 34,0044115 | 0,88% |
| | 21 kHz D | 2660,00 | 2787,00 | 2654,00 | 2752,00 | 2712,00 | 2713,00 | 57,6367938 | 2,12% |
| | 4,3 MHz D | 3800,00 | 3090,00 | 3437,00 | 3547,00 | 3042,00 | 3383,20 | 318,524253 | 9,41% |

Table 4 - Transfer Function Variability between channels and boards.

Since the main source of variability between channels at different frequency and amplitude values is due to the variability between the various analog chains (due to differences between the parameters of the components), the values are reported only for the three analog bands generated by the APU (LF, MF and HF) with different colors for each band, in Table 4, while a color gradient was used for the five different boards.

Table 5, Table 6 and Table 7 refer to the variability between different channels of the same board (respecting the same color gradient as in Table 4). In these tables each row is associated with the reported frequency and a fixed amplitude value of the input signal.

LF channels:

| | | | |
|---|---|---|---|
| 40 Hz | 6,57E+07 | 86938,68 | 0,13% |
| 105 Hz | 4,71E+07 | 661740,13 | 1,40% |
| 40 Hz | 6,63E+07 | 169189,24 | 0,26% |
| 105 Hz | 4,67E+07 | 314165,56 | 0,67% |
| 40 Hz | 6,60E+07 | 513841,74 | 0,78% |
| 105 Hz | 4,65E+07 | 1347550,25 | 2,90% |
| 40 Hz | 6,64E+07 | 156923,55 | 0,24% |
| 105 Hz | 4,64E+07 | 291247,32 | 0,63% |
| 40 Hz | 6,64E+07 | 623718,69 | 0,94% |
| 105 Hz | 4,65E+07 | 418051,03 | 0,90% |

Table 5 - LF Variability of the output between channels with fixed frequency and amplitude

MF channels:

| | | | |
|---|---|---|---|
| 10 kHz | 5,01E+05 | 14956,60 | 2,99% |
| 21 Hz | 3,55E+05 | 11000,00 | 3,10% |
| 100 kHz | 3,48E+05 | 9523,65 | 2,74% |
| 10 kHz | 5,16E+05 | 10807,40 | 2,10% |
| 21 Hz | 3,71E+05 | 14866,07 | 4,01% |
| 100 kHz | 3,57E+05 | 6308,72 | 1,77% |
| 10 kHz | 5,15E+05 | 11928,96 | 2,31% |
| 21 Hz | 3,62E+05 | 18365,73 | 5,08% |
| 100 kHz | 3,54E+05 | 22018,17 | 6,23% |
| 10 kHz | 5,10E+05 | 10084,64 | 1,98% |
| 21 Hz | 3,63E+05 | 12397,58 | 3,41% |
| 100 kHz | 3,49E+05 | 9011,10 | 2,58% |
| 10 kHz | 5,15E+05 | 3209,36 | 0,62% |
| 21 Hz | 3,78E+05 | 4816,64 | 1,27% |
| 100 kHz | 3,70E+05 | 18566,10 | 5,02% |

Table 6 - MF Variability of the output between channels with fixed frequency and amplitude.

HF channels:

| | | | |
|---|---|---|---|
| 500 kHz | 3,87E+03 | 47,44 | 1,23% |
| 21 kHz | 2,68E+03 | 29,63 | 1,11% |
| 4,3 MHz | 3,27E+03 | 291,25 | 8,92% |
| 500 kHz | 3,91E+03 | 19,35 | 0,49% |
| 21 kHz | 2,77E+03 | 39,03 | 1,41% |
| 4,3 MHz | 2,99E+03 | 298,24 | 9,96% |
| 500 kHz | 3,88E+03 | 16,99 | 0,44% |
| 21 kHz | 2,69E+03 | 32,22 | 1,20% |
| 4,3 MHz | 3,22E+03 | 253,25 | 7,86% |
| 500 kHz | 3,89E+03 | 29,21 | 0,75% |
| 21 kHz | 2,73E+03 | 12,42 | 0,45% |
| 4,3 MHz | 3,30E+03 | 277,90 | 8,43% |
| 500 kHz | 3,84E+03 | 14,12 | 0,37% |
| 21 kHz | 2,71E+03 | 7,53 | 0,28% |
| 4,3 MHz | 2,96E+03 | 203,15 | 6,87% |

Table 7 - HF Variability of the output between channels with fixed frequency and amplitude.

It is noteworthy how, in the ULF band, the variability is always less than 1%, except for extreme cases at the edge of the frequency band, where however the variability keeps below 3%.

### 3) Dynamic range and linearity and variability

As the input for all bands, a sine waveform was chosen with a frequency around the center for each band: $MF: 1.5kHz; HF: 500kHz; LF: 40Hz$.

This test is usually intended to verify the linearity of the acquired digital signal over the entire amplitude range for each channel and each input on the pair.

The variability between boards was also tested, and it is shown in Table 8 using different colors for each band, while a color gradient was used for the five different boards. The values of mean amplitude, used for board acceptance, are highlighted in green.

| set-up | | QM1 | QM2 | QM3 | QM4 | QM5 | Mean | STD | % STD/Mean |
|---|---|---|---|---|---|---|---|---|---|
| **DYNAMIC RANGE AND LINEARITY** | | | | | | | | | |
| DC LFA | -6.6 V | -2,11E+09 | -2,11E+09 | -2,11E+09 | -2,11E+09 | -2,11E+09 | -2,11E+09 | 2,85E+06 | 0,13% |
| | -1 V | -3,24E+08 | -3,25E+08 | -3,24E+08 | -3,24E+08 | -3,25E+08 | -3,24E+08 | 5,65E+05 | 0,17% |
| | + 1V | 3,24E+08 | 3,22E+08 | 3,24E+08 | 3,23E+08 | 3,22E+08 | 3,23E+08 | 8,44E+05 | 0,26% |
| | +6.6 V | 2,12E+09 | 2,12E+09 | 2,12E+09 | 2,11E+09 | 2,12E+09 | 2,12E+09 | 3,14E+06 | 0,15% |
| DC LFB | -6.6 V | -2,11E+09 | -2,10E+09 | -2,11E+09 | -2,12E+09 | -2,11E+09 | -2,11E+09 | 4,63E+06 | 0,22% |
| | -1 V | -3,24E+08 | -2,96E+08 | -3,24E+08 | -3,26E+08 | -3,23E+08 | -3,24E+08 | 1,39E+06 | 0,43% |
| | + 1V | 3,23E+08 | 3,56E+08 | 3,22E+08 | 3,24E+08 | 3,24E+08 | 3,23E+08 | 1,02E+06 | 0,32% |
| | +6.6 V | 2,12E+09 | 2,17E+09 | 2,11E+09 | 2,12E+09 | 2,12E+09 | 2,12E+09 | 3,20E+06 | 0,15% |
| DC LFC | -6.6 V | -2,11E+09 | -2,12E+09 | -2,11E+09 | -2,11E+09 | -2,11E+09 | -2,11E+09 | 2,67E+06 | 0,13% |
| | -1 V | -3,24E+08 | -3,25E+08 | -3,24E+08 | -3,25E+08 | -3,24E+08 | -3,24E+08 | 3,86E+05 | 0,12% |
| | + 1V | 3,24E+08 | 3,22E+08 | 3,23E+08 | 3,23E+08 | 3,24E+08 | 3,23E+08 | 6,58E+05 | 0,20% |
| | +6.6 V | 2,12E+09 | 2,12E+09 | 2,12E+09 | 2,12E+09 | 2,12E+09 | 2,12E+09 | 3,00E+06 | 0,14% |
| DC LFD | -6.6 V | -2,11E+09 | -2,12E+09 | -2,12E+09 | -2,12E+09 | -2,10E+09 | -2,11E+09 | 6,65E+06 | 0,31% |
| | -1 V | -3,24E+08 | -3,25E+08 | -3,26E+08 | -3,27E+08 | -3,23E+08 | -3,25E+08 | 1,76E+06 | 0,54% |
| | + 1V | 3,24E+08 | 3,22E+08 | 3,23E+08 | 3,23E+08 | 3,23E+08 | 3,23E+08 | 5,91E+05 | 0,18% |
| | +6.6 V | 2,12E+09 | 2,12E+09 | 2,12E+09 | 2,12E+09 | 2,12E+09 | 2,12E+09 | 3,12E+06 | 0,15% |
| 40 Hz LF A | 1 mVpp | 3,06E+05 | 3,22E+05 | 3,46E+05 | 3,20E+05 | 3,20E+05 | 3,23E+05 | 1,45E+04 | 4,48% |
| | 1Vpp | 3,28E+08 | 3,30E+08 | 3,29E+08 | 3,31E+08 | 3,28E+08 | 3,29E+08 | 1,18E+06 | 0,36% |
| | 13,24 Vpp | 4,18E+09 | 4,21E+09 | 4,21E+09 | 4,19E+09 | 4,19E+09 | 4,20E+09 | 1,24E+07 | 0,29% |
| 40 Hz LF B | 1 mVpp | 3,06E+05 | 3,00E+05 | 3,47E+05 | 3,10E+05 | 3,36E+05 | 3,25E+05 | 1,99E+04 | 6,13% |
| | 1Vpp | 3,29E+08 | 3,31E+08 | 3,29E+08 | 3,31E+08 | 3,30E+08 | 3,30E+08 | 9,16E+05 | 0,28% |
| | 13,24 Vpp | 4,20E+09 | 4,21E+09 | 4,22E+09 | 4,20E+09 | 4,21E+09 | 4,21E+09 | 9,87E+06 | 0,23% |
| 40 Hz LF C | 1 mVpp | 3,03E+05 | 3,31E+05 | 3,38E+05 | 3,12E+05 | 3,32E+05 | 3,23E+05 | 1,49E+04 | 4,62% |
| | 1Vpp | 3,28E+08 | 3,29E+08 | 3,24E+08 | 3,31E+08 | 3,32E+08 | 3,29E+08 | 3,09E+06 | 0,94% |
| | 13,24 Vpp | 4,17E+09 | 4,19E+09 | 4,13E+09 | 4,19E+09 | 4,25E+09 | 4,19E+09 | 4,08E+07 | 0,97% |
| 40 Hz LF D | 1 mVpp | 3,15E+05 | 3,36E+05 | 3,34E+05 | 3,18E+05 | 3,29E+05 | 3,26E+05 | 9,45E+03 | 2,90% |
| | 1Vpp | 3,29E+08 | 3,30E+08 | 3,30E+08 | 3,32E+08 | 3,24E+08 | 3,29E+08 | 2,83E+06 | 0,86% |
| | 13,24 Vpp | 4,18E+09 | 4,21E+09 | 4,23E+09 | 4,21E+09 | 4,14E+09 | 4,20E+09 | 3,28E+07 | 0,78% |
| 1,5 kHz MF 0 (B) | 500 mVpp | 1,22E+06 | 1,26E+06 | 1,26E+06 | 1,27E+06 | 1,27E+06 | 1,26E+06 | 1,88E+04 | 1,50% |
| | 6,5 Vpp | 1,57E+07 | 1,60E+07 | 1,61E+07 | 1,62E+07 | 1,63E+07 | 1,61E+07 | 2,24E+05 | 1,39% |
| 1,5 kHz MF 1 (B) | 500 mVpp | 1,24E+06 | 1,27E+06 | 1,27E+06 | 1,25E+06 | 1,29E+06 | 1,27E+06 | 1,69E+04 | 1,33% |
| | 6,5 Vpp | 1,60E+07 | 1,61E+07 | 1,60E+07 | 1,61E+07 | 1,61E+07 | 1,61E+07 | 6,68E+04 | 0,42% |
| 1,5 kHz MF 2 (C) | 500 mVpp | 1,29E+06 | 1,30E+06 | 1,31E+06 | 1,30E+06 | 1,28E+06 | 1,30E+06 | 1,20E+04 | 0,93% |
| | 6,5 Vpp | 1,65E+07 | 1,66E+07 | 1,67E+07 | 1,66E+07 | 1,64E+07 | 1,66E+07 | 1,24E+05 | 0,75% |
| 500 kHz HF 0 (B) | 5 mVpp | 8,40E+01 | 9,70E+01 | 8,10E+01 | 9,30E+01 | 9,00E+01 | 8,90E+01 | 6,52E+00 | 7,32% |
| | 500 mVpp | 9,61E+03 | 9,70E+03 | 9,65E+03 | 9,62E+03 | 9,57E+03 | 9,63E+03 | 4,95E+01 | 0,51% |
| | 2,97 Vpp | 5,66E+04 | 6,01E+04 | 5,95E+04 | 5,67E+04 | 5,63E+04 | 5,78E+04 | 1,82E+03 | 3,14% |
| 500 kHz HF 1 (D) | 5 mVpp | 8,10E+01 | 1,01E+02 | 9,90E+01 | 9,80E+01 | 9,70E+01 | 9,52E+01 | 8,07E+00 | 8,48% |
| | 500 mVpp | 9,63E+03 | 9,72E+03 | 9,67E+03 | 9,64E+03 | 9,59E+03 | 9,65E+03 | 4,84E+01 | 0,50% |
| | 2,97 Vpp | 5,68E+04 | 6,02E+04 | 5,96E+04 | 5,68E+04 | 5,62E+04 | 5,79E+04 | 1,84E+03 | 3,18% |
| 500 kHz HF 2 (D) | 5 mVpp | 8,70E+01 | 9,20E+01 | 1,08E+02 | 9,80E+01 | 9,10E+01 | 9,52E+01 | 8,17E+00 | 8,58% |
| | 500 mVpp | 9,62E+03 | 9,70E+03 | 9,67E+03 | 9,64E+03 | 9,59E+03 | 9,65E+03 | 4,35E+01 | 0,45% |
| | 2,97 Vpp | 5,67E+04 | 6,01E+04 | 5,96E+04 | 5,68E+04 | 5,63E+04 | 5,79E+04 | 1,82E+03 | 3,14% |

Table 8 - Dynamic range and Linearity for all channels and boards.

Table 9, Table 10 and Table 11 concern to the variability between different channels of the same board (with same color gradient as in Table 8). In these tables, each row is associated with the reported amplitude and at fixed frequency of the input signal.

| LF DC | | | |
|---|---|---|---|
| **Input** | **Mean** | **STD** | **Variabilty** |
| **-6.6 V** | 2,11E+09 | 8,79E+05 | 0,04% |
| **-1 V** | 3,24E+08 | 1,16E+05 | 0,04% |
| **+ 1V** | 3,24E+08 | 1,66E+05 | 0,05% |
| **+6.6 V** | 2,12E+09 | 9,28E+05 | 0,04% |
| **-6.6 V** | 2,11E+09 | 8,86E+06 | 0,42% |
| **-1 V** | 3,18E+08 | 1,47E+07 | 4,63% |
| **+ 1V** | 3,30E+08 | 1,70E+07 | 5,15% |
| **+6.6 V** | 2,13E+09 | 2,33E+07 | 1,09% |
| **-6.6 V** | 2,11E+09 | 4,61E+06 | 0,22% |
| **-1 V** | 3,24E+08 | 1,07E+06 | 0,33% |
| **+ 1V** | 3,23E+08 | 8,10E+05 | 0,25% |
| **+6.6 V** | 2,12E+09 | 4,45E+06 | 0,21% |
| **-6.6 V** | 2,11E+09 | 4,98E+06 | 0,24% |
| **-1 V** | 3,26E+08 | 1,10E+06 | 0,34% |
| **+ 1V** | 3,23E+08 | 4,61E+05 | 0,14% |
| **+6.6 V** | 2,12E+09 | 4,24E+06 | 0,20% |
| **-6.6 V** | 2,11E+09 | 3,36E+06 | 0,16% |
| **-1 V** | 3,24E+08 | 1,20E+06 | 0,37% |
| **+ 1V** | 3,23E+08 | 6,78E+05 | 0,21% |
| **+6.6 V** | 2,12E+09 | 2,20E+06 | 0,10% |

Table 9 – LF variability between channels at different input amplitudes for DC.

In Table 9 the variability is less than 0.5 % except in the case of QM2 suffering from defects introduced during board production.

| LF 40 Hz | | | |
|---|---|---|---|
| **Input** | **Mean** | **STD** | **Variabilty** |
| **1 mVpp** | 3,08E+05 | 5,20E+03 | 1,69% |
| **1Vpp** | 3,29E+08 | 4,83E+05 | 0,15% |
| **13,24 Vpp** | 4,18E+09 | 1,07E+07 | 0,25% |
| **1 mVpp** | 3,08E+05 | 5,20E+03 | 1,69% |
| **1Vpp** | 3,29E+08 | 4,83E+05 | 0,15% |
| **13,24 Vpp** | 4,18E+09 | 1,07E+07 | 0,25% |
| **1 mVpp** | 3,08E+05 | 5,20E+03 | 1,69% |
| **1Vpp** | 3,29E+08 | 4,83E+05 | 0,15% |
| **13,24 Vpp** | 4,18E+09 | 1,07E+07 | 0,25% |
| **1 mVpp** | 3,08E+05 | 5,20E+03 | 1,69% |
| **1Vpp** | 3,29E+08 | 4,83E+05 | 0,15% |
| **13,24 Vpp** | 4,18E+09 | 1,07E+07 | 0,25% |
| **1 mVpp** | 3,08E+05 | 5,20E+03 | 1,69% |
| **1Vpp** | 3,29E+08 | 4,83E+05 | 0,15% |
| **13,24 Vpp** | 4,18E+09 | 1,07E+07 | 0,25% |

Table 10 - LF variability between channels at different input amplitudes for 40 Hz.

| MF 1,5 kHz | | | |
|---|---|---|---|
| **Input** | **Mean** | **STD** | **Variabilty** |
| **500 mVpp** | 1,25E+06 | 3,31E+04 | 2,64% |
| **6,5 Vpp** | 1,61E+07 | 4,14E+05 | 2,57% |
| **500 mVpp** | 3,22E+05 | 1,59E+04 | 4,94% |
| **6,5 Vpp** | 3,30E+08 | 1,10E+06 | 0,33% |
| **500 mVpp** | 3,41E+05 | 6,29E+03 | 1,84% |
| **6,5 Vpp** | 3,28E+08 | 2,62E+06 | 0,80% |
| **500 mVpp** | 3,15E+05 | 4,76E+03 | 1,51% |
| **6,5 Vpp** | 3,31E+08 | 6,55E+05 | 0,20% |
| **500 mVpp** | 3,29E+05 | 6,80E+03 | 2,07% |
| **6,5 Vpp** | 3,29E+08 | 3,21E+06 | 0,98% |

Table 11 – MF variability between channels at different input amplitudes for 1,5 kHz.

| HF 500 kHz | | | |
|---|---|---|---|
| **Input** | **Mean** | **STD** | **Variabilty** |
| **5 mVpp** | 8,40E+01 | 3,00E+00 | 3,57% |
| **500 mVpp** | 9,62E+03 | 9,02E+00 | 0,09% |
| **2,97 Vpp** | 5,67E+04 | 1,03E+02 | 0,18% |
| **5 mVpp** | 9,67E+01 | 4,51E+00 | 4,66% |
| **500 mVpp** | 9,71E+03 | 7,94E+00 | 0,08% |
| **2,97 Vpp** | 6,02E+04 | 8,03E+01 | 0,13% |
| **5 mVpp** | 9,60E+01 | 1,37E+01 | 14,32% |
| **500 mVpp** | 9,66E+03 | 1,24E+01 | 0,13% |
| **2,97 Vpp** | 5,96E+04 | 4,86E+01 | 0,08% |
| **5 mVpp** | 9,63E+01 | 2,89E+00 | 3,00% |
| **500 mVpp** | 9,64E+03 | 1,56E+01 | 0,16% |
| **2,97 Vpp** | 5,68E+04 | 3,99E+01 | 0,07% |
| **5 mVpp** | 9,27E+01 | 3,79E+00 | 4,09% |
| **500 mVpp** | 9,58E+03 | 1,25E+01 | 0,13% |
| **2,97 Vpp** | 5,62E+04 | 3,54E+01 | 0,06% |

Table 12 - HF variability between channels at different input amplitudes for 500 kHz. In the case of QM2, the value at 5 mVpp is chosen at the low amplitude limit, such that, even if the result exceeds 10%, it turns out acceptable considering the non-stringent accuracy requirement in this band.

### 4) Intrinsic noise estimation and variability

This test is intended to assess the intrinsic noise (Standard Deviation, SD) of the acquired digital signal for each channel. This measurement was made by closing the input of each channel of APU_ADAPTER with a 50-ohm cap.

Table 13 shows all the measurements for all the QM boards.

| INTRINSIC NOISE ESTIMATION | | | | | | | | |
|---|---|---|---|---|---|---|---|---|
| Channels | QM1 | QM2 | QM3 | QM4 | QM5 | Mean | STD | % STD/Mean |
| LF A | 3019 | 3146 | 3090 | 3019 | 3197 | 3094,20 | 78,388 | 2,53% |
| LF B | 3075 | 3058 | 3177 | 2978 | 3012 | 3060,00 | 87,447 | 2,86% |
| LF C | 3078 | 3084 | 3009 | 3078 | 2994 | 3048,60 | 43,391 | 1,42% |
| LF D | 3188 | 3104 | 3446 | 3242 | 2964 | 3188,80 | 177,981 | 5,58% |
| Variablity between LF channels --> | | | | | | 3097,90 | 63,622 | 2,05% |
| MF 0 | 140 | 140 | 138 | 140 | 148 | 141,20 | 3,899 | 2,76% |
| MF 1 | 141 | 143 | 144 | 138 | 140 | 141,20 | 2,387 | 1,69% |
| MF 2 | 145 | 144 | 147 | 148 | 145 | 145,80 | 1,643 | 1,13% |
| Variablity between MF channels --> | | | | | | 142,73 | 1,149 | 0,81% |
| HF 0 | 3,00 | 3,23 | 3,31 | 3,34 | 3,16 | 3,21 | 0,136 | 4,23% |
| HF 1 | 4,00 | 3,08 | 3,20 | 3,21 | 3,17 | 3,33 | 0,378 | 11,34% |
| HF 2 | 4,00 | 3,23 | 3,19 | 3,22 | 3,17 | 3,36 | 0,358 | 10,64% |
| Variablity between HF channels --> | | | | | | 3,30 | 0,134 | 4,07% |

Table 13 - Intrinsic Noise Estimation variability for all channels and boards.

To conclude this discussion, all the tests shown in this section demonstrated less than 10% variability between channels for all boards.

For this reason, the calibration measurements shown in the next chapter are relative to only one channel per frequency band.

## 3.5 GAMMA IRRADIATION TESTS

The reason behind radiation testing is that the APU does not have components with space-type qualification nor Space Heritage (SH).

The radiation resistance of the APU EM prototype was tested by gamma irradiation at the Calliope Gamma Irradiation Facility of ENEA Casaccia in June 2021, in order to obtain a space-like heritage.

The Calliope plant is a pool-type irradiation facility equipped with a $60Co$ radio-isotopic source array in a high volume ($7.0m \times 6.0m \times 3.9m$) shielded cell. The source rack has a plane geometry with twenty-five $60Co$ source rods, (Figure $40$). The ionizing radiation consists of two photons of 1.17 and 1.33 $MeV$ energy emitted in coincidence.

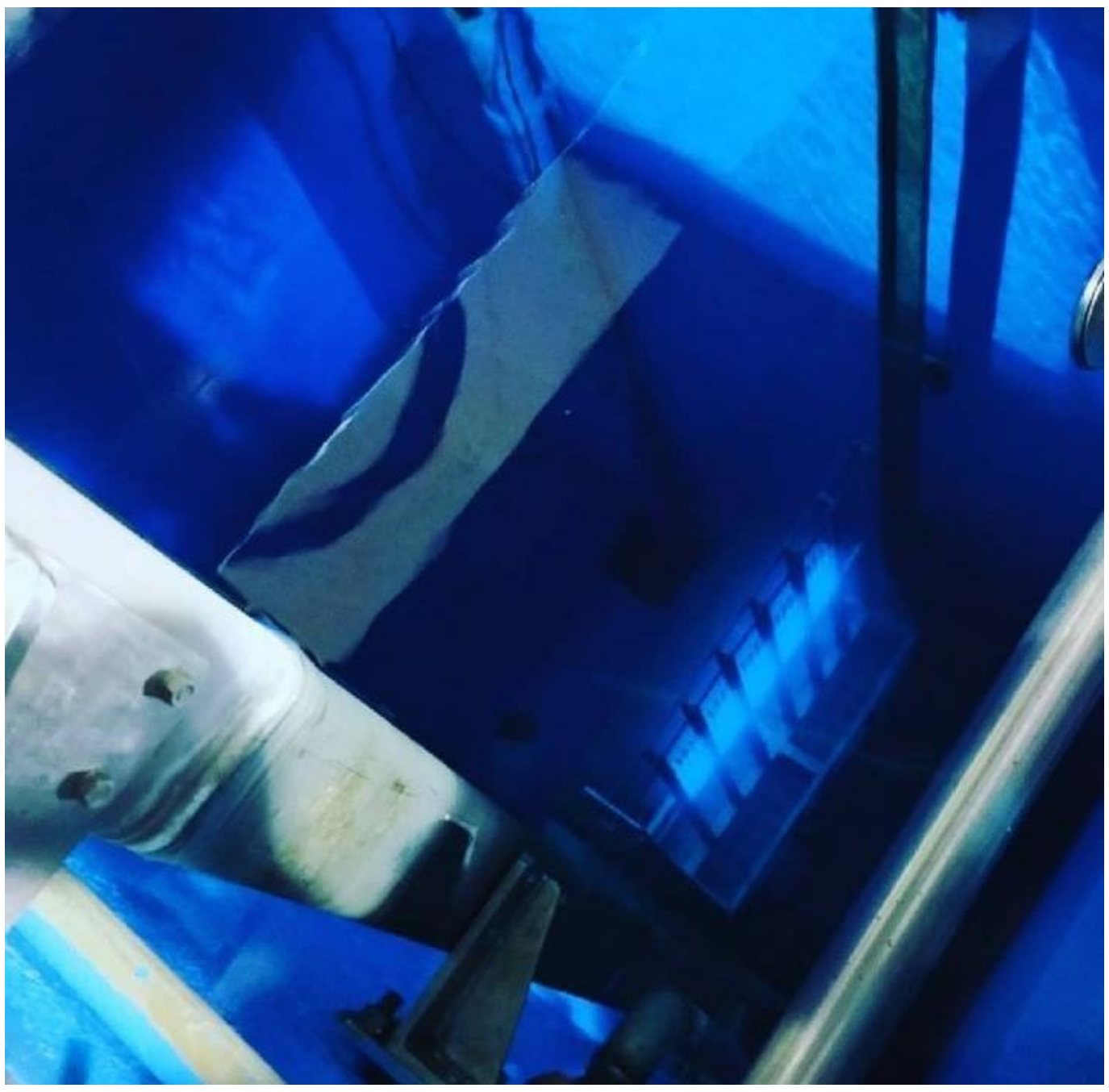

Figure 40 - Calliope rack with $60Co$ sources (pool view) and Cherenkov effect.

The possible tests to characterize an electronic board under ionizing radiations are mainly of 3 types:

- Total ionizing dose (TID).
- Displacement damage dose (DDD).
- Single event effects (SEE).

The physical processes involved in the radiation damage of electronic devices are particularly complex and they depend on several parameters, such as the kind of radiation, energy, and fluence. Broadly speaking, they can be grouped into two major classes:

- **Ionization damage**: caused by electron-hole pairs generated in silicon dioxide ($SiO_2$) and other insulators by ionizing radiation. In $SiO_2$, electrons are much more mobile than holes and they are quickly collected at the positive electrode, even though a fraction of them recombines with holes. The holes, escaping the initial recombination, are relatively immobile and slowly undergo a hopping transport between localized sites in the oxide in the presence of an electric field. Some of them can be trapped, giving rise to the accumulation of positive charge in the oxide, or they can generate interface states at the $SiO_2$/Si interface, affecting the device operation.
- **(Non-ionizing) Bulk damage or Displacement Damage Dose** (DDD): caused by collisions of energetic protons, neutrons, ions, and electrons, which transfer sufficient energy to knock out a Si atom from its lattice position. A vacancy interstitial pair (called Frenkel defect) is generated this way, and it migrates until a stable defect is formed by association with other defects, impurities, or dopants. The impact of DDD is major for opto-electronic devices only, as remarked by the current ECSS standard.

Ionizing radiation effects on semiconductor devices are classified into two major types:

- **Total Ionizing Dose** (TID). When we talk about TID, we are generically talking about long-term effects that introduce permanent damage in devices, such as threshold shift in MOSFETs, increase in leakage current (which directly affect power consumption), timing change etc. Threshold shifting for example can be due to trapping of charges in the oxide. CMOS-Bulk Devices (IC's) experience "latch up" due to a parasitic four-layer PNPN path, inherent in most unhardened devices. These parasitic four-layer devices act like a Silicon Control Rectifier (SCR), which, once latched, cannot be turned off without shutting off the power.

- **Single event effects** (SEEs). Single-event upsets (SEUs), single-event transients (SETs) and single-event latch-ups (SELs) result from the highly localized deposition of energy by single particles or their reaction products, when the energy deposition is sufficient to cause observable effects. For example, A single high energy particle may trigger snapback, if the field across the drain region is sufficiently high. Snapback is due to the prospect of a parasitic bipolar transistor existing between the drain and source region of a MOS transistor, which amplifies the avalanche current that results from the impact of a cosmic-ray heavy ion. This results in an extremely high current between the drain and source region of the transistor, with subsequent localized heating.

To study the aging of the APU board and to verify system-level functionalities, a TID type test was performed, keeping the system under test (SUT) powered and active for the entire duration of the measurements. Only an overall dose check was performed on the whole system, with no test on individual electronic components.

Although probably present, SEE-type events were not kept under control.

Figure 41 and Figure 42 show the test setup, the assembly of a suitable protective shield (Pb bricks and slabs; Figure 43) around the critical part, (specifically the APU_ADAPTER and the Xilinx evaluation board), and the layout of the instrumentation and boards.

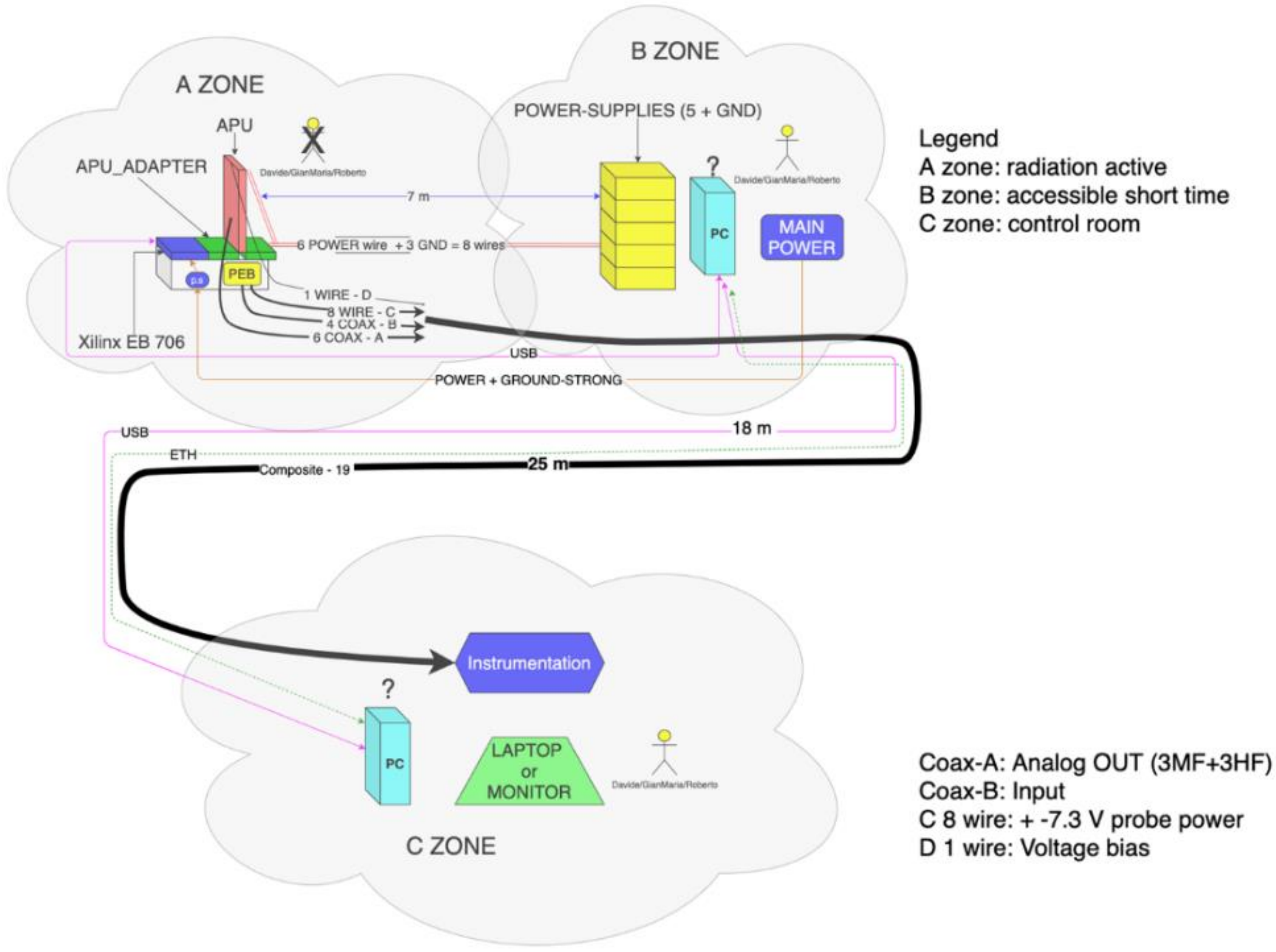

Figure 41 – Synthetic scheme of the set-up used for APU irradiation. This sketch allowed us to organize the experimental set-up, since suitable to understand the effective spatial arrangement of the instruments used for the tests.

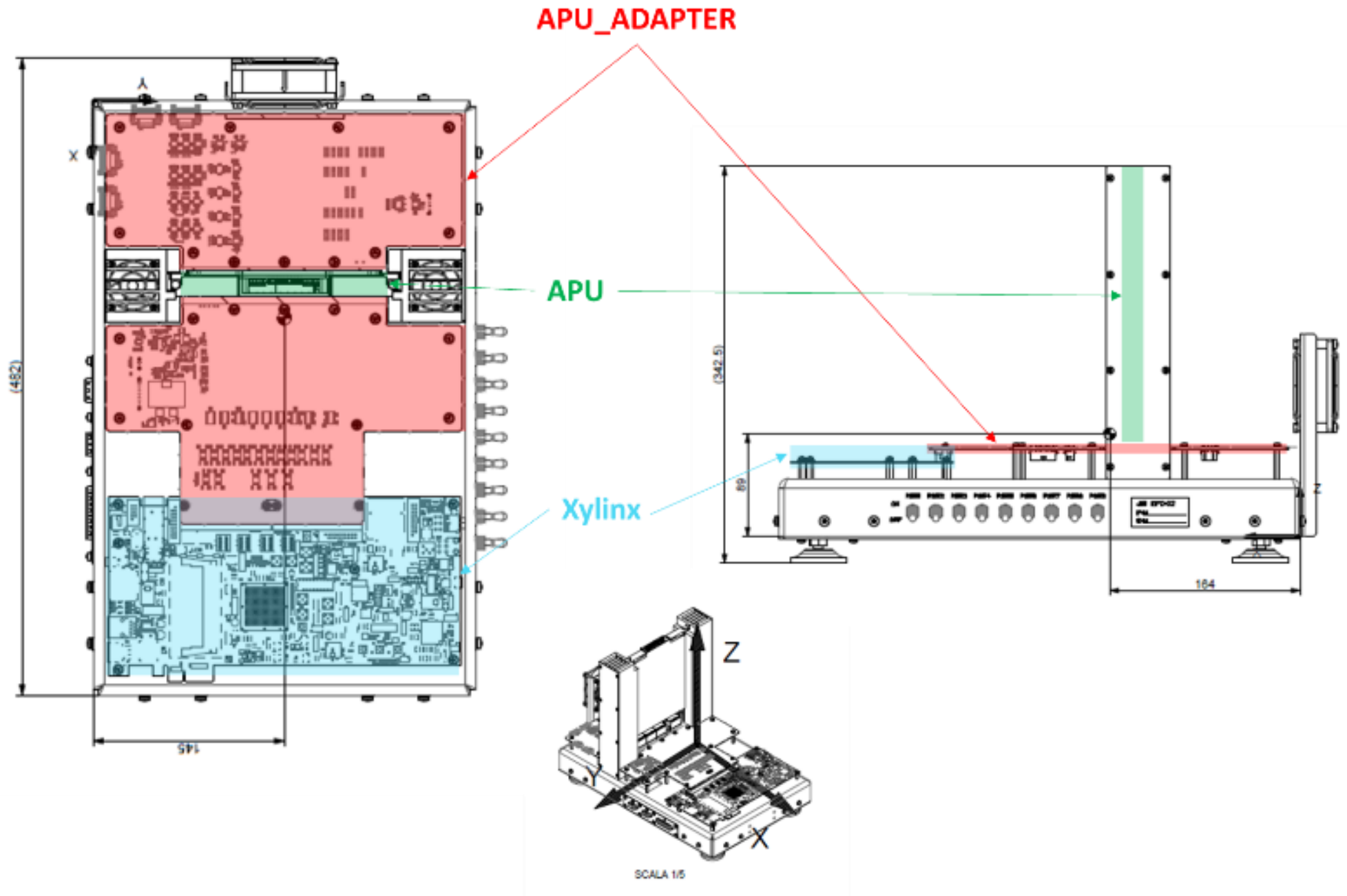

Figure 42 – ATE Electronic boards in the irradiation test.

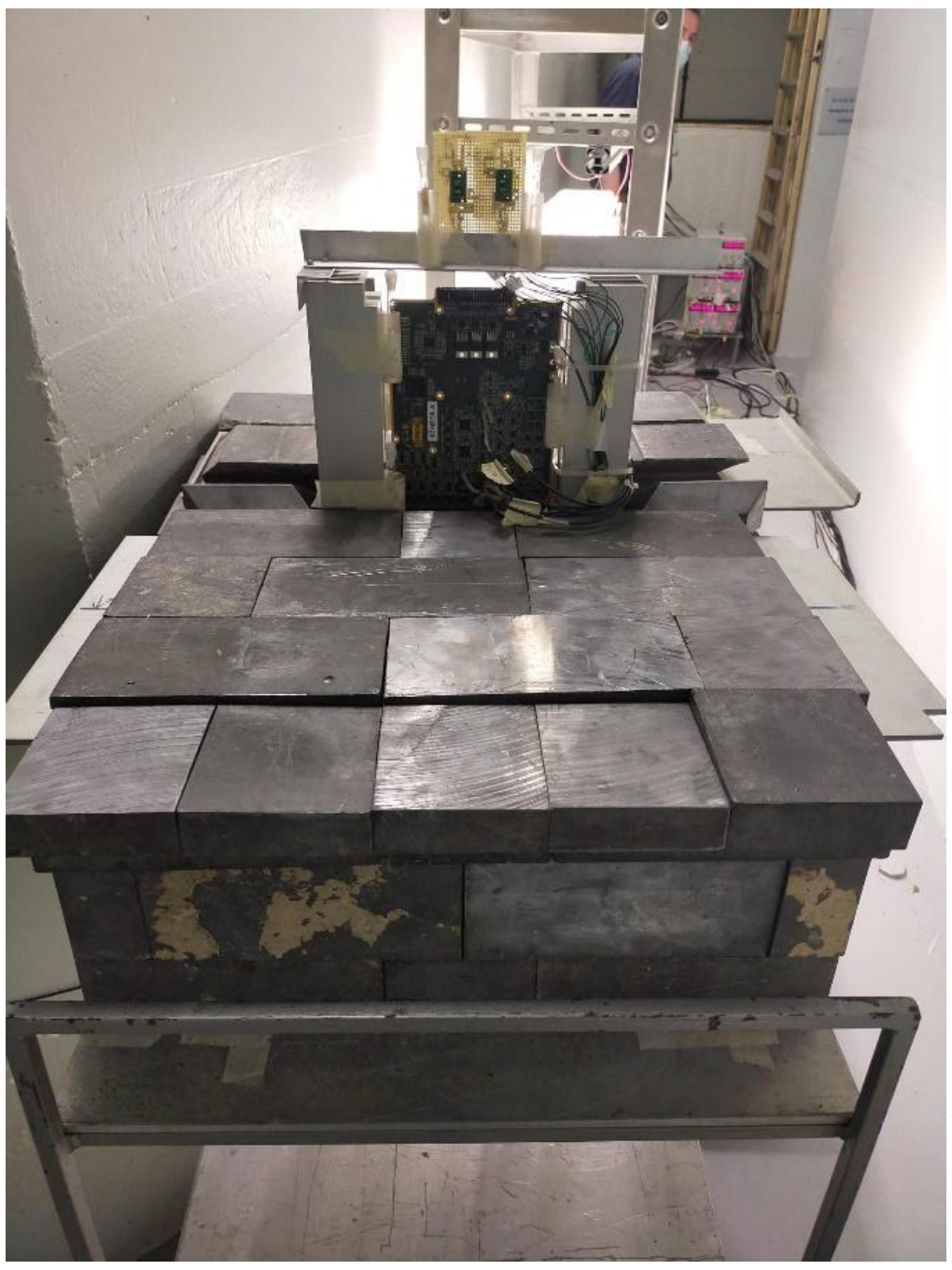

Figure 43 - Photo of the lead screen used for irradiation tests.

The APU (green in Figure 42) was placed within the irradiated area at a specific location. In this location, the total dose received by the APU board (using multiple dosimetric methods) was 10 $krad$, with total daily dose of about 1 $krad$ ($\sim 42\ rad/h$); the radiation non-uniformity on the SUT was $< 20\%$. The APU was the only board exposed, in order to ensure that any failure due to radiation was attributable to the APU alone. For this reason all the other boards were suitably shielded with lead, (see Figure 43).

The tests aimed to verify:

- APU electrical functionality: power supplies and controls.
- APU Performance: channel offset; transfer function; dynamic range and linearity; intrinsic noise.

The results obtained in these tests were satisfactory and met the expected requirements.

It is also worth noting that the test was very conservative with respect to the radiation environment that the satellite will encounter in its trajectory (LEO orbit, about 500 km altitude and about 97° inclination). Considering the CSES-02 mission profile and the position of APU board within the satellite, the expected TID for the board in 6 years of mission is less than 0.5 $krad$.

# CHAPTER 4

## CALIBRATION AND PERFORMANCE TESTS IN THE PLASMA CHAMBER

*The results of the calibration with signal generators at INFN-Tor Vergata and the characterization of the instrument in the Plasma Chamber at INAF -IAPS will be shown in this Chapter. The variability of the calibration factors will be discussed as well, considering various control parameters.*

### 4.1 CALIBRATION MEASUREMENTS

Definition of a calibration factor (CF): any CF provides a correspondence between the potential difference applied to a probe pair and the digital output from the channels of the instrument.

In the tests described in this section, the signals were applied directly to the probes via ohmic contact. Here, the calibration procedures will be described and the evaluation of the first calibration factors for the EFD-02 QM analog electronics will be shown. The system under test consisted of 2 EFPs, the APU board and "DSP" subsystems.

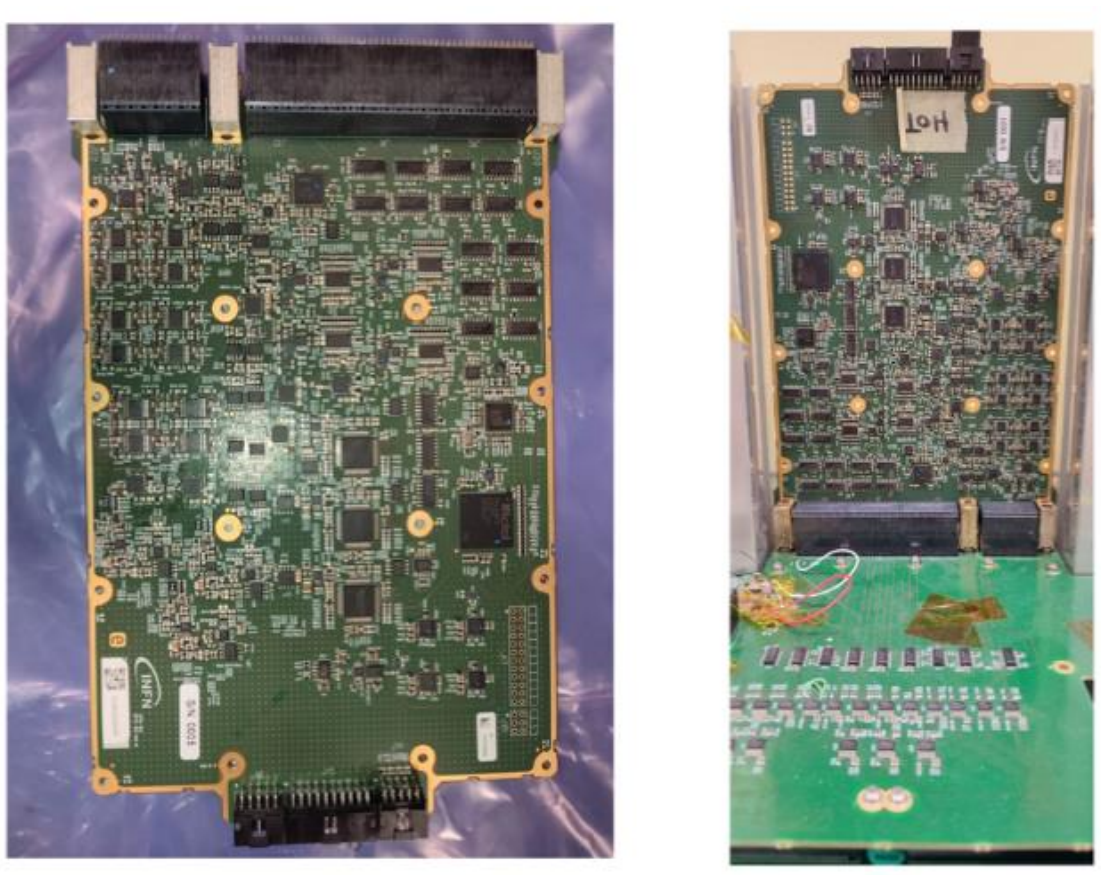

Figure 44 - QM APU (left side) -and QM APU mounted on APU_ADAPTER (right side).

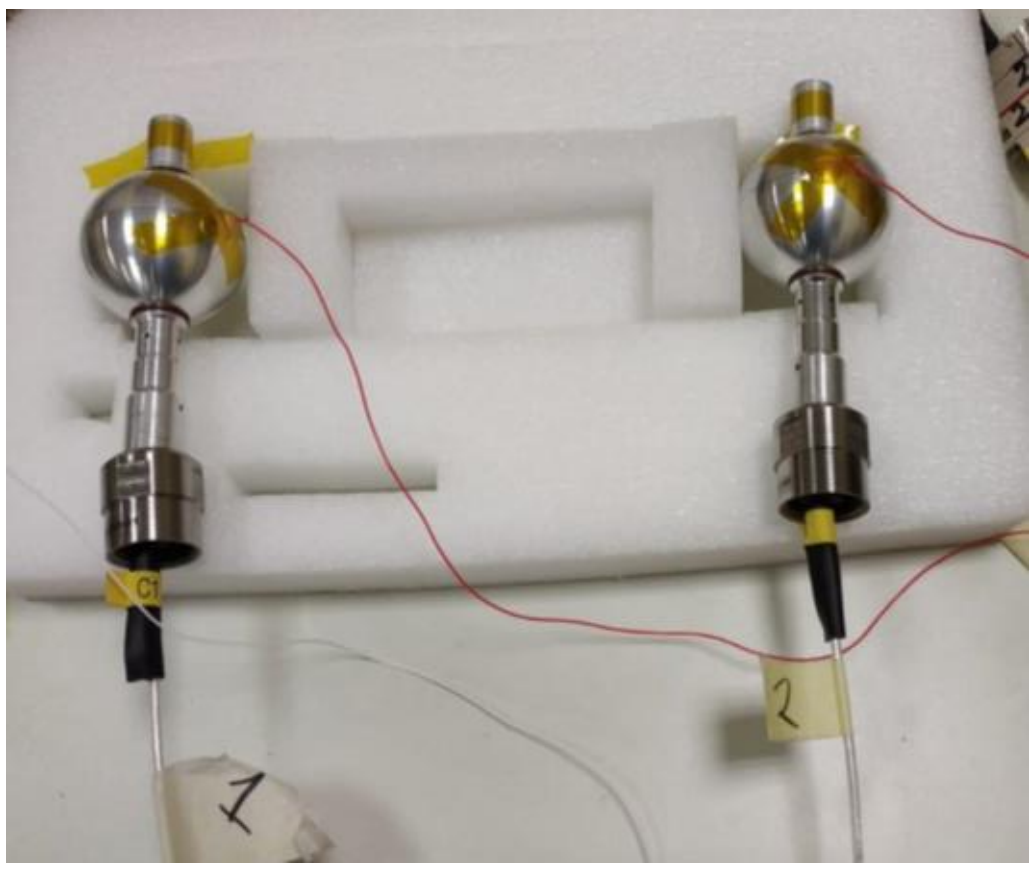

Figure 45 - EFPs used for calibration tests.

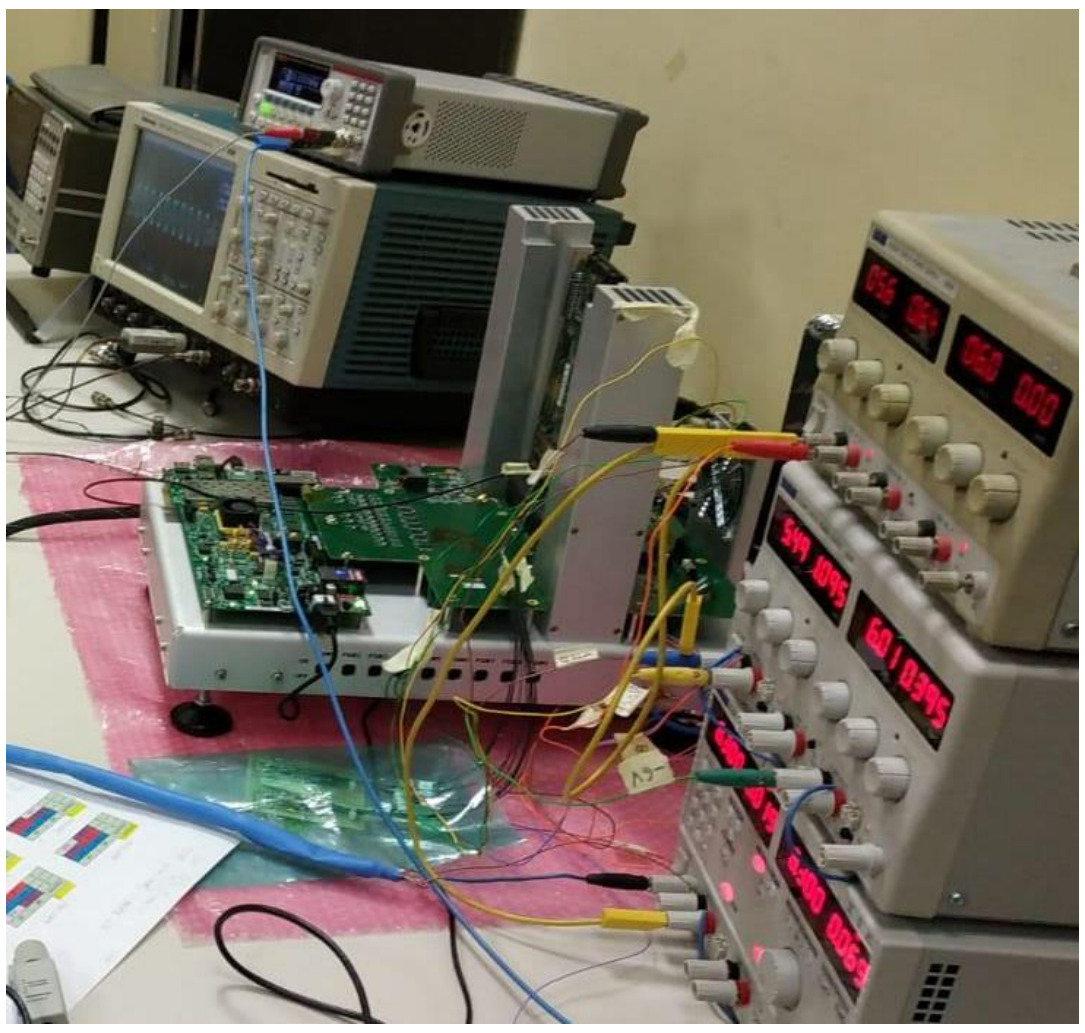

Figure 46 - Set up used for calibration measurements. This picture shows the APU board mounted on the APU_ADAPTER and the Xilinx evaluation board which allowed the controls and DSP.

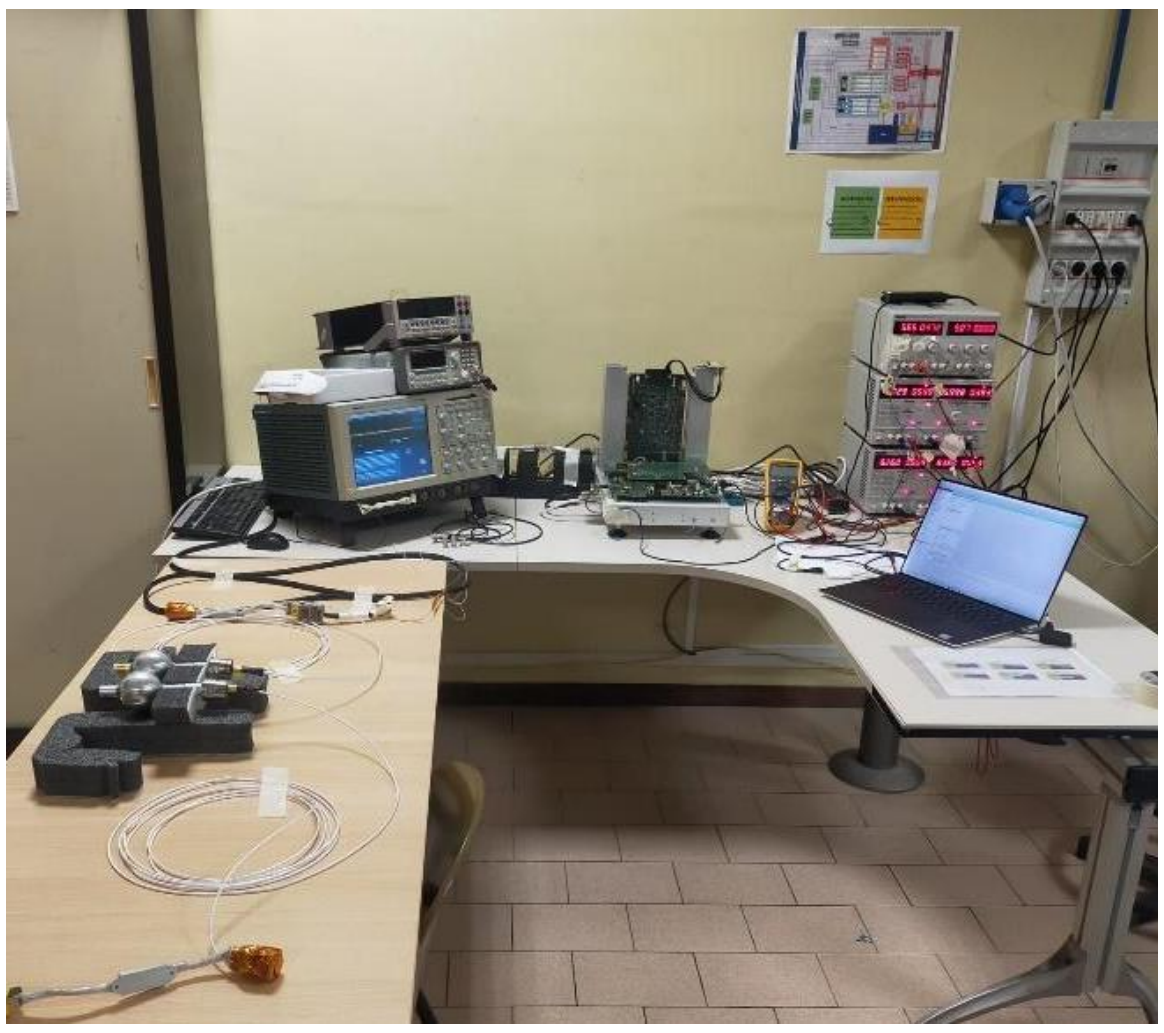

Figure 47- Arrangement of the setup for calibration measurements at INFN-Tor Vergata.

The CF in ($V/Digit$) is obtained from the ratio between the value of either the electric potential applied to the corresponding EFP and the numerical value of the raw-level digitized data (for what concerns signals in the ULF band, (A, B, C, D); or the potential difference applied to pairs of probes (e.g., A-B) for signals in all other bands (ELF, VLF, VLFe and HF).

Considering possible differences or variability at different levels of the entire analog electronic chain, a set of calibration factors was measured for each channel.

The calibration coefficients may vary due to parameter variation, boundary conditions or electronic component tolerances. The measured DC value is obtained as the mean over the overall data sample. Since the analog chain is the same, only one channel was chosen as representative for calibration (ULFA, ELF0, VLF0, VLFe0, HF0). So, after the standard deviation was taken as a reference to assess the variability between channels (see chapter 3), the system was calibrated as shown below. The main factors that can generate this variability are:

- Different values of potential, due to the non-complete linearity of the amplification chains up to digitization, which implies the need to evaluate multiple calibration factors as a function of signal amplitude.

- Different frequencies (including DC, in the case of ULF channels) due to the possible "ripple" of the mid-band transfer function, which once again implies the need to directly assess (or estimate) multiple calibration factors as a function of frequency.

One pair of probes was used to determine the CFs. The selected switch matrix is the one shown in Table 14, and chosen as the default one for the measurements.

| *DEFAULT* | | |
|---|---|---|
| **Channel** | **MF** | **HF** |
| 0 | B - A | B - A |
| 1 | B - C | D - C |
| 2 | D - C | D - A |

Table 14 – Differences between the probes chosen as the default switch matrix.

Out of the four available (A, B, C or D), A and B sensors were selected as the probe pair to test. The input signal (sine or DC voltage) was applied directly to Probe A, maintaining the other Probe at GND. All measurements were made at a nominal injected bias current of $+125nA$ , whose value is representative of a realistic situation. The signal applied to each channel was a pure sine waveform of given amplitude and frequency. An additional DC level was generated in the case of DC calibration for the ULF channel.

The configuration used for calibration was the APU_ADAPTER system. The input signals were accurately measured with a precision oscilloscope. The length of the output data stream for the measurement of each channel was about 8 k-samples.

The sine amplitude value acquired and measured after digitization is the peak-to-peak value obtained offline from the manipulation with a MATLAB algorithm of digitized data.

This algorithm has been devised for the calculation of the average peak-to-peak value within an acquisition time window, considering both the sampling frequency of the band under test and the frequency of the signal sent, and it has been optimized to be immune to the aliasing phenomenon.

To further reduce errors due to sampling, the measurements were made selecting those acquisitions in which the sample was taken as close as possible to the maximum and minimum values of the sine wave sent.

Although not influential for the purposes of calibration, the EFD sensitivity values measured after calibration are reported below, in the form of standard deviation of the waveforms (for the bands of greatest scientific interest) relating to the intrinsic noise generated by the various analog chains.

$$ULF\ noise = \approx 7\mu V$$

$$ELF\ noise = \approx 50\ \mu V$$

$$\mathrm{V}LF\ noise = \approx 360\ \mu V$$

$$\mathrm{V}LF\mathrm{e}\ noise = \approx 480\ \mu V.$$

## 4.2 DC AND ULF LINEARITY AND CONVERSION FACTORS

First, the instrumental zero offset introduced by the DC voltage generator set at 0 V was measured. This contribution is an addition to the zero offset of the channel under calibration, with the Probe connected to GND.

This offset introduced by the generator has already been removed from the measured values, together with the peak-to-peak measurements in the ULF band.

Several measurements were carried out at 30 Hz because this frequency value was chosen as the ULF band center for the study of the transfer function, which will be shown later, (*Figure 49*).

To facilitate an overview of the variability of the calibration factor as a function of both frequency and amplitude, a surface plot is shown in the next *Figure 48*. Here, it must be stressed that a logarithmic scale is selected for amplitude values of the input signal in order to test the whole dynamic range of the band under investigation. Signal amplitude is monitored up to saturation, and also upper limits of band frequency are reached: this is why a major non-uniformity of the CF can be appreciated in the upper right side of the graph.

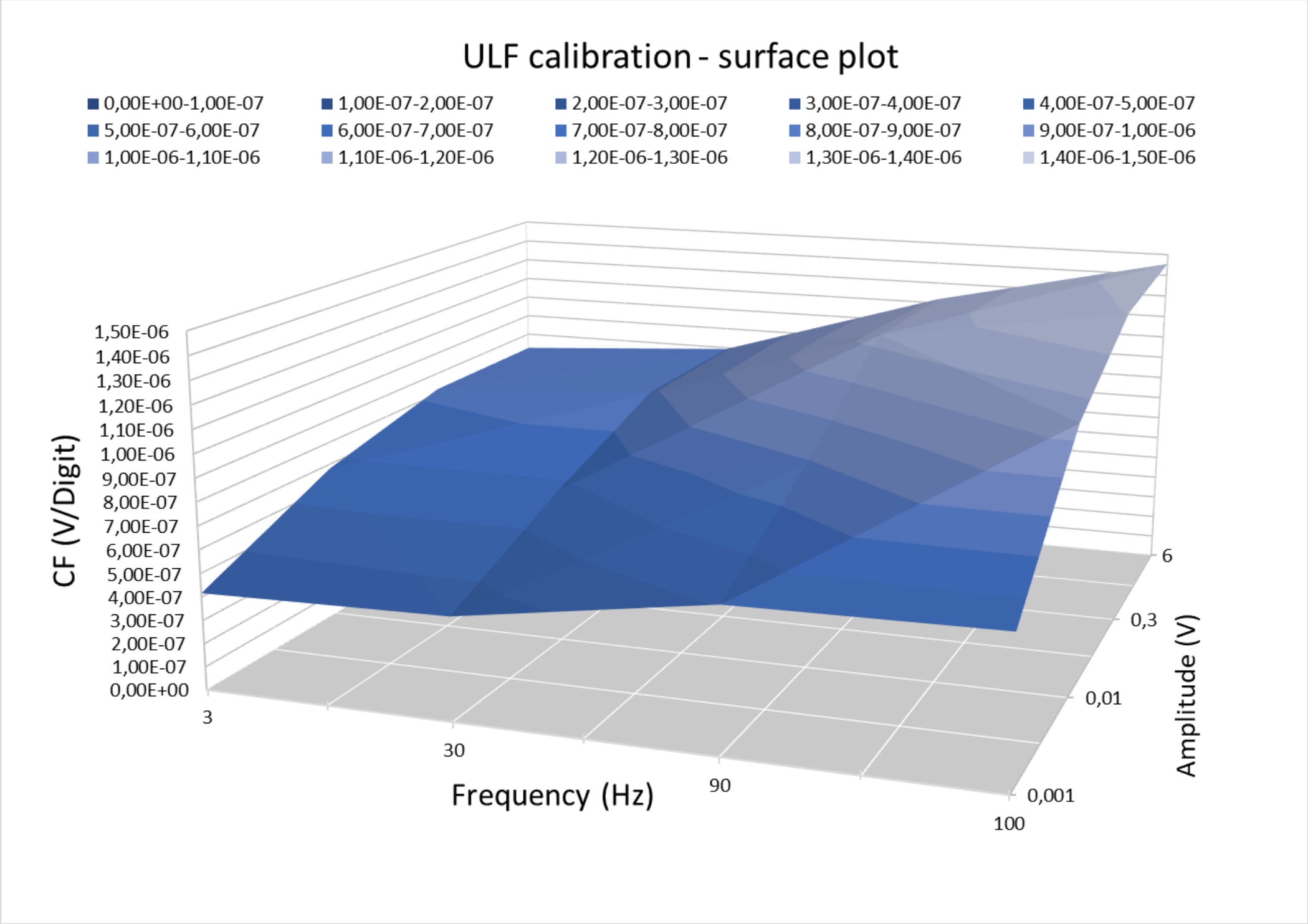

Figure 48 - Calibration factors surface plot vs. frequency and amplitude.

Figure 49 shows the input in the entire DC dynamic range versus output data. The slope coefficient of the trend line represents the best conversion factor value for amplitude changes, excluding those with the smallest value of the input signal where the correspondent measurements are not considered sufficiently accurate for the calibration.

It is noteworthy that the linear trend perfectly fits with the experimental data.

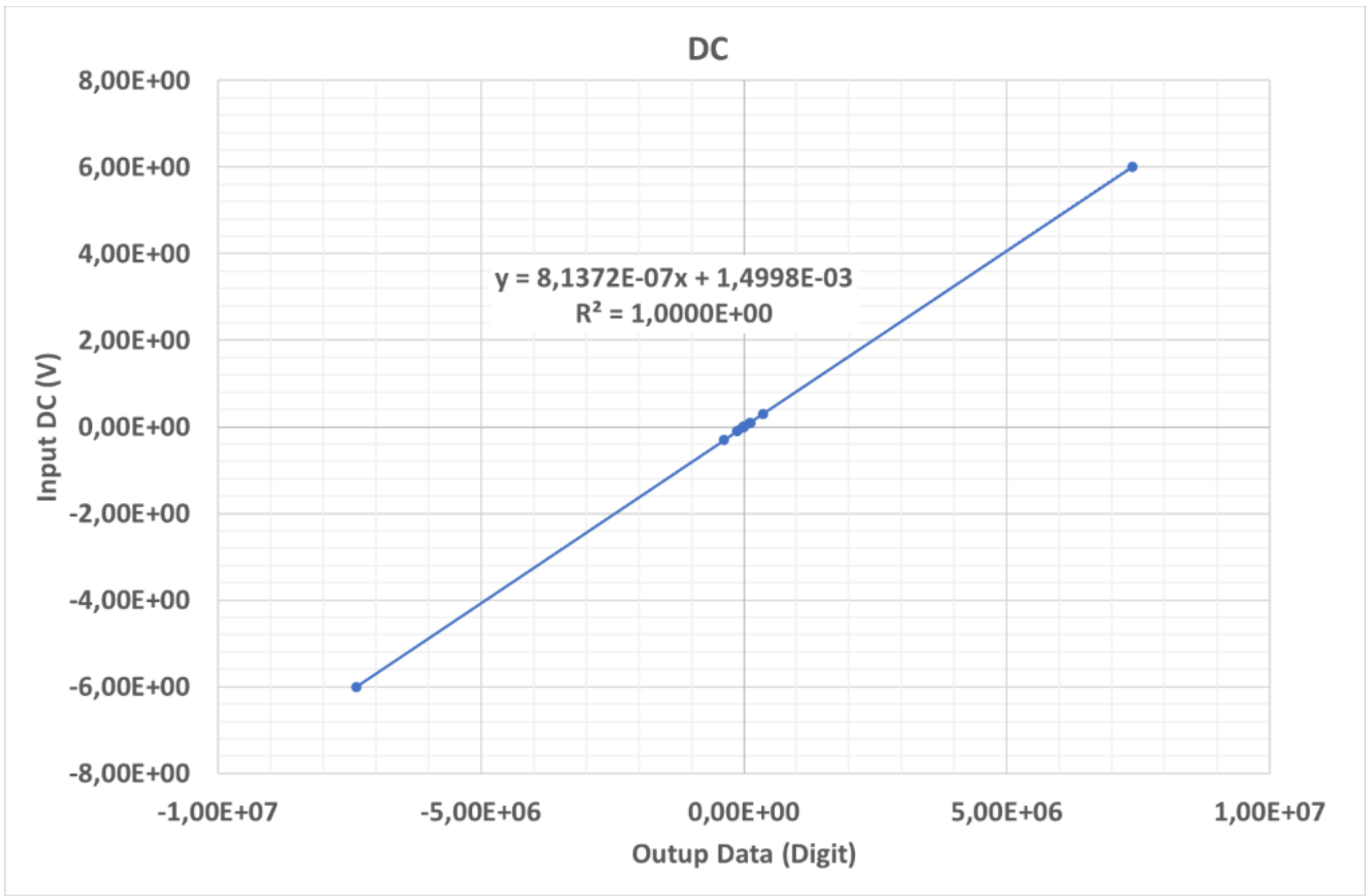

Figure 49 - DC Voltage vs. Data. The Conversion Factor is the best representative value for all DC dynamic range.

In addition, the value of the average CF is reported in Table 15 together with the value obtained from linear regression.

| ***CF***<br>***(Linear regression)*** | ***CF***<br>***(Mean)*** |
|---|---|
| $8{,}137 \cdot 10^{-7}\ V/Digit$ | $8{,}114 \cdot 10^{-7}\ V/Digit$ |

Table 15 - Best representative value for all DC dynamic range.

## 4.3 ULF Transfer Function

To study the behavior of the ULF chain as the frequency varies, a fixed amplitude input signal of 300 $mV_{pp}$ was chosen, monitored (and held constant with a tolerance of $\pm 1\ mV$ ) by the oscilloscope during each test. The results are shown in the following Table 16 and related Figure 50.

| TF | nominal | 300 mVpp | | |
|---|---|---|---|---|
| freq (Hz) | input (V) | digit_pp | Decibel | CF (V/dig) |
| 0 | 0,2998 | 365183 | 0,000 | 8,21E-07 |
| 3 | 0,2956 | 359629 | -0,133 | 8,22E-07 |
| 5 | 0,2953 | 346580 | -0,454 | 8,52E-07 |
| 10 | 0,2962 | 327991 | -0,933 | 9,03E-07 |
| 15 | 0,2967 | 350198 | -0,364 | 8,47E-07 |
| 20 | 0,2966 | 372414 | 0,170 | 7,96E-07 |
| 30 | 0,2967 | 334297 | -0,768 | 8,88E-07 |
| 40 | 0,2966 | 378358 | 0,308 | 7,84E-07 |
| 60 | 0,2961 | 376622 | 0,268 | 7,86E-07 |
| 70 | 0,2959 | 308475 | -1,466 | 9,59E-07 |
| 90 | 0,2956 | 239497 | -3,664 | 1,23E-06 |
| 100 | 0,2964 | 211171 | -4,758 | 1,40E-06 |
| 110 | 0,2967 | 39612 | -19,294 | 7,49E-06 |
| 115 | 0,2952 | 6204 | -35,397 | 4,76E-05 |

Table 16 - ULF Transfer Function data.

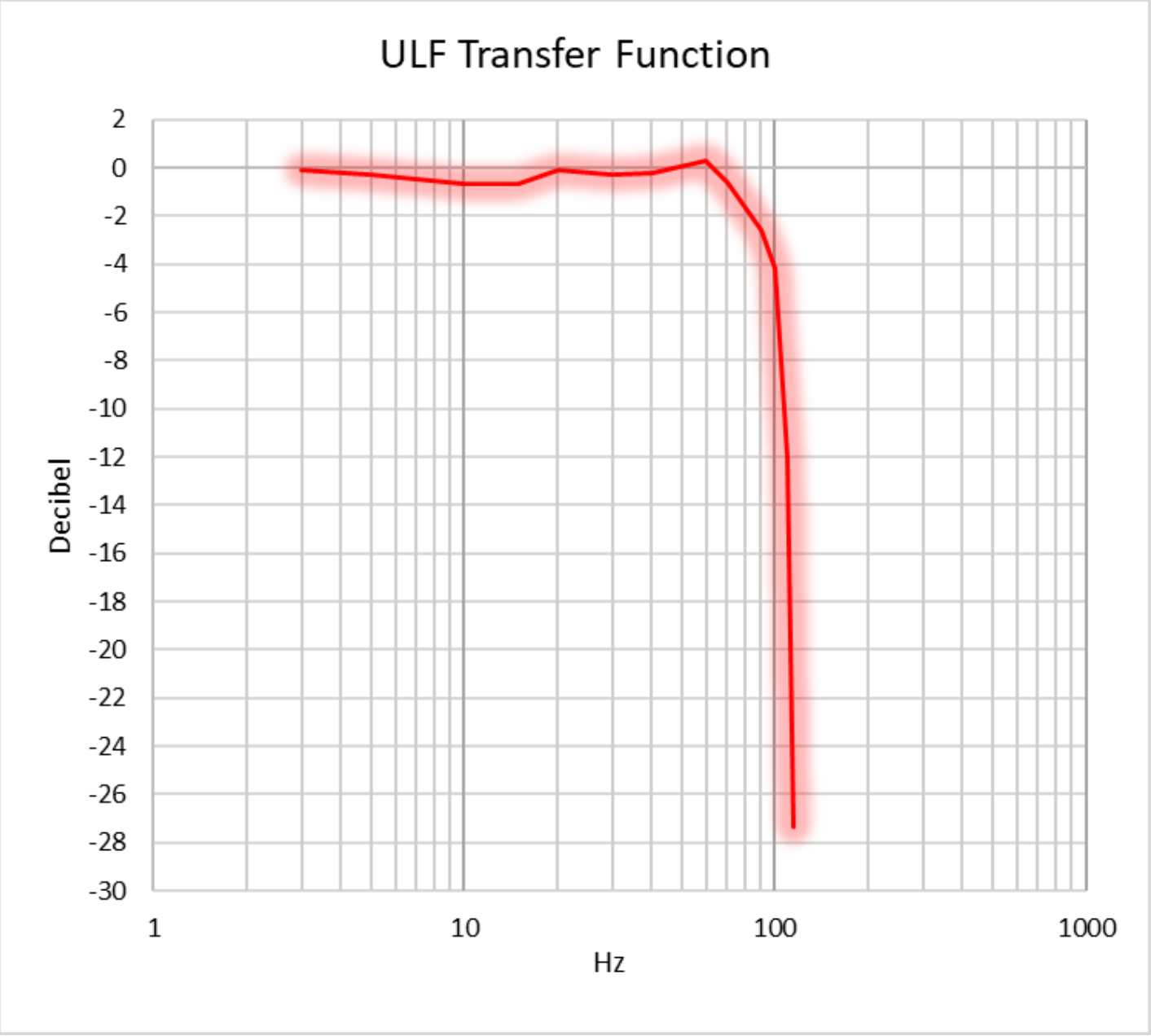

Figure 50 - ULF Transfer Function with sine input of 300 mVpp.

## 4.4 ELF CHANNEL 0 (PAIR B-A) CONVERSION FACTORS

The signal in the ELF chain was sent only to probe B while probe A was connected to GND. The ELF channel 0 was chosen and the difference between B and A was selected.

In Figure 51 the surface plot of the conversion factor variability is shown.

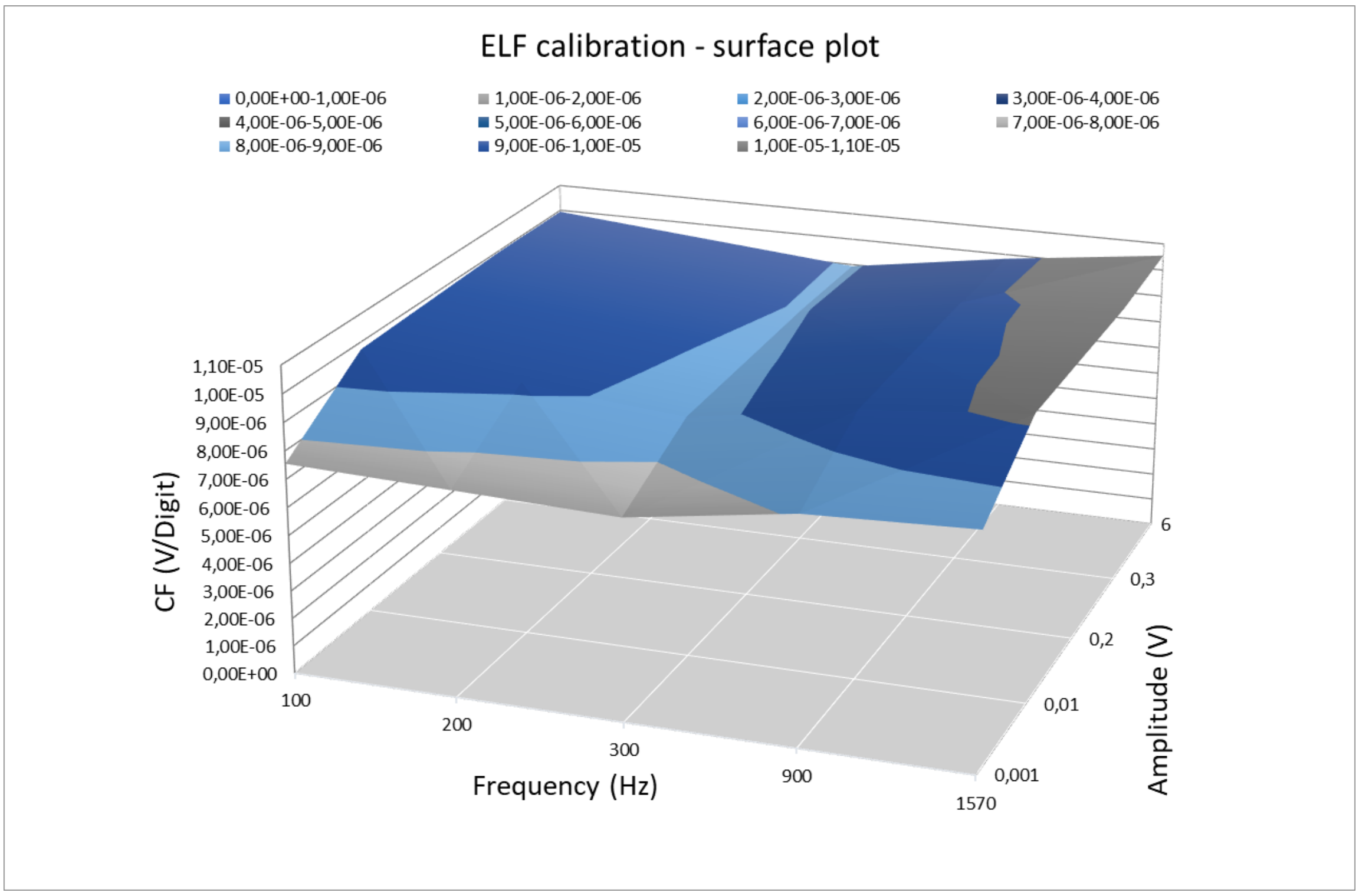

Figure 51 - ELF (B-A) Conversion Factors surface plot vs. Frequency and Amplitude.

The mean CF value in the ELF band is reported in Table 17.

| ***CF (Mean)*** |
| --- |
| $9{,}7128 \cdot 10^{-6}\ \ V/digit$ |

Table 17 - ELF mean CF.

## 4.5 ELF TRANSFER FUNCTION

For the ELF chain, a fixed amplitude input signal of 297 $mV_{pp}$ was chosen, monitored (and held constant with a tolerance of $\pm 1\ mV$ ) by the oscilloscope during each test. The results are shown in the following Table 18 and related Figure 52.

| TF | nominal | 300 mVpp | | |
|---|---|---|---|---|
| freq (Hz) | input (V) | digit_pp | Decibel | CF (V/dig) |
| 12 | 0,297 | 151 | -46,449 | 1,97E-03 |
| 16 | 0,298 | 1995 | -24,030 | 1,49E-04 |
| 19 | 0,299 | 21228 | -3,491 | 1,41E-05 |
| 30 | 0,297 | 28649 | -0,887 | 1,04E-05 |
| 60 | 0,297 | 30920 | -0,224 | 9,61E-06 |
| 100 | 0,298 | 29949 | -0,501 | 9,95E-06 |
| 200 | 0,297 | 31728 | 0,000 | 9,36E-06 |
| 300 | 0,298 | 33371 | 0,439 | 8,93E-06 |
| 350 | 0,298 | 31168 | -0,155 | 9,56E-06 |
| 500 | 0,298 | 33265 | 0,411 | 8,96E-06 |
| 900 | 0,298 | 30253 | -0,413 | 9,85E-06 |
| 1570 | 0,298 | 29068 | -0,761 | 1,03E-05 |
| 2000 | 0,297 | 28712 | -0,868 | 1,03E-05 |
| 2022 | 0,297 | 27008 | -1,399 | 1,10E-05 |
| 2050 | 0,294 | 22964 | -2,808 | 1,28E-05 |
| 2100 | 0,297 | 15248 | -6,365 | 1,95E-05 |
| 2200 | 0,296 | 3988 | -18,014 | 7,42E-05 |

Table 18 – ELF Transfer Function data.

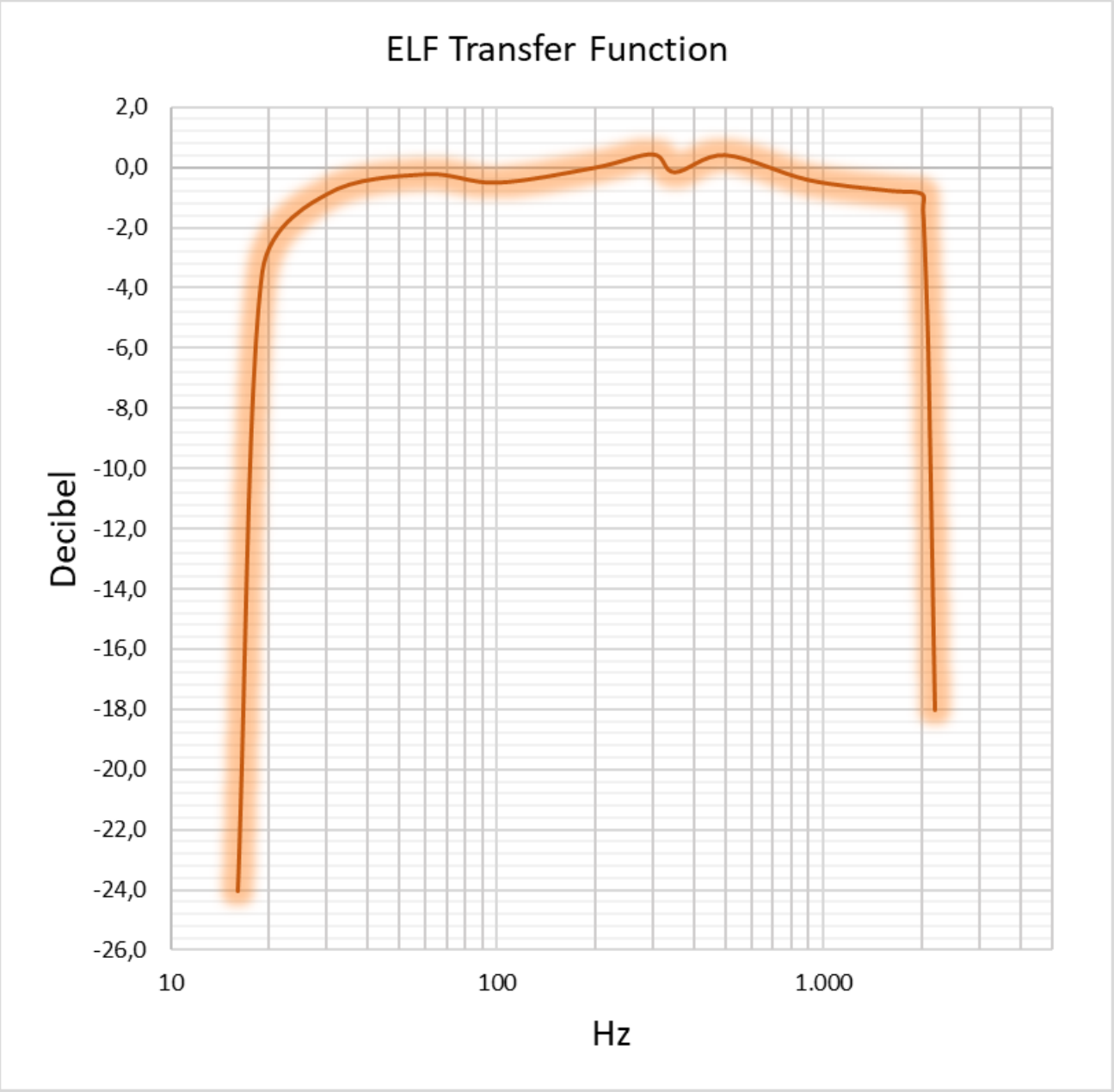

Figure 52 - ELF Transfer Function with sine input of 300 mVpp

## 4.6 VLF CHANNEL 0 (PAIR B-A) CONVERSION FACTORS

The signal was sent only to probe B, while probe A was connected to GND. The VLF channel 0 was chosen, and the difference between B and A was selected. In Figure 53, the surface plot of the conversion factor variability is shown. Again, in this graph a logarithmic scale is selected for amplitude values of the input signal in order to test the whole dynamic range of the band under investigation. Signal amplitude is monitored up to saturation, and also upper limits of band frequency are reached: this is why a major non-uniformity of the CF can be appreciated in the upper right side of the graph.

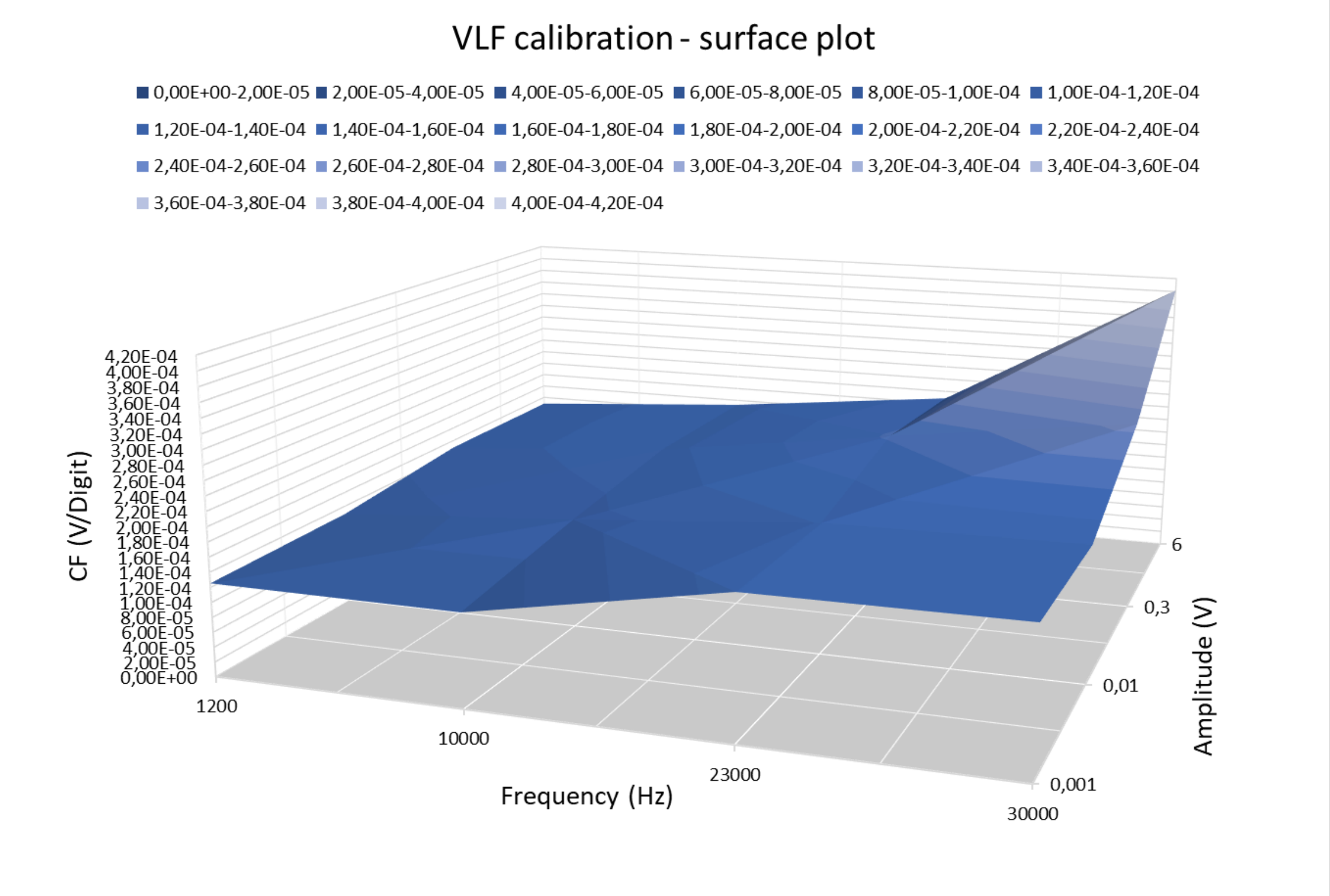

Figure 53 - VLF (B-A) Conversion Factors surface plot vs. Frequency and vs. Amplitude.

For completeness, the mean CF value in the VLF band is reported in Table 19 while the average values of the calibration coefficients for the FFTs at specific frequency values (minimum, medium, maximum) are reported in Table 20.

| $CF\ (Mean)$ |
|---|
| $1{,}8 \cdot 10^{-4}\ \ V/digit$ |

Table 19 - VLF mean CF.

| $Frequency\ (Hz)$ | $Mean\ CF\ (mV/digit)$ |
|---|---|
| $1{,}5 \cdot 10^3$ | 0,1530 |
| $10 \cdot 10^3$ | 0,1759 |
| $29 \cdot 10^3$ | 0,2069 |

Table 20-– Mean CF values for FFTs in VLFe band.

## 4.7 VLF Transfer Function

For the VLF chain a fixed amplitude input signal of 298 $mV_{pp}$ was chosen, monitored (and held constant with a tolerance of $\pm 1\ mV$ ) by the oscilloscope during each test. The results are shown in the following Table 21 and related Figure 54.

| TF | nominal | 300 mVpp | | |
|---|---|---|---|---|
| freq (Hz) | input (V) | digit_pp | Decibel | CF (V/dig) |
| 100 | 0,298 | 101 | -25,936 | 2,95E-03 |
| 400 | 0,298 | 1155 | -4,771 | 2,58E-04 |
| 500 | 0,298 | 1493 | -2,541 | 2,00E-04 |
| 1000 | 0,298 | 1991 | -0,041 | 1,50E-04 |
| 1200 | 0,296 | 1987 | 0,000 | 1,49E-04 |
| 10000 | 0,294 | 1685 | -1,373 | 1,74E-04 |
| 23000 | 0,296 | 1332 | -3,474 | 2,22E-04 |
| 30000 | 0,297 | 1115 | -5,048 | 2,66E-04 |

Table 21 - VLF Transfer Function data.

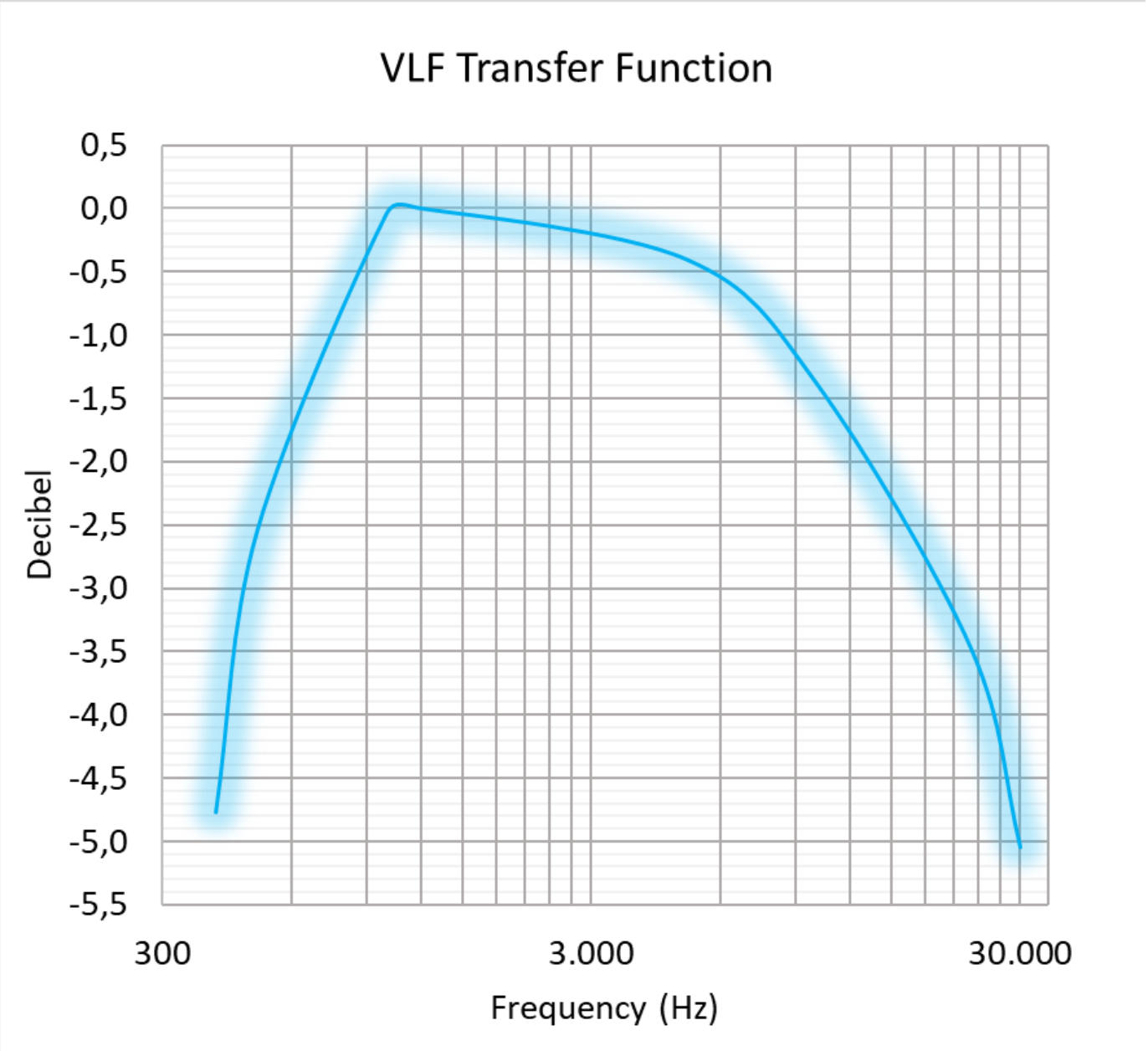

Figure 54 – VLF Transfer Function with sine input of 300 $mVpp$.

## 4.8 VLFe channel 0 (pair B-A) Conversion Factors

The signal was sent only to probe B, while probe A was connected to GND. The VLF channel 0 was chosen, and the difference between B and A was selected.

In Figure 55, the surface plot of the conversion factor variability is shown.

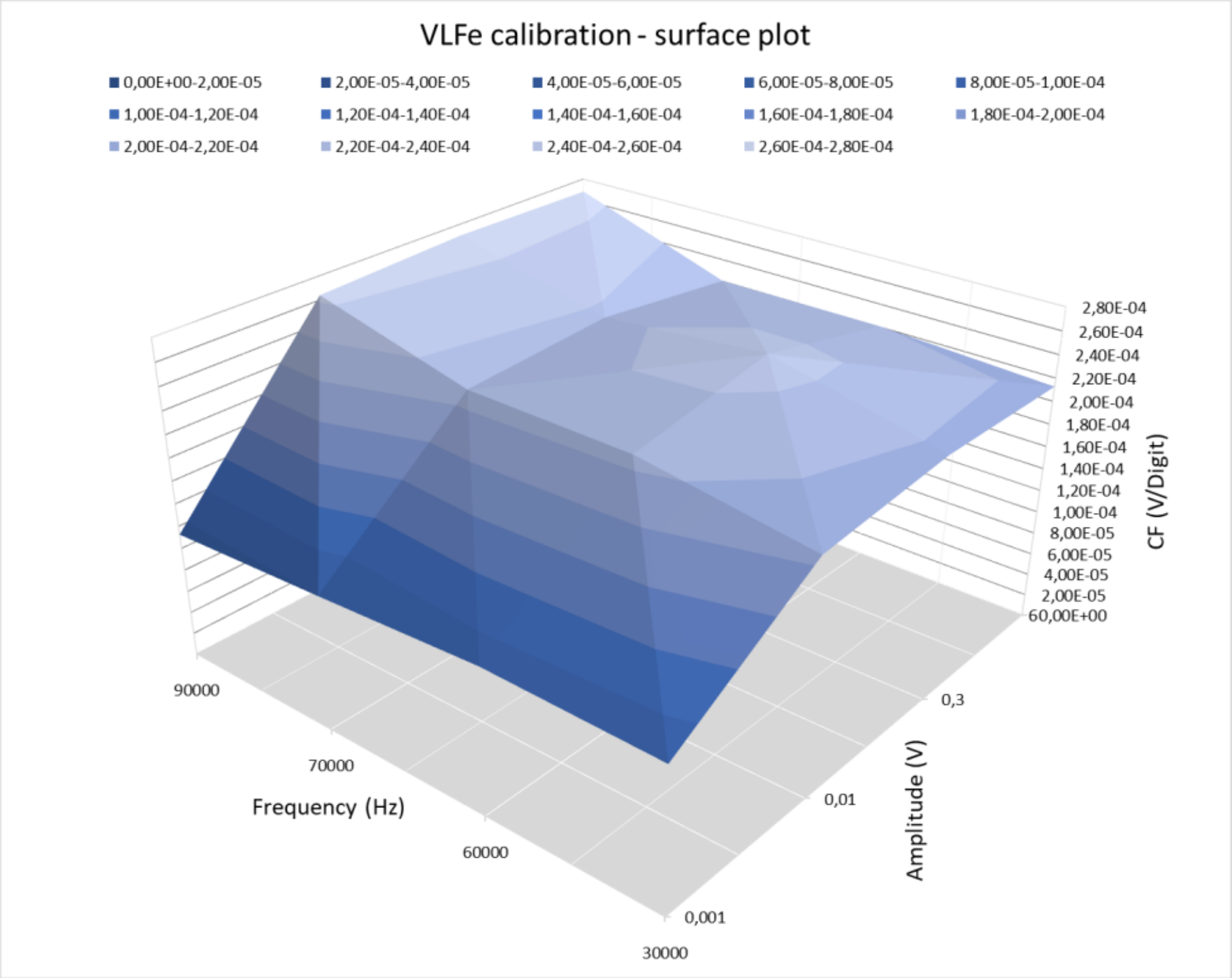

Figure 55 - VLFe (B-A) Conversion Factors surface plot vs. Frequency and vs. Amplitude.

In this graph it is evident how the calibration factor drops for very small values of the input signal because in addition to the greater difficulty in accurately measuring the peak-to-peak value of the input sine wave, the signal-to-noise ratio is lower, especially when approaching frequency values close to the cut-off frequency of the band.

| $CF\ (Mean)$ |
|---|
| $2{,}4 \cdot 10^{-4}\ \ V/digit$ |

Table 22- VLFe mean CF.

| $Frequency\ (Hz)$ | $Mean\ CF\ (mV/digit)$ |
|---|---|
| $22 \cdot 10^3$ | 0,1846 |
| $50 \cdot 10^3$ | 0,2925 |
| $90 \cdot 10^3$ | 0,2526 |

Table 23– Mean CF values for FFTs in VLFe band.

For completeness, the mean CF value in the VLFe band is reported in Table 22 while the average values of the calibration coefficients for the FFTs at specific frequency values (minimum, medium, maximum) are reported in Table 23.

## 4.9 VLFe Transfer Function

For the VLFe chain, a fixed-amplitude input signal of $298\,mV_{pp}$ was chosen, monitored (and held constant with a tolerance of $\pm 1\ mV$ ) by the oscilloscope during each test. The results are shown in the following Table 24 and related Figure 56.

| TF | nominal | 300 mV | | |
|---|---|---|---|---|
| freq (Hz) | input (V) | digit_pp | Decibel | CF (V/dig) |
| 20500 | 0,298 | 65 | -26,438 | 4,58E-03 |
| 20700 | 0,298 | 363 | -11,498 | 8,21E-04 |
| 20800 | 0,298 | 627 | -6,751 | 4,75E-04 |
| 21000 | 0,298 | 1126 | -1,666 | 2,65E-04 |
| 30000 | 0,298 | 1364 | 0,000 | 2,18E-04 |
| 60000 | 0,298 | 1252 | -0,744 | 2,38E-04 |
| 70000 | 0,298 | 1261 | -0,682 | 2,36E-04 |
| 90000 | 0,298 | 1100 | -1,868 | 2,71E-04 |
| 92000 | 0,298 | 982 | -2,854 | 3,03E-04 |
| 95000 | 0,298 | 877 | -3,836 | 3,40E-04 |
| 100000 | 0,298 | 605 | -7,061 | 4,93E-04 |

Table 24 - VLFe Transfer Function data.

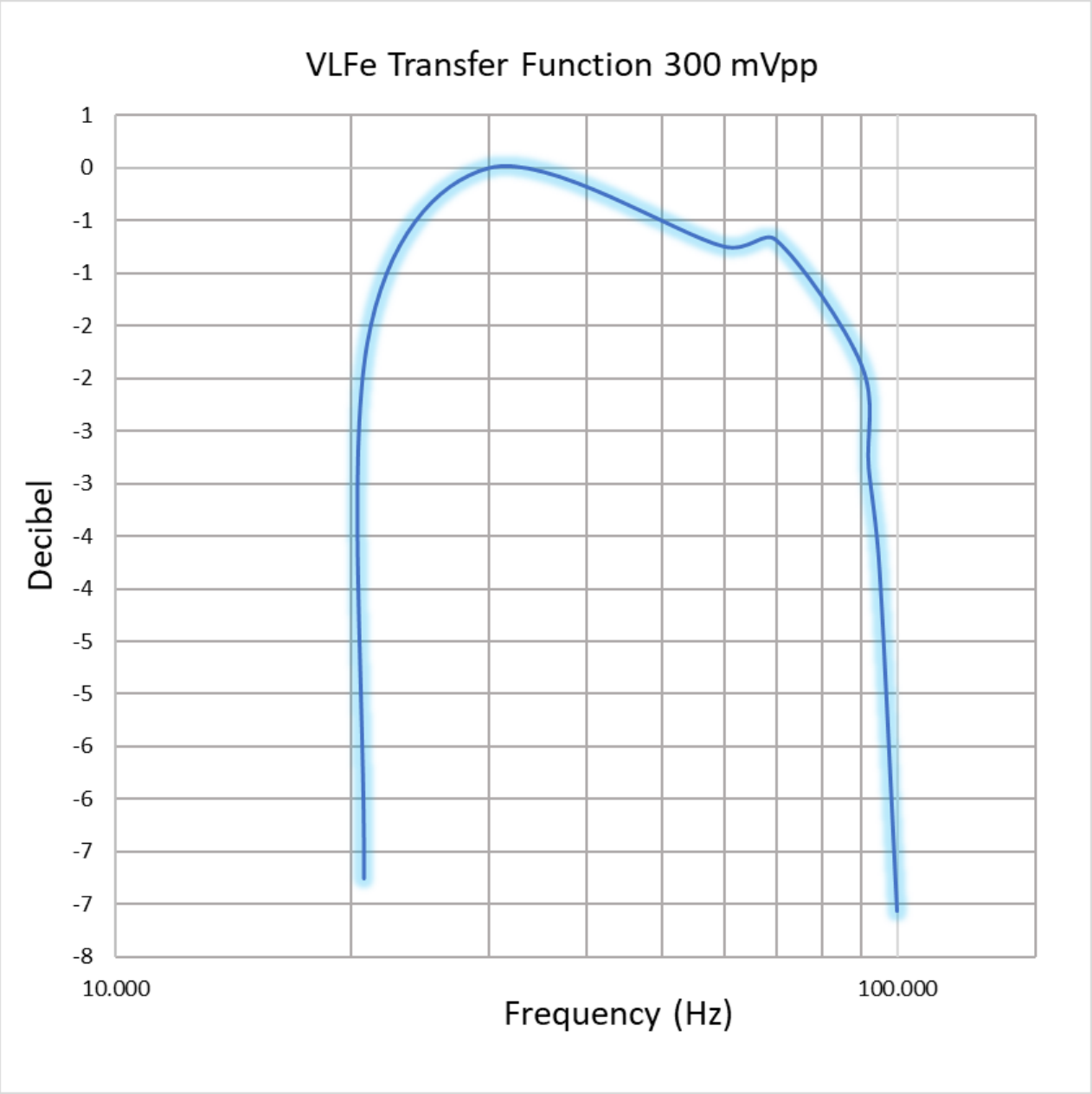

Figure 56 - VLFe Transfer Function with sine input of 300 mVpp.

## 4.10 HF CHANNEL 0 (PAIR B-A) CONVERSION FACTORS

Since during flight the HF band will transmit only FFT amplitude values, the next set of measurements concerns the peak value of the FFT. See Table 25 and the CF Variability in the surface plot shown in Figure 57.

| ***Frequency* (*Hz*)** | ***Mean CF* (*mV*/*digit*)** |
|---|---|
| $2{,}5 \cdot 10^4$ | 1,5147 |
| $5 \cdot 10^5$ | 1,1997 |
| $3{,}5 \cdot 10^6$ | 0,9428 |

Table 25 - HF - FFT Mean Calibration factors.

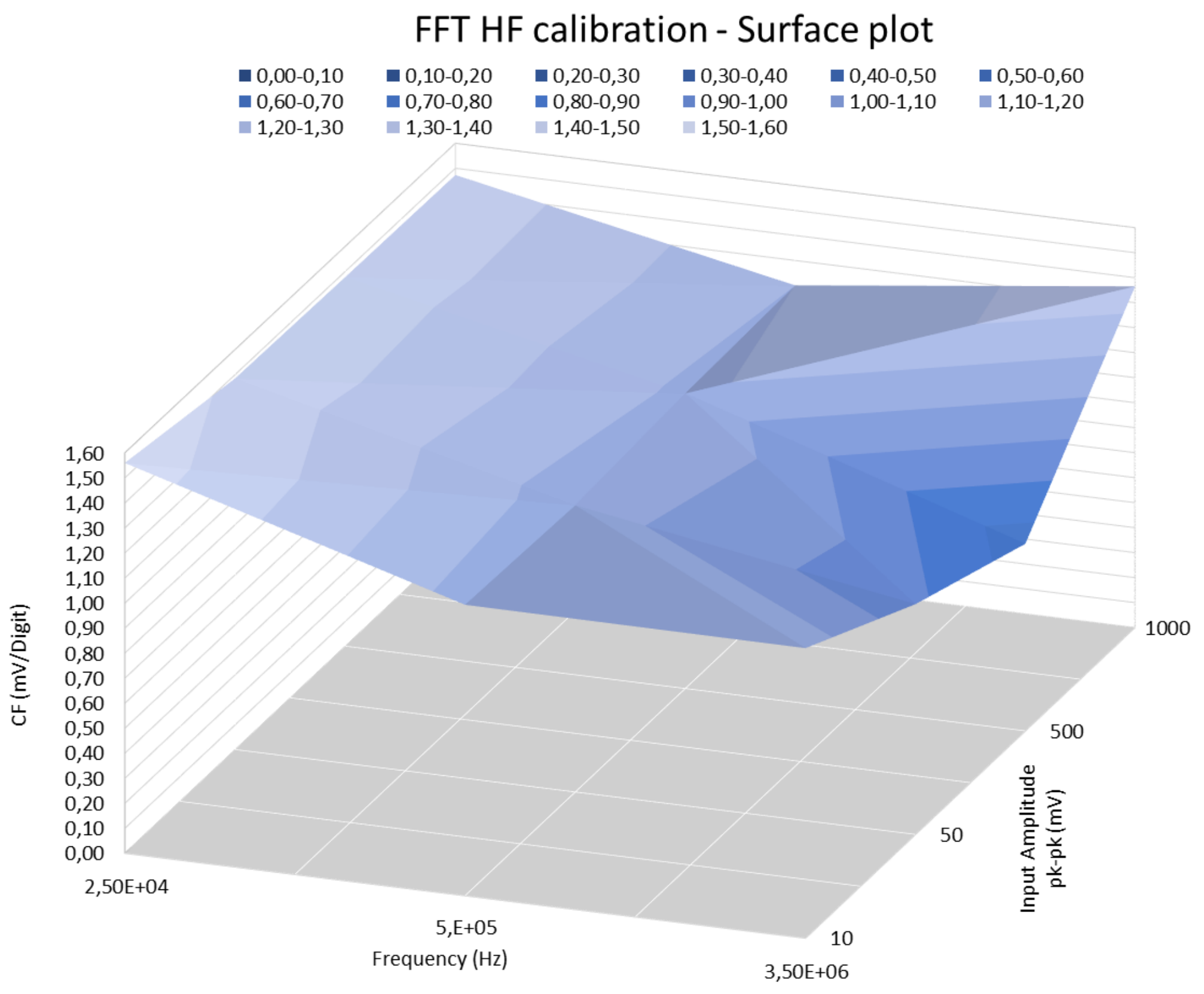

Figure 57 - HF (B-A) FFT Conversion Factors surface plot vs. Frequency and vs. Amplitude.

## 4.11 HF Transfer Function

For the HF chain, a fixed-amplitude input signal of 500 $mV_{pp}$ was chosen, monitored (and held constant with a tolerance of $\pm 1\ mV$ ) by the oscilloscope during each test. The results are shown in the following Table *26* and related Figure 58. The presence of the little overshoot at the frequencies greater than 2 MHz is caused by the resonance of the sensor input circuit (Probe), which has a non-abatable parasitic capacitance in the order of a few pF to few tens of pF, acting together with the external capacitance to produce the peak.

| TF | nominal | 500 mVpp | | |
|---|---|---|---|---|
| freq (MHz) | input (V) | digit_pp | Decibel | CF (V/dig) |
| 0,025 | 0,490 | 369 | 0,00E+00 | 1,33E-03 |
| 0,5 | 0,493 | 416 | 9,88E-01 | 1,19E-03 |
| 1 | 0,495 | 462 | 1,86E+00 | 1,07E-03 |
| 1,5 | 0,496 | 670 | 5,08E+00 | 7,40E-04 |
| 2 | 0,494 | 826 | 6,93E+00 | 5,98E-04 |
| 2,5 | 0,495 | 815 | 6,79E+00 | 6,07E-04 |
| 3 | 0,483 | 706 | 5,76E+00 | 6,84E-04 |
| 3,5 | 0,483 | 608 | 4,46E+00 | 7,94E-04 |
| 3,8 | 0,457 | 427 | 1,87E+00 | 1,07E-03 |
| 4 | 0,455 | 147 | -7,35E+00 | 3,10E-03 |

Table 26 - HF Transfer Function data.

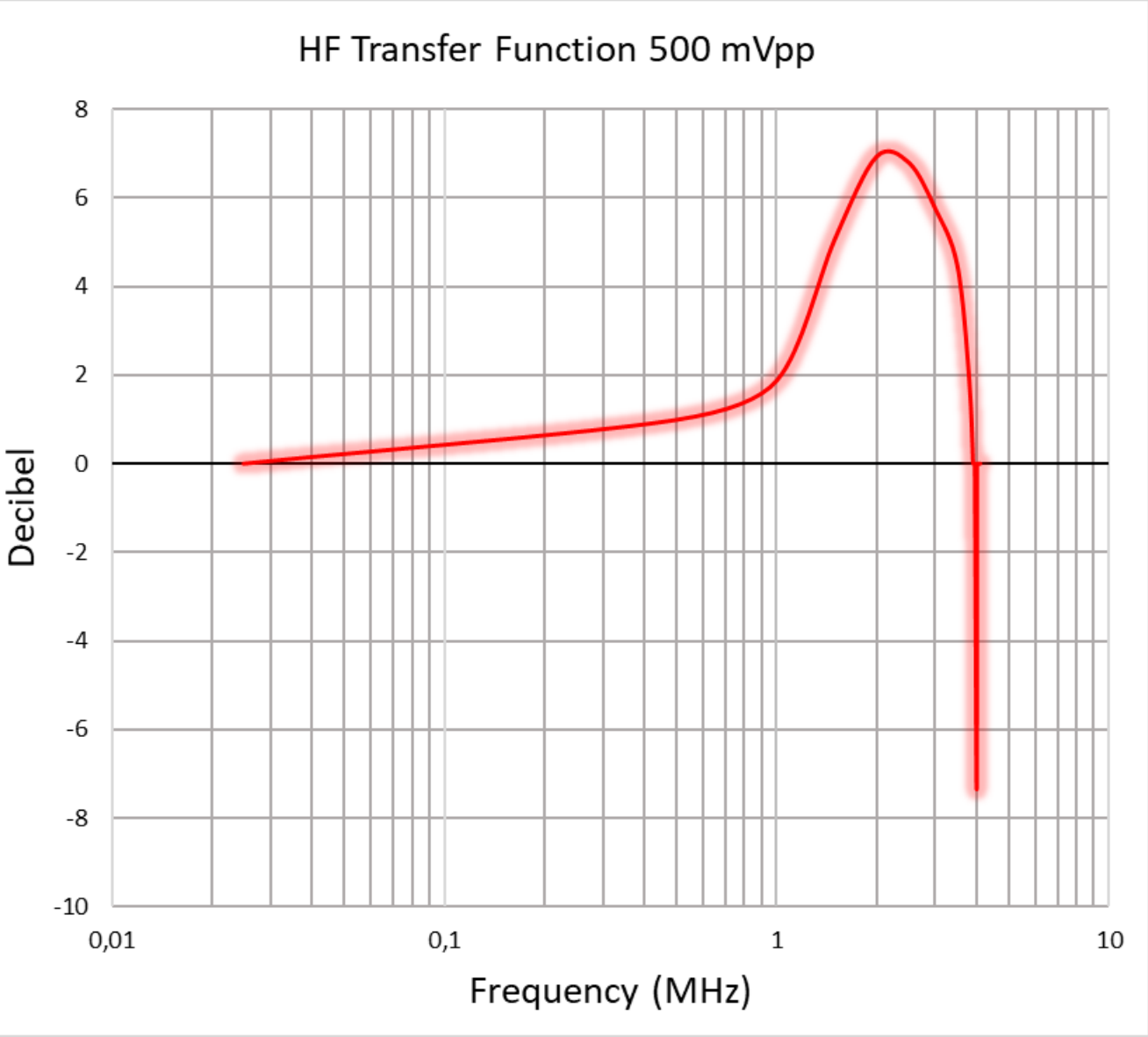

Figure 58 - HF Transfer Function with sine input of 500 mVpp.

From the measurements shown in this paragraph, it is evident that the main cause of variability for the calibration factor is linked to the frequency of the signal, and therefore to the transfer function shape, for any channel under examination.

The variability between channels has been studied in chapter 3 and does not show evidence worthy of note, while the measurements show a slight dependence on the amplitude of the input signal. This dependence is linked both to the oscilloscope losing its measurement accuracy when input values are too small (~1 mV), and to saturation and phenomena related to the slew rate when those same values are too large.

## 4.12 PLASMA CHAMBER MEASUREMENTS

The system, composed of the EFPs and the APU at the QM stage, was tested in the Plasma Chamber facility at INAF-IAPS, Figure 59, to evaluate its performance in a flight environment as similar as possible to the real case [77].

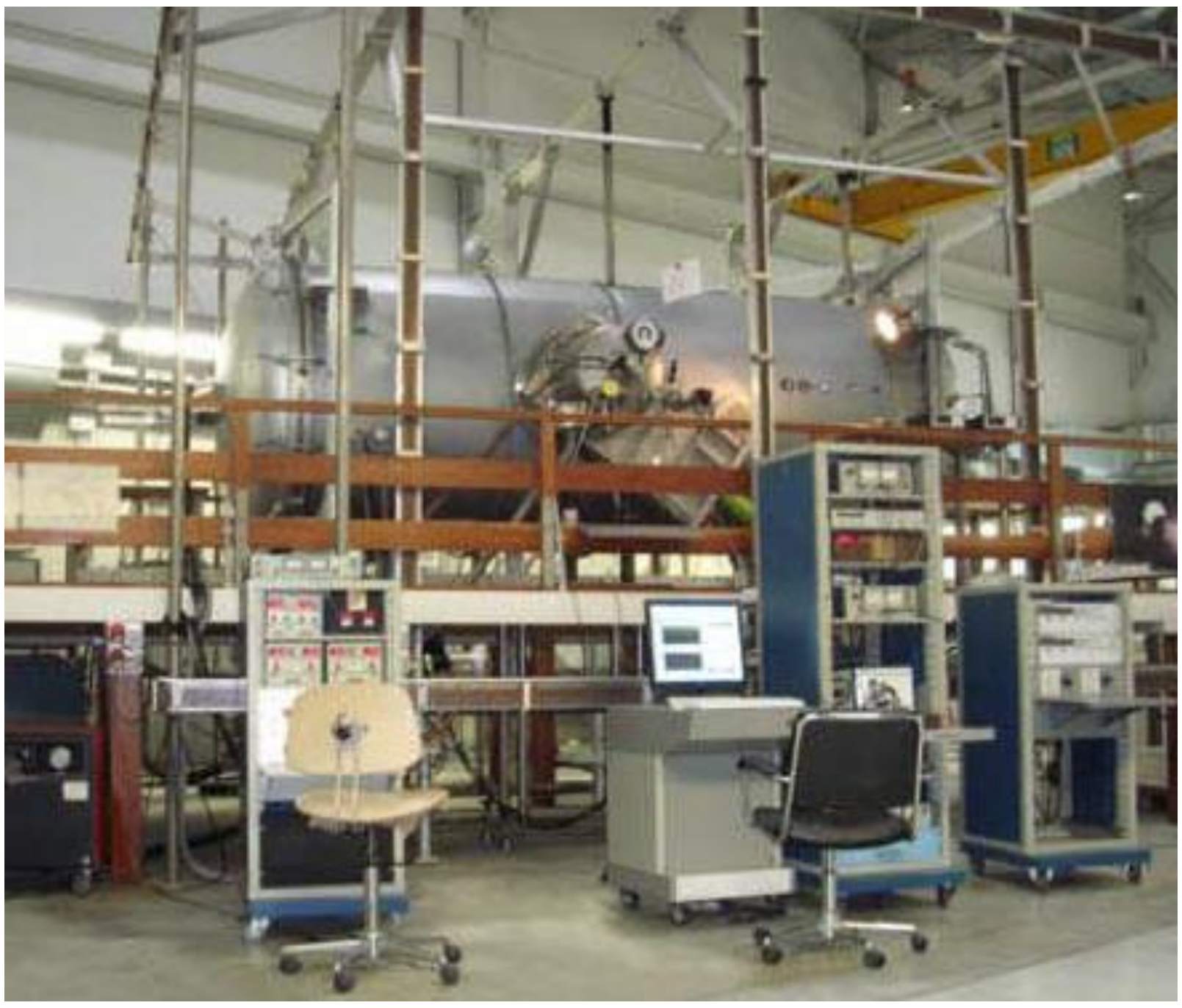

Figure 59 – IAPS Plasma Chamber facility.

Various tests have been conceived to outline the instrument capabilities when operating under different conditions encountered along the spacecraft (S/C) orbit.

To evaluate the effects of the magnetic field component, $B_x$, parallel to the plasma beam (i.e., the typical conditions along a Sun-synchronous orbit ), we measured the EFD floating potential ($V_f$) for three different magnetic field values ($B_x = 0.01G$, value of the minimum residual field in the chamber, $B_x = 0.25G$ and $B_x = 0.45G$), representing the typical excursion of Bx across the latitudinal range spanned by the satellite. The other components were kept at their minimum values (≤ 0.01G).

Such observations were compared to the theoretical $V_{f_{th}}$ values recovered from the Orbit Motion Limited (OML) theory. The theoretical $V_{f_{th}}$ values were computed using input plasma density and electron temperature ($T_e$) values retrieved from Langmuir probe measurements on board the same satellite [78].

The OML expected results together with the measured ones are shown in Figure 60.

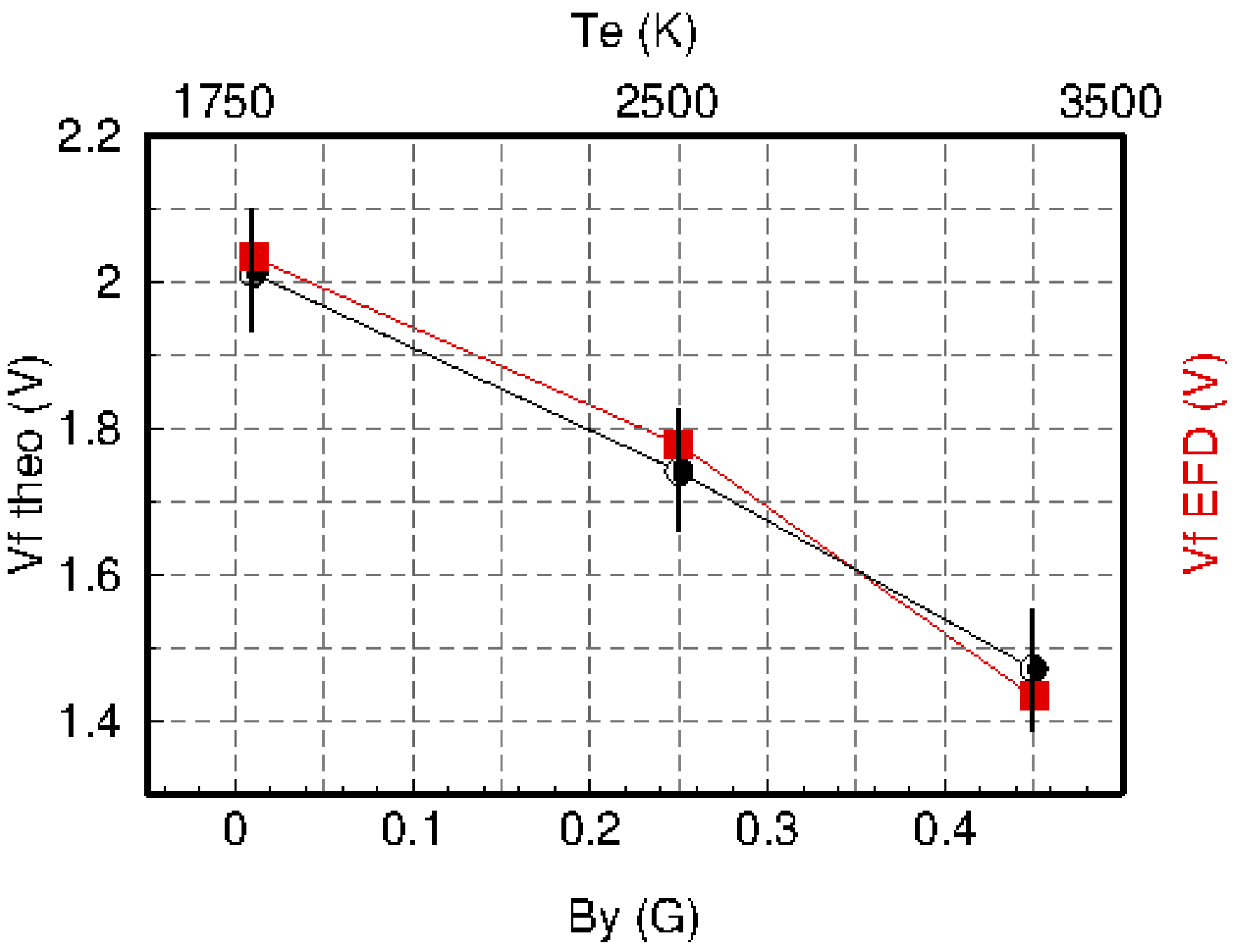

Figure 60 - Theoretical values (black circles) of EFD are displayed together with the measured floating potential (red squared markers) for three different values of magnetic field component parallel to the plasma beam in the Plasma Chamber. The error bar amplitude of the theoretical values is obtained by the propagation of the diagnostic measurement's uncertainties, while the error bar amplitude of the measured potential by EFD is obtained by the environmental noise and it is not appreciable on this scale.

The observed $V_f$ decrease for increasing B is an expected behavior caused by the smaller electron gyroradius and the relevant forced motion of the electrons around the probe surface, which, in turn, reduces the electron saturation current ([68], and reference therein). In addition, an increased electron temperature ($T_e$) is observed, as expected from the reduced slope of the characteristic curve of the sensors.

Moreover, the $V_f$ reduction is also coherent with the rule of thumb establishing that $V_f$ is a few $k_B T$ lower than the plasma potential $V_{pl}$ ([79], and reference therein). Then, when $V_{pl}$ is constant, $V_f$ decreases for increasing $T_e$ (as shown in the top horizontal axis in (Figure 60).

In conclusion, these performance tests have assessed EFD-02's intrinsic noise level to be largely lower than environmental noise observed in the plasma, which ensures the accurate evaluation of small perturbations in particle collection on the probe surface as the ones induced by variations in the Earth's magnetic field along the orbit.

# CONCLUSIONS

The focus of this doctoral thesis has been to show the fundamental work accomplished to obtain a correct and accurate ground calibration of the EFD-02 instrument scheduled to fly by the end of 2023 on board the second satellite of the CSES mission.

In the initial part of the thesis, all the main scientific studies that motivated the CSES mission have been summarized, including the most accredited theories that link seismic events to possible earthquake precursors. The scientific results of missions prior to CSES, and their limitations, have been described in detail.

The cutting-edge characteristics of EFD-02 in measuring the electric field at high precision have been thoroughly described. The capabilities of the instrument are outstanding even in a very noisy environment like an ionospheric plasma, thanks to its high sensitivity, especially in the band of greatest scientific interest - the ULF band - in which approximately a value of $1\,\mu V/m$ is reached. With such a performance, it will be possible to study the variations of the local electric field in space with great sensitivity, and in combination with data from the CSES/EFD-01 twin .

The variability of EFD-02 calibration coefficients has been widely discussed, both across the various channels due to intrinsic differences in the APU analog circuitry and those, unavoidable, related to the frequency and amplitude of the input signal injected into the probes.

Thanks to the precise calibration of the instrument made in laboratory, it has been also possible in the final part of the thesis to demonstrate how the collection of charges in the plasma chamber varies by applying known values of the magnetic field, thus recreating conditions comparable to those the instrument will find in flight on board the CSES satellite.

Due to the various delays introduced by the pandemic situation, in this thesis it has been not possible to study and characterize the instrument during the flight phase and data taking in space; nevertheless, the calibration made here, and the related treatment, will become essential in the next-coming commissioning phase, when it will be possible to measure electric fields produced by known anthropic and non-anthropic sources. Those measurements used as reference points would have no meaning without the calibration performed during these three years of PhD course.

# GLOSSARY

**Acoustic gravity waves (AGW): they** consist of relatively high-frequency acoustic and low-frequency internal gravity (IG) branches. AGWs have typical periods of $100s \leq \tau < 1day$ and are strongly affected by the Earth's gravitational field. Such waves have typical wavelengths around $\lambda \cong 10km$ and propagation velocities around $v_p \cong 30\, m/s$. Various methods of measuring the atmospheric parameters indicate the presence of AGWs in a large range of heights extending from the troposphere to $z \leq 500km$. The most complete survey of such observations in the troposphere has been provided by Gossard and Hooke (1975).

**Bandwidth**: the bandwidth of an amplifier is the range of frequencies for which the amplifier gives "satisfactory performance". The definition of "satisfactory performance" may be different for different applications. However, a common and well-accepted metric is the half-power points (i.e., the frequencies where the power goes down by half of its peak value) on the output vs. frequency curve. Therefore, bandwidth can be defined as the difference between the lower and upper half power points. This is also known as the –3 dB bandwidth.

**Brightness Temperature (BT):** or radiance temperature, is the temperature at which a black body in thermal equilibrium with its surroundings should be to duplicate the observed intensity of a grey body object at a frequency $\nu$. This concept is used in radio astronomy, planetary science, and materials science.

**Conversion factor (CF):** it is a parameter that provides a correspondence between a potential difference applied to a Probe and the digital raw values of the scientific data from the channels of the instrument.

**Coronal Mass Ejections (CME)**: it is a significant release of plasma from the Sun's corona into the heliosphere. CMEs are often associated with solar flares and other forms of solar activity, but a broadly accepted theoretical understanding of these relationships has not been established.

**Cosmic rays**: these high-energy particles arriving from outer space are mainly (89%) protons – the lightest and most common element in the universe – but they also include nuclei of helium (10%) and heavier nuclei (1%), all the way up to uranium. When they arrive at Earth, they collide with the nuclei of atoms in the upper atmosphere, creating more particles, mainly pions. The charged pions can swiftly decay, emitting particles called muons. Unlike pions, these do not interact strongly with matter, and can travel through the atmosphere to penetrate the ground. The rate of muons arriving at the surface of the Earth is such that about one per second passes through a volume the size of a person's head.

**Dynamic Range (DR)**: The output dynamic range is the range, usually given in dB, between the smallest and largest useful output levels. The lowest useful level is limited by output noise, while the largest is limited most often by distortion. The ratio of these two values is quoted as the amplifier dynamic range.

**Evaluation Board**: sometimes interchangeable with the term "development board", this expression implies that you can use the board to develop an application. Firmware developers often use development boards to test their code in isolation, followed by testing with other peripherals.

**Field line resonance:** it is the resonant coupling between an isotropic mode and an anisotropic mode in a magnetized plasma. Field line resonances allow us to understand many features of ultra-low frequency oscillations in the terrestrial magnetosphere, since resonant mode coupling is the current paradigm to explain geomagnetic pulsations.

**Flight model (FM)**: the flight model is the flight end item that is configured. The QM is tested for a larger range of loads as compared to the FM, because for qualification, the loads are always designed with a factor of safety. Hence, QM is tested more rigorously but the final FM will be the one orbiting in the space, since testing the latter to the limits may induce micro-cracks in the structure resulting in earlier failure.

**Gyroradius**: the gyroradius $\rho$ of a particle of charge $e$ and mass $m$ in a magnetic field of strength B is one of the fundamental parameters used in plasma physics. Its exact classical definition for a single particle of perpendicular velocity $v_\perp$ is $\rho_e = \frac{v_\perp}{\omega_{ce}}$, written here for an electron with cyclotron frequency $\omega_{ce} = \frac{eB}{m_e}$ (see any textbook on plasma physics, e.g.,Baumjohann and Treumann, 2012).

**Linearity**: an ideal amplifier would be a totally linear device, but real amplifiers are only linear within limits. When the input signal to the amplifier is increased, the output also increases until a point is reached where some part of the amplifier becomes saturated and cannot produce any more output; this is called clipping, and it results in distortion.

**Low Earth orbit (LEO)**: it is an orbit that is relatively close to Earth's surface. It is normally at an altitude of less than 1000 km but could be as low as 160 km above Earth – which is low compared to other orbits, but still very far above Earth's surface. Satellites in this orbit travel at a speed of around 7.8 km per second; at this speed, a satellite takes approximately 90 minutes to circle Earth, meaning it travels around Earth about 16 times a day.

**Outgoing Longwave Radiation (OLR):** it is a measure of the amount of energy emitted to space by the Earth's surface, oceans, and atmosphere. As such, it is a critical component of the Earth's radiation budget.

**Peak ground acceleration (PGA):** it is equal to the maximum ground acceleration that occurred during earthquake shaking at a location. PGA is equal to the amplitude of the largest absolute acceleration recorded on an accelerogram at a site during a particular earthquake.

**Plasma Chamber**: it is an experimental apparatus capable to reproduce a large-volume ionospheric environment, which is particularly suitable to perform studies on a variety of plasma physics phenomena. It consists of a large volume vacuum chamber equipped with a plasma source (e.g., of Kaufman type), either or ionospheric type or solar. The former one  produces a plasma (e.g., from Argon or another neutral gas) with parameters (i.e., electron density and temperature) very close to the values encountered in the daytime ionosphere at F layer altitudes. A variation of plasma density in the experimental region, could be obtained by varying the source current discharge and neutral gas flow rate.

**Plasma sheath**: since electrons are much lighter than ions, they can escape from plasma at a much faster speed than ions if there is no confining potential barrier. Once electrons are mostly depleted from the boundary interface between plasma and electrodes or samples, a region with only positive ions and neutrals will be formed. This usually dark boundary region is called a plasma sheath. Positive charges in plasma sheath can push more ions to diffuse out of plasma.

**Power spectrum density (PSD)**: it is the measure of the power content of a signal versus frequency. A PSD is typically used to characterize random broadband signals. The amplitude of the PSD is normalized by the spectral resolution employed to digitize the signal.

**Probe:** it is the sensor that allows to make a measurement of potential from which to derive the electric field. It consists of a sensor located at the tip of a conductive boom. It is equipped with cylindrical conducting stubs bootstrapped at the electrode potential.

**Qualification model (QM)**: this model fully mirrors all features of the flight model design, and it is used for complete functional and environmental qualification testing both for hardware and software validation before launch.

**Remote sensing:** the science and technology through which characteristics and properties of targets on the Earth can be identified and determined from a distance. It has so far provided systematic, dedicated, and repetitive observations of the Earth's surface (atmosphere, water, land, living species, vegetation, pollution, climate) from global to local scales. Satellite observations have contributed to the spectacular improvement of the accuracy of weather forecasts over the last few decades. It has the means to respond and facilitate environmental management, in order to make sound and evidence-based decisions in relation to Earth's resources at a global scale, and across different continents, nations, and domains.

**Signal / Noise ratio**: it is a measure of the strength of the desired signal relative to background noise (undesired signal). S/N can be determined by using a fixed formula that returns the ratio of the two levels, showing whether and how noise is impacting the desired signal. The higher the ratio, the better the signal quality.

**Silicon Control Rectifier**: A silicon-controlled rectifier, or semiconductor-controlled rectifier, is a four-layer solid-state current-controlling device with three terminals. They have anode and cathode terminals like a conventional diode, and a third control terminal, referred to as the Gate. SCRs are unidirectional devices, i.e., they conduct current only in one direction like a diode or rectifier. SCRs are triggered only by currents going into the gate. The SCR combines the rectifying features of diodes and the On - Off control features of transistors. The name "silicon-controlled rectifier" is General Electric's trade name for a type of thyristor. SCRs are mainly used in devices where the control of high power, possibly at high voltage, is needed. The ability to switch large currents on and off makes the SCR suitable for use in medium to high-voltage AC power control applications, such as lamp dimming, regulators and motor control. In addition, unintentional SCRs can form in integrated circuits and, when these SCRs get triggered, circuit malfunction, or even reliability problems and damage, can result.

**Solar flare**: it is an intense localized eruption of electromagnetic radiation in the Sun's atmosphere. Flares occur in active regions and are often, but not always, accompanied by coronal mass ejections, solar particle events, and other solar phenomena. The occurrence of solar flares varies across the 11-year solar cycle.

**Strong motion duration**: earthquake duration is the total time of ground shaking from the arrival of seismic waves until the return to ambient conditions. Much of this time is at low shaking levels, which have negligible effect on seismic structural response and on earthquake damage potential. As a result, a parameter termed “strong motion duration” has been defined by several investigators to be used for the purpose of evaluating seismic response and assessing the potential for structural damage due to earthquakes. Only the portion of an earthquake record which has sufficiently high acceleration amplitude, energy content, or some other parameters significantly affects seismic response.

**Surface latent heat flux (SLHF):** it is the transfer of latent heat (resulting from evaporation, condensation, and other moisture phase changes) between the Earth's surface and the atmosphere through the effects of turbulent air motion. Evaporation from the Earth's surface represents a transfer of energy from the surface to the atmosphere. The mean latent heat flux is the accumulated flux divided by the length of the accumulation period, which depends on the data extracted.

**Total Electron Content (TEC):** it is an important descriptive quantity for the ionosphere of the Earth. TEC is the total number of electrons integrated between two points, along a tube of one meter squared cross section.

**Transient Luminous Event (TLE):** Sometimes called upper atmospheric lightning or ionospheric lightning, TLEs are short-lived electrical-breakdown phenomena or electrically induced forms of luminous plasma that occur well above the altitudes of normal lightning and Cumulonimbus (Cb) clouds.

**Whistler wave:** it is a very low frequency (VLF) electromagnetic (radio) wave generated by lightning. Frequencies of terrestrial whistlers are 1 kHz to 30 kHz, with a maximum amplitude usually at 3 kHz to 5 kHz. They are produced by lightning strikes (mostly intracloud and return-path) where the impulse travels along the Earth's magnetic field lines from one hemisphere to the other. They undergo dispersion of several kHz due to the slower velocity of the lower frequencies through the plasma environments of the ionosphere and magnetosphere. Thus, they are perceived as a descending tone which can last for a few seconds.

# BIBLIOGRAPHY

1. PICOZZA P., CONTI L. AND SOTGIU A. (2021), "Looking for Earthquake Precursors from Space: A Critical Review.", Front. Earth Sci. 9:676775. doi: 10.3389/feart.2021.676775

2. TANIMOTO, T., HEKI, K., AND ARTRU-LAMBIN, J. (2015). "Interaction of Solid Earth, Atmosphere, and Ionosphere.", Treatise Geophys. 4. 421–443. doi: 10.1016/B978-0-444-53802-4.00083-X

3. GELLER, R. J. (1997). "Earthquake Prediction: A Critical Review.", Geophys. J. Int. 131 (3), 425–450. doi:10.1111/j.1365-246X.1997.tb06588.x

4. GELLER, R. J., JACKSON, D. D., KAGAN, Y. Y., AND MULARGIA, F. (1997). "Earthquakes Cannot Be Predicted". Science 275 (5306), 1616. doi:10.1126/science.275.5306.1616

5. HOUGH, S. (2020). "The Great Quake Debate: The Crusader, the Skeptic, and the Rise of Modern Seismology." Seattle: University of Washington Press

6. KANAMORI, H. (2003). "Earthquake prediction: An overview". Int. Geophys. 81, 1205–1216. doi:10.1016/S0074-6142(03)80186-9

7. PULINETS, S., AND OUZOUNOV, D. (2011). "Lithosphere-Atmosphere-Ionosphere Coupling (LAIC) Model - a Unified Concept for Earthquake Precursors Validation." J. Asian Earth Sci. 41 (4–5), 371–382. doi: 10.1016/j.jseaes.2010.03.005

8. FREUND, F. T., TAKEUCHI, A., LAU, B. W. S., AL-MANASEER, A., FU, C. C., BRYANT, N. A., ET AL. (2007). "Stimulated Infrared Emission from Rocks: Assessing a Stress Indicator." eEarth 2 (1), 7–16. www.electronic-earth.net/2/7/2007/. doi:10.5194/ee-2-7-2007

9. KUO, C. L., HUBA, J. D., JOYCE, G., AND LEE, L. C. (2011). "Ionosphere Plasma Bubbles and Density Variations Induced by Pre-earthquake Rock Currents and Associated Surface Charges." J. Geophys. Res. 116 (10), a–n. doi:10.1029/2011JA016628

10. KUO, C. L., LEE, L. C., AND HUBA, J. D. (2014). "An Improved Coupling Model for the Lithosphere-Atmosphere-Ionosphere System." J. Geophys. Res. Space Phys. 119 (4), 3189–3205. doi:10.1002/2013JA019392

11. MIGULIN, V., LARKINA, V., MOLCHANOV, O., NALIVAIKO, A., GOKHBERG, M., LIPEROVSKY, V., ET AL. (1982). "Detection of the Earthquake Effects on the VLF-ELF Noises in the Upper Ionosphere." Preprint IZMIRAN, Institute of Terrestrial

Magnetism, Ionosphere and Radio Wave Propagation (IZMIRAN), USSR Academy of Sciences, Moscow 25 (390), 28

12. WAKITA, H., NAKAMURA, Y., NOTSU, K., NOGUCHI, M., AND ASADA, T. (1980). "Radon Anomaly: A Possible Precursor of the 1978 Izu-Oshima-Kinkai Earthquake." Science 207 (4433), 882–883. doi:10.1126/science.207.4433.882

13. TENG, T.-L., SUN, L.-F., AND MCRANEY, J. K. (1981). "Correlation of Groundwater Radon Anomalies with Earthquakes in the Greater Palmdale Bulge Area." Geophys. Res. Lett. 8 (5), 441–444. doi:10.1029/GL008i005p00441

14. CICERONE, R. D., EBEL, J. E., AND BRITTON, J. (2009). "A Systematic Compilation of Earthquake Precursors." Tectonophysics. 476, 371–396. doi:10.1016/ j. tecto.2009.06.008

15. CONTI, L., PICOZZA, P., AND SOTGIU, A. (2021). "A Critical Review of Ground Based Observations of Earthquake Precursors. "Front. Earth Sci. Sec. Geohazards Georisks. doi:10.3389/feart.2021.676766

16. KOIKE, K., YOSHINAGA, T., SUETSUGU, K., KASHIWAYA, K., AND ASAUE, H. (2015). "Controls on Radon Emission from Granite as Evidenced by Compression Testing to Failure." Geophys. J. Int. 203 (1), 428–436. doi:10.1093/gji/ggv290

17. PIERCE, E. T. (1976). "Atmospheric Electricity and Earthquake Prediction." Geophys. Res. Lett. 3 (3), 185–188. doi:10.1029/GL003i003p00185

18. SOROKIN, V. M., ISAEV, N. V., YASCHENKO, A. K., CHMYREV, V. M., AND HAYAKAWA, M. (2005). Strong DC "Electric Field Formation in the Low Latitude Ionosphere over Typhoons." J. Atmos. Solar-Terrestrial Phys. 67 (14), 1269–1279. doi: 10.1016/j.jastp.2005.06.014

19. KOREPANOV, V., HAYAKAWA, M., YAMPOLSKI, Y., AND LIZUNOV, G. (2009). "AGW as a Seismo-Ionospheric Coupling Responsible Agent." Phys. Chem. Earth, Parts A/B/C 34 (6–7), 485–495. doi: 10.1016/j.pce.2008.07.014

20. CHAKRABORTY, S., SASMAL, S., CHAKRABARTI, S. K., AND BHATTACHARYA, A. (2018). "Observational Signatures of Unusual Outgoing Longwave Radiation (OLR) and Atmospheric Gravity Waves (AGW) as Precursory Effects of May 2015 Nepal Earthquakes." J. Geodynamics 113, 43–51. doi: 10.1016/j.jog.2017.11.009

21. ZHANG, X., ZHAO, S., SONG, R., AND ZHAI, D. (2019). "The Propagation Features of LF Radio Waves at Topside Ionosphere and Their Variations Possibly Related to Wenchuan Earthquake in 2008." Adv. Space Res. 63 (11), 3536–3544. doi: 10.1016/j.asr.2019.02.008

22. ALESHINA, M. E., VORONOV, S. A., GAL'PER, A. M., KOLDASHOV, S. V., AND MASLENNIKOV, L. V. (1992). "Correlation between Earthquake Epicenters and Regions of High-Energy Particle Precipitations from the Radiation belt." Cosmic Res. 30 (1), 65–68.

23. GALPERIN, Y. I., GLADYSHEV, V. A., DZHORDZHIO, N. V., LARKINA, V. I., AND MOGILEVSKIJ, M. M. (1992). "Energetic Particles Precipitation from the Magnetosphere above the Epicenter of Approaching Earthquake." Kosmicheskie Issledovaniya 30 (1), 89–106.

24. PULINETS, S. A., OUZOUNOV, D., KARELIN, A. V., BOYARCHUK, K. A., AND POKHMELNYKH, L. A. (2006). "The Physical Nature of thermal Anomalies Observed before strong Earthquakes." Phys. Chem. Earth, Parts A/B/C 31 (4–9), 143–153. doi: 10.1016/j.pce.2006.02.042

25. LAGOUTTE, D., BROCHOT, J. Y., DE CARVALHO, D., ELIE, F., HARIVELO, F., HOBARA, Y., ET AL. (2006). "The DEMETER Science Mission Centre." Planet. Space Sci. 54 (5), 428–440. doi: 10.1016/j.pss.2005.10.014

26. ZHANG, X., SHEN, X., ZHAO, S., YAO, L., OUYANG, X., AND QIAN, J. (2014). "The Characteristics of Quasistatic Electric Field Perturbations Observed by DEMETER Satellite before Large Earthquakes." J. Asian Earth Sci. 79 (PA), 42–52. doi: 10.1016/j.jseaes.2013.08.026

27. PÍSA, D., PARROT, M., AND SANTOLÍK, O. (2011). "Ionospheric Density Variations Recorded before the 2010 Mw 8.8 Earthquake in Chile." J. Geophys. Res. Space Phys. 116 (8), 8309. doi:10.1029/2011JA016611

28. LI, M., AND PARROT, M. (2012). "Real Time Analysis of the Ion Density Measured by the Satellite DEMETER in Relation with the Seismic Activity." Nat. Hazards Earth Syst. Sci. 12 (9), 2957–2963. doi:10.5194/nhess-12-2957-2012

29. ZHU, TAO & ZHOU, JIANGUO & WANG, HONGQIANG. (2013). "Electromagnetic emissions during dilating fracture of a rock." Journal of Asian Earth Sciences. 73. 252–262. 10.1016/j.jseaes.2013.05.004

30. EDWARD L. AFRAIMOVICH, ELVIRA I. ASTAFYEVA, "TEC anomalies—Local TEC changes prior to earthquakes or TEC response to solar and geomagnetic activity changes?", Earth Planets Space, 60, 961–966, 2008

31. https://www.uninettunouniversity.net/it/p1_cses-limadou.aspx

32. SGRIGNA, V. & BUZZI, A. & CONTI, LIVIO & PICOZZA, PIERGIORGIO & STAGNI, C. & ZILPIMIANI, D. (2008). "The ESPERIA satellite project for detecting seismo-associated effects in the topside ionosphere. First instrumental tests in space." Earth Planets and Space. 60. 463-475. 10.1186/BF03352813

33. M. PIERSANTI ET AL., "Magnetospheric–Ionospheric–Lithospheric Coupling Model. 1: Observations during the 5 August 2018 Bayan Earthquake", MDPI, remote sensing, 11 October 2020

34. PULINETS, S.A.; OUZOUNOV, D.P. "Lithosphere–atmosphere–ionosphere coupling (LAIC) model—A unified concept for earthquake precursors validation." J. Asian Earth Sci. 2011, 41, 371–382. doi: 10.1016/j.jseaes.2010.03.005

35. PULINETS, S.A.; OUZOUNOV, D.P.; KARELIN, A.V.; DAVIDENKO, D.V. "Physical bases of the generation of short-term earthquake precursors: A complex model of ionization-induced geophysical processes in the lithosphere-atmosphere-ionosphere-magnetosphere system." Geomagn. Aeron. 2015, 55, 521–538. doi:10.1134/S0016793215040131

36. SOROKIN, V.M.; YASCHENKO, A.K.; HAYAKAWA, M. "Formation mechanism of the lower-ionospheric disturbances by the atmosphere electric current over a seismic region." J. Atmos. Sol.-Terr. Phys. 2006, 68, 1260–1268. doi: 10.1016/j.jastp.2006.03.005

37. HAYAKAWA, M.; KASAHARA, Y.; NAKAMURA, T.; HOBARA, Y.; ROZHNOI, A.; SOLOVIEVA, M.; MOLCHANOV, O.; KOREPANOV, V. "Atmospheric gravity waves as a possible candidate for seismo-ionospheric perturbation." J. Atmos. Electr. 2011, 31, 129–140. doi:10.1541/jae.31.129

38. MIYAKI, K.; HAYAKAWA, M.; MOLCHANOV, O.A. "The role of gravity waves in the lithosphere-ionosphere coupling, as revealed from the sub ionospheric LF propagation data. In Seismo Electromagnetics: Lithosphere-Atmosphere-Ionosphere Coupling." Hayakawa, M., Molchanov, O.A., Eds.; TERRAPUB: Tokyo, Japan, 2002; pp. 229–232

39. 11. MOLCHANOV, O.A.; HAYAKAWA, M.; MIYAKI, K. "VLF/LF sounding of the lower ionosphere to study the role of atmospheric oscillations in the lithosphere-ionosphere coupling." Adv. Polar Upper Atmos. Res. 2001, 15,146–158

40. MUTO, F.; KASAHARA, Y.; HOBARA, Y.; HAYAKAWA, M.; ROZHNOI, A.; SOLOVIEVA, M.; MOLCHANOV, O.A. "Further study on the role of atmospheric gravity waves on the seismo-ionospheric perturbations as detected by sub ionospheric VLF/LF propagation." Nat. Hazards Earth Syst. Sci. 2009, 9, 1111–1118.doi:10.5194/nhess-9-1111-2009

41. FREUND, F. "Time-resolved study of charge generation and propagation in igneous rocks." J. Geophys. Res. 2000, 105, 11001–11019. doi:10.1029/1999JB900423.

42. FREUND, F. "Pre-earthquake signals: Underlying physical processes." J. Asian Earth Sci. 2011, 41, 383–400. doi: 10.1016/j.jseaes.2010.03.009

43. PULINETS, S.A.; BOYARCHUK, K. "Ionospheric Precursors of Earthquakes." Springer: Berlin/Heidelberg, Germany, 2004

44. LIPEROVSKY, V.A.; POKHOTELOV, O.A.; MEISTER, C.-V.; LIPEROVSKAYA, E.V. "Physical models of coupling in the lithosphere-atmosphere-ionosphere system before earthquakes." Geomagn. Aeron. 2008, 48, 795–806. doi:10.1134/S0016793208060133

45. OYAMA, K.-I.; DEVI, M.; RYU, K.; CHEN, C.H.; LIU, J.Y.; LIU, H.; BANKOV, L.; KODAMA, T. "Modifications of the ionosphere prior to large earthquakes: Report from the ionospheric precursor study group." Geosci. Lett. 2016, 3, 6. doi:10.1186/s40562-016-0038-3

46. CAMPOS, "L.M.B.C. On three-dimensional acoustic-gravity waves in model non-isothermal atmospheres." Wave Motion 1983, 5, 1–14. doi:10.1016/0165-2125(83)90002-1

47. ACHENBACH, J.D. "Wave Propagation in Elastic Solids." Elsevier: New York, NY, USA, 1984

48. LAMB, H. "On Waves in an Elastic Plate." Proc. R. Soc. Lond. Ser. A 1917, 93, 114–128

49. PIERSANTI, M.; CESARONI, C.; SPOGLI, L.; ALBERTI, T. "Does TEC react to a sudden impulse as a whole? The 2015 Saint Patrick's Day storm event." Adv. Space Res. 2017, 60, 1807–1816. doi: 10.1016/j.asr.2017.01.021

50. BERUBE, D.; MOLDWIN, M.B.; AHN, M. "Computing magnetospheric mass density from field line resonances in a realistic magnetic field geometry." J. Geophys. Res. 2006, 111, A08206. doi:10.1029/2005JA011450. Remote Sens. 2020, 12, 3299 25 of 25

51. MENK, F.W.; KALE, Z.; SCIFFER, M.; ROBINSON, P.; WATERS, C.L.; GREW, R.; CLILVERD, M.; MANN, I. "Remote sensing the plasmasphere, plasmapause, plumes and other features using ground-based magnetometers." J. Space Weather Space Clim. 2014, 4, A34. doi:10.1051/swsc/2014030.

52. RANKIN, R.; TIKHONCHUK, V.T. "Dispersive shear Alfvén waves on model Tsyganenko magnetic field lines." Adv. Space Res. 2001, 28, 1595. doi:10.1016/S0273-1177(01)00481-1

53. SINGER, H.J.; SOUTHWOOD, D.J.; WALKER, R.J.; KIVELSON, M.G. "Alfven wave resonances in a realistic magnetospheric magnetic field geometry." J. Geophys. Res. 1981, 86, 4589. doi:10.1029/JA086iA06p04589

54. VELLANTE, M.; PIERSANTI, M.; HEILIG, B.; REDA, J.; DEL CORPO, A. (2014), "Magnetospheric plasma density inferred from field line resonances: Effects of using different magnetic field models." In Proceedings of the 2014 XXXIth URSI General Assembly and Scientific Symposium (URSI GASS), Beijing, China, 16–23 August 2014; pp. 1–4. doi:10.1109/URSIGASS.2014.6929941

55. WATERS, C.L.; SAMSON, J.C.; DONOVAN, E.F. "Variation of plasma trough density derived from magnetospheric field line resonances." J. Geophys. Res. 1996, 101, 24737–24745. doi:10.1029/96JA01083

56. WATERS, C.L.; KABIN, K.; RANKIN, R.; DONOVAN, E.; SAMSON, J.C. "Effects of the magnetic field model and wave polarization on the estimation of proton number densities in the magnetosphere." Planet. Space Sci. 2006, doi: 10.1016/j.pss.2006.04.041

57. WARNER, M.R., ORR, D. "Time of flight calculations for high latitude geomagnetic pulsations." Planet. Space Sci. 1979, 27, 679. doi:10.1016/0032-0633(79)90165-X

58. PIERSANTI, M.; VILLANTE, U.; WATERS, C.; COCO, I. "The 8 June 2000 ULF wave activity: A case study." J. Geophysics. Res. 2012, 117, A02204. doi:10.1029/2011JA016857

59. GOKHBERG, M.B.; NEKRASOV, A.K.; SHALIMOV, S.L. "Influence of unstable outputs of greenhouse gases on the ionosphere in seismically active regions." Izv. Phys. Solid Earth 1996, 32, 679–682

60. M. PIERSANTI ET AL., "Magnetospheric–Ionospheric–Lithospheric Coupling Model. 1: Observations during the 5 August 2018 Bayan Earthquake", MDPI, remote sensing, 11 October 2020

61. P. DIEGO ET AL.: "Plasma and Fields Evaluation at the Chinese Seismo-Electromagnetic Satellite for EFD Measurements", DOI 10.1109/ACCESS.2017.2674019, April 24, 2017

62. R. L. BOYD, "Langmuir probes on spacecraft, in Plasma Diagnostics." W. Lochte-Holtgreven, Ed. Amsterdam, The Netherlands: North Holland, 1968

63. D. BADONI ET AL., "An electric eld detector for high-performance measurements of the electric eld in the ionosphere." in Proc. 34th Int. Cosmic Ray Conf. (ICRC), The Hague, The Netherlands, Jul./Aug. 2015, p. 588

64. HAYAKAWA, M.; KASAHARA, Y.; NAKAMURA, T.; HOBARA, Y.; ROZHNOI, A.; SOLOVIEVA, M.; MOLCHANOV, O.; KOREPANOV, V. "Atmospheric gravity waves as a possible candidate for seismo-ionospheric perturbation." J. Atmos. Electr. 2011, 31, 129–140. doi:10.1541/jae.31.129

65. MIYAKI, K.; HAYAKAWA, M.; MOLCHANOV, O.A. "The role of gravity waves in the lithosphere-ionosphere coupling, as revealed from the sub-ionospheric LF propagation data. In Seismo Electromagnetics: Lithosphere-Atmosphere-Ionosphere Coupling." Hayakawa, M., Molchanov, O.A., Eds.; TERRAPUB: Tokyo, Japan, 2002; pp. 229–232

66. GOUSHEVA, M., DANOV, D., HRISTOV, P., AND MATOVA, M. (2009). "Ionospheric Quasi-Static Electric Field Anomalies during Seismic Activity in August-September 1981." Nat. Hazards Earth Syst. Sci. 9 (1), 3–15. doi:10.5194/nhess-9-3-2009

67. GOUSHEVA, M., DANOV, D., HRISTOV, P., AND MATOVA, M. (2008). "Quasi-static Electric fields Phenomena in the Ionosphere Associated with Pre- and post-Earthquake Effects." Nat. Hazards Earth Syst. Sci. 8 (1), 101–107. doi:10.5194/nhess-8-101-2008

68. CHUNG ET AL., "Electric Probes in Stationary and Flowing Plasmas: Theory and Application", 1975, Springer-Verlag New York Inc

69. DIEGO, P.; HUANG, J.; PIERSANTI, M.; BADONI, D.; ZEREN, Z.; YAN, R.; REBUSTINI, G.; AMMENDOLA, R.; CANDIDI, M.; GUAN, Y.-B.; ET AL. "The Electric Field Detector on Board the China Seismo Electromagnetic Satellite—In-Orbit Results and Validation." Instruments 2021, 5, 1. https://dx.doi.org/10.3390/instruments5010001

70. PIERSANTI, M.; PEZZOPANE, M.; ZHIMA, Z.; DIEGO, P.; XIONG, C.; TOZZI, R.; PIGNALBERI, A.; D'ANGELO, G.; BATTISTON, R.; HUANG, J.; et al. "Can an impulsive variation of the solar wind plasma pressure trigger a plasma bubble? A case study based on CSES, SWARM and THEMIS data." Adv. Space Res. 2020

71. http://cses.roma2.infn.it/

72. DE SANTIS CRISTIAN, AND RICCIARINI SERGIO, "The High Energy Particle Detector (HEPD-02) for the second China Seismo-Electromagnetic Satellite (CSES-02)", doi = "10.22323/1.395.0058","Proceedings of 37th International Cosmic Ray Conference, PoS (ICRC2021)",2021,"395","058"

73. KUNITSYN, VIACHESLAV E.; TERESHCHENKO, EVGENI D. (2003). "Ionospheric Tomography. Physics of Earth and Space Environments." Springer. p. 276. ISBN 9783540004042

74. DIEGO, P., ET AL., "Plasma and Fields Evaluation at the Chinese Seismo-Electromagnetic Satellite for Electric Field Detector Measurements." IEEE Access 2017, 5, 3824–3833

75. http://www.iaps.inaf.it/2020/09/30/cses-limadou-2

76. FRANÇOIS LEFEVRE, ELISABETH BLANC, JEAN-LOUIS PINÇON, TARANIS TEAM. "TARANIS-a Satellite Project Dedicated to the Physics of TLEs and TGFs." coupling of thunderstorms and lightning discharges to near-earth space: proceedings of the workshop", 2009, Corte, France. pp.3-7, ff10.1063/1.3137711ff. ffinsu-02926960

77. http://www.iaps.inaf.it/2019/04/24/laboratorio-di-fisica-del-plasma-e-camera-del-plasma

78. Diego, P., et al., "Plasma and Fields Evaluation at the Chinese Seismo-Electromagnetic Satellite for Electric Field Detector Measurements." IEEE Access 2017, 5, 3824–3833

79. Chen F. F., 2003, "Langmuir Probe Diagnostic, Lecture Notes - Mini-Course on Plasma Diagnostics.", IEEE-ICOPS meeting, Jeju, Korea, June 5, 2003

80. "Types of Orbits". *marine.rutgers.edu. Archived from* the original *on 22 August* 2019. Retrieved 24 June 2017

81. D. Badoni, R. Ammendola, I. Bertello, P. Cipollone, L. Conti, C. De Santis, P. Diego, G. Masciantonio, P. Picozza, R. Sparvoli, P. Ubertini, G. Vannaroni,"A high-performance electric field detector for space missions "Planetary and Space Science, Volume 153, 2018, Pages 107-119, ISSN 0032-0633, https://doi.org/10.1016/j.pss.2018.01.013

82. J.J. Berthelier, M. Godefroy, F. Leblanc, M. Malingre, M. Menvielle, D. Lagoutte, J.Y. Brochot, F. Colin, F. Elie, C. Legendre, P. Zamora, D. Benoist, Y. Chapuis, J. Artru, R. Pfaff, "ICE, the electric field experiment on DEMETER, Planetary and Space Science", Volume 54, Issue 5, 2006, Pages 456-471, ISSN 0032-0633, https://doi.org/10.1016/j.pss.2005.10.016

83. https://cses.web.roma2.infn.it/?page_id=203

84. Rebustini, G. and Ammendola, R. and Badoni, D. and De Santis, C. and Diego, P., A. Parmentier (2022) Calibration of the Electric Field Detector for the CSES-02 satellite. Il nuovo Cimento C, 45 (6). pp. 1-4. ISSN 1826-9885

85. http://demeter.cnrs-orleans.fr/

86. Matteo Martucci et al., CSES/Limadou Collaboration, "New results on protons inside the South Atlantic Anomaly, at energies between 40–250 MeV in the period 2018-2020, from the CSES-01 satellite mission" DOI: 10.1103/PhysRevD.105.062001

87. S. Bartocci et al., "Deep learning-based event reconstruction for the Limadou High-Energy Particle Detector", Physical Review D 105, 022004 –published 21 January 2022, https://link.aps.org/doi/10.1103/PhysRevD.105.022004

88. G. Ambrosi et al., "The electronics of the High-Energy Particle Detector on board the CSES-01 satellite", Nuclear Instruments and Methods in Physics Research Section A: Accelerators, Spectrometers, Detectors and Associated

Equipment, Volume 1013, 11 October 2021, 165639. https://doi.org/10.1016/j.nima.2021.165639.

89. PALMA F, ET AL., "The August 2018 Geomagnetic Storm Observed by the High-Energy Particle Detector on Board the CSES-01 Satellite", Appl. Sci. (2021) 11, 5680, DOI 10.3390/app11125680

90. M. MARTUCCI ET AL., "Trapped proton fluxes estimation inside the South Atlantic Anomaly using the NASA AE9/AP9/SPM radiation models along the China Seismo-Electromagnetic Satellite orbit", Applied Sciences, April2021, 11(8), 3465; https://doi.org/10.3390/app11083465

91. A. SOTGIU ET AL., "Control and data acquisition software of the high-energy particle detector on board the China Seismo-Electromagnetic Satellite space mission", Wiley Online Library, December 2020, https://doi.org/10.1002/spe.2947

92. G. AMBROSI ET AL., "Beam test calibrations of the HEPD detector on board the China Seismo-Electromagnetic Satellite", Nuclear Inst. and Methods in Physics Research, A 974 (2020), https://doi.org/10.1016/j.nima.2020.164170

93. P. PICOZZA ET AL., "Scientific Goal and In-orbit performance of the High-Energy Particle Detector on board the CSES", The Astrophysical Journal Supplement Series, 243:16 (17pp), 2019 July, https://doi.org/10.3847/1538-4365/ab276c.

94. Z. ZHIMA, ET AL., (2021), "Storm-Time Features of the Ionospheric ELF/VLF Waves and Energetic Electron Fluxes Revealed by the China Seismo-Electromagnetic Satellite", Appl. Sci. 11:2617.

95. BARTOCCI, S. ET AL. (2020), "Galactic Cosmic-Ray Hydrogen Spectra in the 40-250 MeV Range Measured by the High-energy Particle Detector (HEPD) on board the CSES-01 Satellite between 2018 and 2020". The Astrophysical Journal. 901. 8. 10.3847/1538-4357/abad3e.